\documentclass[twocolumn,tighten]{aastex63}

\usepackage{graphicx}
\usepackage{amsmath}
\usepackage{comment}
\usepackage{array,multirow,color,tablefootnote}
\usepackage{xcolor}

\newcommand{\rsun}{\ensuremath{R_\sun}}
\newcommand{\msun}{\ensuremath{M_\sun}}
\newcommand{\lsun}{\ensuremath{L_\sun}}
\newcommand{\rjup}{\ensuremath{R_{\rm Jup}}}
\newcommand{\mjup}{\ensuremath{M_{\rm Jup}}}
\newcommand{\teff}{\ensuremath{T_{\rm eff}}}
\newcommand{\teq}{\ensuremath{T_{\rm eq}}}
\newcommand{\loggstar}{\ensuremath{\log{g_{*}}}}
\newcommand{\feh}{\ensuremath{{\rm [Fe/H]}}}

\newcommand{\loggpl}{\ensuremath{\log{g_{p}}}}
\newcommand{\kms}{\ensuremath{\rm km\,s^{-1}}}
\newcommand{\ms}{\ensuremath{\rm m\,s^{-1}}}
\newcommand{\vsini}{\ensuremath{v\sin{i_{*}}}}

\newcommand{\tess}{TESS}

\newcommand{\Ks}{\ensuremath{K_s}}
\newcommand{\masy}{\ensuremath{\rm mas\,yr^{-1}}}

\newcommand{\mss}{\ensuremath{\rm m\,s^{-2}}}
\newcommand{\gcmc}{\ensuremath{\rm g\,cm^{-3}}}

\shorttitle{The TOI-2109 System}
\shortauthors{Wong et~al.}

\begin{document}

\title{TOI-2109\MakeLowercase{b}: An Ultrahot Gas Giant on a 16 hr Orbit}

\author[0000-0001-9665-8429]{Ian~Wong}
\altaffiliation{NASA Postdoctoral Program Fellow}
\affiliation{Department of Earth, Atmospheric and Planetary Sciences, Massachusetts Institute of Technology, Cambridge, MA 02139, USA}
\affiliation{NASA Goddard Space Flight Center, 8800 Greenbelt Rd, Greenbelt, MD 20771, USA}
\correspondingauthor{Ian Wong}
\email{iwong@mit.edu}

\author[0000-0002-1836-3120]{Avi~Shporer}
\affiliation{Department of Physics and Kavli Institute for Astrophysics and Space Research, Massachusetts Institute of Technology, Cambridge, MA 02139, USA}

\author[0000-0002-4891-3517]{George~Zhou}
\affiliation{Centre for Astrophysics, University of Southern Queensland, West Street, Toowoomba, QLD 4350, Australia}

\author[0000-0003-4269-3311]{Daniel~Kitzmann}
\affiliation{Center for Space and Habitability, University of Bern, Bern, Switzerland}

\author[0000-0002-9258-5311]{Thaddeus~D.~Komacek}
\altaffiliation{51 Pegasi b Fellow}
\affiliation{Department of Astronomy, University of Maryland, College Park, MD 20742, USA}
\affiliation{Department of the Geophysical Sciences, The University of Chicago, Chicago, IL 60637, USA}

\author[0000-0003-2278-6932]{Xianyu~Tan}
\affiliation{Atmospheric, Oceanic and Planetary Physics, Department of Physics, University of Oxford, Oxford OX1 3PU, UK}

\author[0000-0003-1001-0707]{Ren\'{e}~Tronsgaard}
\affiliation{DTU Space, National Space Institute, Technical University of Denmark, Elektrovej 328, DK-2800 Kgs. Lyngby, Denmark}

\author[0000-0003-1605-5666]{Lars~A.~Buchhave}
\affiliation{DTU Space, National Space Institute, Technical University of Denmark, Elektrovej 328, DK-2800 Kgs. Lyngby, Denmark}

\author[0000-0003-2527-1475]{Shreyas~Vissapragada}
\affiliation{Division of Geological and Planetary Sciences, California Institute of Technology, 1200 East California Blvd, Pasadena, CA 91125, USA}

\author[0000-0002-0371-1647]{Michael~Greklek-McKeon}
\affiliation{Division of Geological and Planetary Sciences, California Institute of Technology, 1200 East California Blvd, Pasadena, CA 91125, USA}

\author[0000-0001-8812-0565]{Joseph~E.~Rodriguez}
\affiliation{Department of Physics and Astronomy, Michigan State University, East Lansing, MI 48824, USA}

\author[0000-0003-2086-7712]{John~P.~Ahlers}
\altaffiliation{NASA Postdoctoral Program Fellow}
\affiliation{Exoplanets and Stellar Astrophysics Laboratory, Code 667, NASA Goddard Space Flight Center, Greenbelt, MD 20771, USA}
\affiliation{GSFC Sellers Exoplanet Environments Collaboration}

\author[0000-0002-8964-8377]{Samuel~N.~Quinn}
\affiliation{Center for Astrophysics ${\rm \mid}$ Harvard {\rm \&} Smithsonian, 60 Garden Street, Cambridge, MA 02138, USA}

\author[0000-0001-9800-6248]{Elise~Furlan}
\affiliation{NASA Exoplanet Science Institute, Caltech/IPAC, Mail Code 100-22, 1200 E. California Blvd., Pasadena, CA 91125, USA}

\author[0000-0002-2532-2853]{Steve~B.~Howell}
\affiliation{NASA Ames Research Center, Moffett Field, CA 94035, USA}

\author[0000-0001-6637-5401]{Allyson~Bieryla} 
\affiliation{Center for Astrophysics ${\rm \mid}$ Harvard {\rm \&} Smithsonian, 60 Garden Street, Cambridge, MA 02138, USA}

\author[0000-0003-1907-5910]{Kevin~Heng}
\affiliation{Center for Space and Habitability, University of Bern, Bern, Switzerland}
\affiliation{Department of Physics, Astronomy and Astrophysics Group, University of Warwick, Coventry CV4 7AL, UK}
\affiliation{University Observatory Munich, Ludwig Maximilian University, Scheinerstrasse 1, Munich D-81679, Germany}

\author[0000-0002-5375-4725]{Heather~A.~Knutson}
\affiliation{Division of Geological and Planetary Sciences, California Institute of Technology, 1200 East California Blvd, Pasadena, CA 91125, USA}

\author[0000-0001-6588-9574]{Karen~A.~Collins}
\affiliation{Center for Astrophysics ${\rm \mid}$ Harvard {\rm \&} Smithsonian, 60 Garden Street, Cambridge, MA 02138, USA}

\author[0000-0001-9504-1486]{Kim~K.~McLeod}
\affiliation{Department of Astronomy, Wellesley College, Wellesley, MA 02481, USA}

\author{Perry~Berlind}
\affiliation{Center for Astrophysics ${\rm \mid}$ Harvard {\rm \&} Smithsonian, 60 Garden Street, Cambridge, MA 02138, USA}

\author[0000-0002-2546-9708]{Peyton~Brown} 
\affiliation{Vanderbilt University, 2201 West End Ave, Nashville, TN 37235, USA}

\author[0000-0002-2830-5661]{Michael~L.~Calkins}
\affiliation{Center for Astrophysics ${\rm \mid}$ Harvard {\rm \&} Smithsonian, 60 Garden Street, Cambridge, MA 02138, USA}

\author[0000-0002-6424-3410]{Jerome~P.~de~Leon}
\affiliation{Department of Astronomy, The University of Tokyo, 7-3-1 Hongo, Bunkyo-ku, Tokyo 113-0033, Japan}

\author[0000-0002-2341-3233]{Emma~Esparza-Borges}
\affiliation{Instituto de Astrof\'isica de Canarias, E-38205 La Laguna, Tenerife, Spain}
\affiliation{Departamento de Astrof\'isica, Universidad de La Laguna, E-38206 La Laguna, Tenerife, Spain}

\author[0000-0002-9789-5474]{Gilbert~A.~Esquerdo}
\affiliation{Center for Astrophysics ${\rm \mid}$ Harvard {\rm \&} Smithsonian, 60 Garden Street, Cambridge, MA 02138, USA}

\author[0000-0002-4909-5763]{Akihiko~Fukui}
\affiliation{Instituto de Astrof\'isica de Canarias, E-38205 La Laguna, Tenerife, Spain}
\affiliation{Komaba Institute for Science, The University of Tokyo, 3-8-1 Komaba, Meguro, Tokyo 153-8902, Japan}

\author[0000-0002-4503-9705]{Tianjun~Gan}
\affiliation{Department of Astronomy and Tsinghua Centre for Astrophysics, Tsinghua University, Beijing 100084, China}

\author[0000-0002-5443-3640]{Eric~Girardin}
\affiliation{Grand Pra Observatory, 1984 Les Haud{\`e}res, Switzerland}

\author[0000-0003-2519-6161]{Crystal~L.~Gnilka}
\affiliation{NASA Ames Research Center, Moffett Field, CA 94035, USA}
\affiliation{NASA Exoplanet Science Institute, Caltech/IPAC, Mail Code 100-22, 1200 E. California Blvd., Pasadena, CA 91125, USA}

\author[0000-0002-5658-5971]{Masahiro~Ikoma}
\affiliation{National Astronomical Observatory of Japan, 2-21-1 Osawa, Mitaka, Tokyo 181-8588, Japan}

\author[0000-0002-4625-7333]{Eric~L.~N.~Jensen}
\affiliation{Department of Physics and Astronomy, Swarthmore College, Swarthmore, PA 19081, USA}

\author[0000-0003-0497-2651]{John~Kielkopf}
\affiliation{Department of Physics and Astronomy, University of Louisville, Louisville, KY 40292, USA}

\author[0000-0001-9032-5826]{Takanori~Kodama}
\affiliation{Komaba Institute for Science, The University of Tokyo, 3-8-1 Komaba, Meguro, Tokyo 153-8902, Japan}

\author{Seiya~Kurita}
\affiliation{Department of Earth and Planetary Science, Graduate School of Science, The University of Tokyo, 7-3-1 Hongo, Bunkyo-ku, Tokyo 113-0033, Japan}

\author[0000-0002-9903-9911]{Kathryn~V.~Lester}
\affiliation{NASA Ames Research Center, Moffett Field, CA 94035, USA}

\author[0000-0003-0828-6368]{Pablo~Lewin}
\affiliation{The Maury Lewin Astronomical Observatory, Glendora, CA 91741, USA}

\author[0000-0001-8134-0389]{Giuseppe~Marino}
\affiliation{Wild Boar Remote Observatory, Florence, Italy}
\affiliation{Gruppo Astrofili Catanesi, Catania, Italy}

\author[0000-0001-9087-1245]{Felipe~Murgas}
\affiliation{Instituto de Astrof\'isica de Canarias, E-38205 La Laguna, Tenerife, Spain}
\affiliation{Departamento de Astrof\'isica, Universidad de La Laguna, E-38206 La Laguna, Tenerife, Spain}

\author[0000-0001-8511-2981]{Norio~Narita}
\affiliation{Instituto de Astrof\'isica de Canarias, E-38205 La Laguna, Tenerife, Spain}
\affiliation{Komaba Institute for Science, The University of Tokyo, 3-8-1 Komaba, Meguro, Tokyo 153-8902, Japan}
\affiliation{Japan Science and Technology Agency, PRESTO, 3-8-1 Komaba, Meguro, Tokyo 153-8902, Japan}
\affiliation{Astrobiology Center, 2-21-1 Osawa, Mitaka, Tokyo 181-8588, Japan}

\author[0000-0003-0987-1593]{Enric~Pall{\'e}}
\affiliation{Instituto de Astrof\'isica de Canarias, E-38205 La Laguna, Tenerife, Spain}
\affiliation{Departamento de Astrof\'isica, Universidad de La Laguna, E-38206 La Laguna, Tenerife, Spain}


\author[0000-0001-8227-1020]{Richard~P.~Schwarz}
\affiliation{Patashnick Voorheesville Observatory, Voorheesville, NY 12186, USA}


\author[0000-0002-3481-9052]{Keivan~G.~Stassun} 
\affiliation{Department of Physics and Astronomy, Vanderbilt University, Nashville, TN 37235, USA}
\affiliation{Department of Physics, Fisk University, Nashville, TN 37208, USA}

\author[0000-0002-6510-0681]{Motohide~Tamura}
\affiliation{Department of Astronomy, The University of Tokyo, 7-3-1 Hongo, Bunkyo-ku, Tokyo 113-0033, Japan}
\affiliation{National Astronomical Observatory of Japan, 2-21-1 Osawa, Mitaka, Tokyo 181-8588, Japan}
\affiliation{Astrobiology Center, 2-21-1 Osawa, Mitaka, Tokyo 181-8588, Japan}

\author[0000-0002-7522-8195]{Noriharu~Watanabe}
\affiliation{Department of Multi-Disciplinary Sciences, Graduate School of Arts and Sciences, The University of Tokyo, 3-8-1 Komaba, Meguro, Tokyo 153-8902, Japan}


\author[0000-0001-5578-1498]{Bj{\" o}rn~Benneke}
\affiliation{Department of Physics and Institute for Research on Exoplanets, Universit{\' e} de Montr{\' e}al, Montr{\' e}al, QC, Canada}


\author[0000-0003-2058-6662]{George~R.~Ricker}
\affiliation{Department of Physics and Kavli Institute for Astrophysics and Space Research, Massachusetts Institute of Technology, Cambridge, MA 02139, USA}

\author[0000-0001-9911-7388]{David~W.~Latham}
\affiliation{Center for Astrophysics ${\rm \mid}$ Harvard {\rm \&} Smithsonian, 60 Garden Street, Cambridge, MA 02138, USA}

\author[0000-0001-6763-6562]{Roland~Vanderspek}
\affiliation{Department of Physics and Kavli Institute for Astrophysics and Space Research, Massachusetts Institute of Technology, Cambridge, MA 02139, USA}

\author[0000-0002-6892-6948]{Sara~Seager}
\affiliation{Department of Earth, Atmospheric and Planetary Sciences, Massachusetts Institute of Technology, Cambridge, MA 02139, USA}
\affiliation{Department of Physics and Kavli Institute for Astrophysics and Space Research, Massachusetts Institute of Technology, Cambridge, MA 02139, USA}
\affiliation{Department of Aeronautics and Astronautics, Massachusetts Institute of Technology, Cambridge, MA 02139, USA}

\author[0000-0002-4265-047X]{Joshua~N.~Winn}
\affiliation{Department of Astrophysical Sciences, Princeton University, Princeton, NJ 08544, USA}

\author[0000-0002-4715-9460]{Jon~M.~Jenkins}
\affiliation{NASA Ames Research Center, Moffett Field, CA 94035, USA},


\author[0000-0003-1963-9616]{Douglas~A.~Caldwell}
\affiliation{NASA Ames Research Center, Moffett Field, CA 94035, USA}
\affiliation{SETI Institute, 189 Bernardo Ave, Suite 200, Mountain View, CA 94043, USA}

\author[0000-0003-0241-2757]{William~Fong}
\affiliation{Department of Physics and Kavli Institute for Astrophysics and Space Research, Massachusetts Institute of Technology, Cambridge, MA 02139, USA}

\author[0000-0003-0918-7484]{Chelsea~X.~Huang}
\altaffiliation{Juan Carlos Torres Fellow}
\affiliation{Department of Physics and Kavli Institute for Astrophysics and Space Research, Massachusetts Institute of Technology, Cambridge, MA 02139, USA}

\author[0000-0002-4510-2268]{Ismael~Mireles}
\affiliation{Department of Physics and Astronomy, University of New Mexico, 210 Yale Blvd NE, Albuquerque, NM 87106, USA}

\author[0000-0001-5347-7062]{Joshua~E.~Schlieder}
\affiliation{NASA Goddard Space Flight Center, 8800 Greenbelt Rd, Greenbelt, MD 20771, USA}

\author[0000-0001-7842-3714]{Bernie~Shiao}
\affiliation{Space Telescope Science Institute, 3700 San Martin Drive, Baltimore, MD 21218, USA}

\author{Jesus~Noel~Villase{\~n}or}
\affiliation{Department of Physics and Kavli Institute for Astrophysics and Space Research, Massachusetts Institute of Technology, Cambridge, MA 02139, USA}

\begin{abstract}
We report the discovery of an ultrahot Jupiter with an extremely short orbital period of $0.67247414\,\pm\,0.00000028$ days ($\sim$16 hr). The $1.347 \pm 0.047$ \rjup\ planet, initially identified by the Transiting Exoplanet Survey Satellite (TESS) mission, orbits TOI-2109 (TIC 392476080) --- a $\teff \sim 6500$ K F-type star with a mass of $1.447 \pm 0.077$ \msun, a radius of $1.698 \pm 0.060$ \rsun, and a rotational velocity of $\vsini = 81.9 \pm 1.7$ \kms. The planetary nature of TOI-2109b was confirmed through radial velocity measurements, which yielded a planet mass of $5.02 \pm 0.75$ \mjup. Analysis of the Doppler shadow in spectroscopic transit observations indicates a well-aligned system, with a sky-projected obliquity of $\lambda = 1\overset{\circ}{.}7 \pm 1\overset{\circ}{.}7$. From the \tess\ full-orbit light curve, we measured a secondary eclipse depth of $731 \pm 46$ ppm, as well as phase-curve variations from the planet's longitudinal brightness modulation and ellipsoidal distortion of the host star. Combining the \tess-band occultation measurement with a \Ks-band secondary eclipse depth ($2012 \pm 80$ ppm) derived from ground-based observations, we find that the dayside emission of TOI-2109b is consistent with a brightness temperature of $3631 \pm 69$ K, making it the second hottest exoplanet hitherto discovered. By virtue of its extreme irradiation and strong planet--star gravitational interaction, TOI-2109b is an exceptionally promising target for intensive follow-up studies using current and near-future telescope facilities to probe for orbital decay, detect tidally driven atmospheric escape, and assess the impacts of H$_2$ dissociation and recombination on the global heat transport.
\end{abstract}

\section{Introduction}
\label{sec:intro}

Studies of exoplanet demographics show that hot Jupiters (i.e., short-period gas giants) are extremely rare, occurring around just $\sim$0.5\% of Sun-like stars \citep[e.g.,][]{howard2012,wright2012,masuda2017,zhou2019}. However, despite their relative scarcity, these objects have played an outsized role in developing our current understanding of exoplanet atmospheres, which in turn has significantly shaped theories of planet formation, evolution, and dynamics. Their large size in relation to their host stars and high temperatures enable a broad range of intensive studies that extend far beyond the rudimentary measurements of planet mass and radius. Over the past two decades, a wide arsenal of observational techniques has been leveraged to probe the atmospheric properties of hot Jupiters in ever-increasing detail, including the longitudinal and vertical temperature distribution, the chemical composition on both global and local scales, the prevalence of condensate clouds and photochemical hazes, and the underlying physical processes driving heat transport across the atmosphere (see, for example, the reviews in \citealt{crossfield2019} and \citealt{madhusudhan2019}). 

In recent years, the subset of hot Jupiters located at the most extreme end of the observed temperature range has attracted special attention. These so-called ultrahot Jupiters, with dayside temperatures exceeding $\sim$2500 K \citep[e.g.,][]{bell2018,parmentier2018}, are characterized by a number of distinct physical and dynamical properties that set them apart from the rest of the hot gas-giant population. Some notable examples of ultrahot Jupiters include KELT-9b (the hottest known exoplanet; \citealt{kelt9}), WASP-12b \citep{wasp12}, and WASP-33b \citep{colliercameron2010}. The intense stellar irradiation of ultrahot Jupiters is sufficient to dissociate most molecular species found in exoplanet atmospheres, including H$_2$, resulting in a dayside hemisphere primarily composed of atomic and ionic gases \citep[e.g.,][]{arcangeli2018,bell2018,hoeijmakers2018}. The enhanced short-wavelength opacity from refractory elements (e.g., Fe and Mg) and dissociated H$^-$, along with the concomitant destruction of molecules responsible for radiative cooling (e.g., H$_2$O), is expected to create high-altitude temperature inversions across the dayside atmospheres of ultrahot Jupiters \citep[e.g.,][]{kitzmann2018,lothringer2018,parmentier2018}, as well as largely featureless near-infrared emission spectra \citep[e.g.,][]{arcangeli2018,kreidberg2018,mansfield2018}.

Theoretical and numerical modeling of ultrahot Jupiter atmospheres has further demonstrated that the dissociation of H$_2$ on the dayside and its recombination on the cooler nightside greatly amplify the efficiency of day--night heat circulation, thereby dampening the temperature contrast between the two hemispheres \citep[e.g.,][]{bell2018,tan2019}. The large-scale atmospheric dynamics of an exoplanet can be directly probed by measuring the brightness of the object across a full orbit, from which the longitudinal temperature distribution and global energy budget can be deduced \citep[e.g.,][]{cowanagol,parmentier2017}. Such phase-curve observations have been carried out at near-infrared wavelengths for a sizable fraction of the known ultrahot Jupiters orbiting bright stars \citep[e.g.,][]{bell2021}, including KELT-9b \citep{mansfield2020}, WASP-33b \citep{zhang2018}, and WASP-103b \citep{kreidberg2018}. The results of these studies have generally corroborated the prediction of relatively modest day--night temperature contrasts when compared to cooler hot Jupiters.

The high temperatures of ultrahot Jupiters make them uniquely amenable to visible-light phase-curve studies as well, and previous works have taken advantage of the near-continuous long-baseline temporal coverage of Kepler and \tess\ to carry out systematic phase-curve analyses \citep[e.g.,][]{esteves2013,esteves2015,wong2020year1,wong2021year2}. At these wavelengths, the large masses and close-in orbits of ultrahot Jupiters induce additional synchronous variations in the host stars' brightness that are detectable in high-quality time-series photometry. The amplitudes of these signals, which stem from the tidal distortion of the stellar surface and the Doppler shifting of the star's spectrum, provide information about the mutual planet--star gravitational interaction and the astrophysical properties of the host star \citep[e.g.,][]{faigler2011,faigler2015,shporer2017}.


The close proximity of ultrahot Jupiters to their host stars and the correspondingly powerful gravitational forces can lead to significant deformations of the planets' equilibrium shapes \citep[e.g.,][]{budaj2011,li2010} and, in the most extreme scenarios, mass loss through atmospheric stripping \citep[e.g.,][]{jackson2016}, which has been observed in a few systems \citep[e.g.,][]{haswell2012,yan2018,bell2019}. In addition, the strong planet--star tidal interaction in ultrahot Jupiter systems can drive rapid orbital decay that may be discernible within decade-long timescales \citep[e.g.,][]{rasio1996,sasselov2003}, as in the case of WASP-12b \citep{m16,patra2017,patra2020,yee2020,turner2021}. The measurement of orbital decay provides another window into the astrophysical properties of the host star.

The previous discussion underscores how observations of ultrashort-period gas giants can offer a wealth of information about both the planets and the host stars. While future advances in telescope capabilities will allow for comparably in-depth explorations of smaller and cooler exoplanets, ultrahot Jupiters will continue to be among the most fruitful candidates for impactful efforts at characterization, providing crucial insights into the nature of planets at their most extreme.


\begin{deluxetable}{lcc}[t]
\tablewidth{0pc}
\setlength{\tabcolsep}{10pt}
\renewcommand{\arraystretch}{1.0}
\tabletypesize{\footnotesize}
\tablecaption{
    Target Information
    \label{tab:info}
}
\tablehead{ \vspace{-0.2cm} \\
    \multicolumn{1}{l}{Parameter} &
    \multicolumn{1}{c}{Value}    &
    \multicolumn{1}{c}{Source}   
}
\startdata
TIC & 392476080 & TIC V8\tablenotemark{\scriptsize a}\\
R.A. & 16$^{\rm h}$52$^{\rm m}$45$^{\rm s}$ & Gaia DR2\tablenotemark{\scriptsize b} \\
Decl. & $+16^{\circ}34'48''$  & Gaia DR2\tablenotemark{\scriptsize b} \\
$\mu_{\rm ra}$ (\masy)  & $-$8.449 $\pm$ 0.043 & Gaia DR2\tablenotemark{\scriptsize b} \\
$\mu_{\rm dec}$ (\masy) & $-$9.257 $\pm$ 0.051 & Gaia DR2\tablenotemark{\scriptsize b} \\
Parallax (mas) & 3.788 $\pm$ 0.039 & Gaia DR2\tablenotemark{\scriptsize b} \\
Distance (pc)  & 262.04 $\pm$ 2.73 & Gaia DR2\tablenotemark{\scriptsize b} \\
Epoch & 2015.5 & Gaia DR2\tablenotemark{\scriptsize b}\\
$B_T$ (mag)    & 10.731 $\pm$ 0.032 & Tycho-2\tablenotemark{\scriptsize c} \\
$V_T$ (mag)    & 10.268 $\pm$ 0.029 & Tycho-2\tablenotemark{\scriptsize c} \\
$G_{\mathrm{BP}}$ (mag) & 10.3638 $\pm$ 0.0011 & Gaia DR2\tablenotemark{\scriptsize b} \\
$G$ (mag) & 10.11376 $\pm$ 0.00034 & Gaia DR2\tablenotemark{\scriptsize b} \\
$G_{\mathrm{RP}}$ (mag) & 9.73916 $ \pm $ 0.00094 & Gaia DR2\tablenotemark{\scriptsize b} \\
\tess\ (mag) & 9.7857 $\pm$ 0.0061 & TIC V8\tablenotemark{\scriptsize a}\\
$J$ (mag)    & 9.382 $\pm$ 0.024 & 2MASS\tablenotemark{\scriptsize d} \\
$H$ (mag)    & 9.129 $\pm$ 0.026 & 2MASS\tablenotemark{\scriptsize d} \\
$K$ (mag)    & 9.070 $\pm$ 0.021 & 2MASS\tablenotemark{\scriptsize d} \\
$W1$ (mag) & 9.059 $\pm$ 0.023 & WISE\tablenotemark{\scriptsize e} \\
$W2$ (mag) & 9.093 $\pm$ 0.020 & WISE\tablenotemark{\scriptsize e} \\
$W3$ (mag) & 9.062 $\pm$ 0.030 & WISE\tablenotemark{\scriptsize e} \\
\enddata
\textbf{Notes.}
\vspace{-0.15cm}\tablenotetext{\textrm{a}}{\cite{stassun18a}.}
\vspace{-0.15cm}\tablenotetext{\textrm{b}}{\cite{gaia18}.}
\vspace{-0.15cm}\tablenotetext{\textrm{c}}{\cite{tycho}.}
\vspace{-0.15cm}\tablenotetext{\textrm{d}}{\cite{cutri03}.}
\vspace{-0.15cm}\tablenotetext{\textrm{e}}{\cite{cutri13}.}
\vspace{-1cm}
\end{deluxetable}

In this paper, we describe a newly discovered transiting ultrahot Jupiter --- TOI-2109b --- which has the shortest orbital period of any known gas-giant exoplanet at the time of this writing. This target was initially identified as a planet candidate from data obtained by the \tess\ mission \citep{ricker2014}. We have carried out an intensive year-long campaign of follow-up observations to confirm and characterize the planet. The layout of the paper is as follows. Section~\ref{sec:obs} summarizes the body of observations, which includes the \tess\ light curve, ground-based transit and secondary eclipse photometry, radial velocity monitoring of the orbit, high-angular-resolution imaging, and spectroscopic transit observations. Stellar characterization of the host star is described in Section~\ref{sec:star}, and the results of our fits to the various data sets are presented in Section~\ref{sec:ana}. In Section~\ref{sec:dis}, we delve into the broader implications of our discovery, with a focus on the planet's atmospheric properties, the planet--star tidal interaction, orbital decay, and prospects for further atmospheric study with current and near-future facilities. We conclude with a brief summary in Section~\ref{sec:sum}.

\section{Observations}
\label{sec:obs}

\begin{figure*}[t]
\centering
\includegraphics[width=0.75\linewidth]{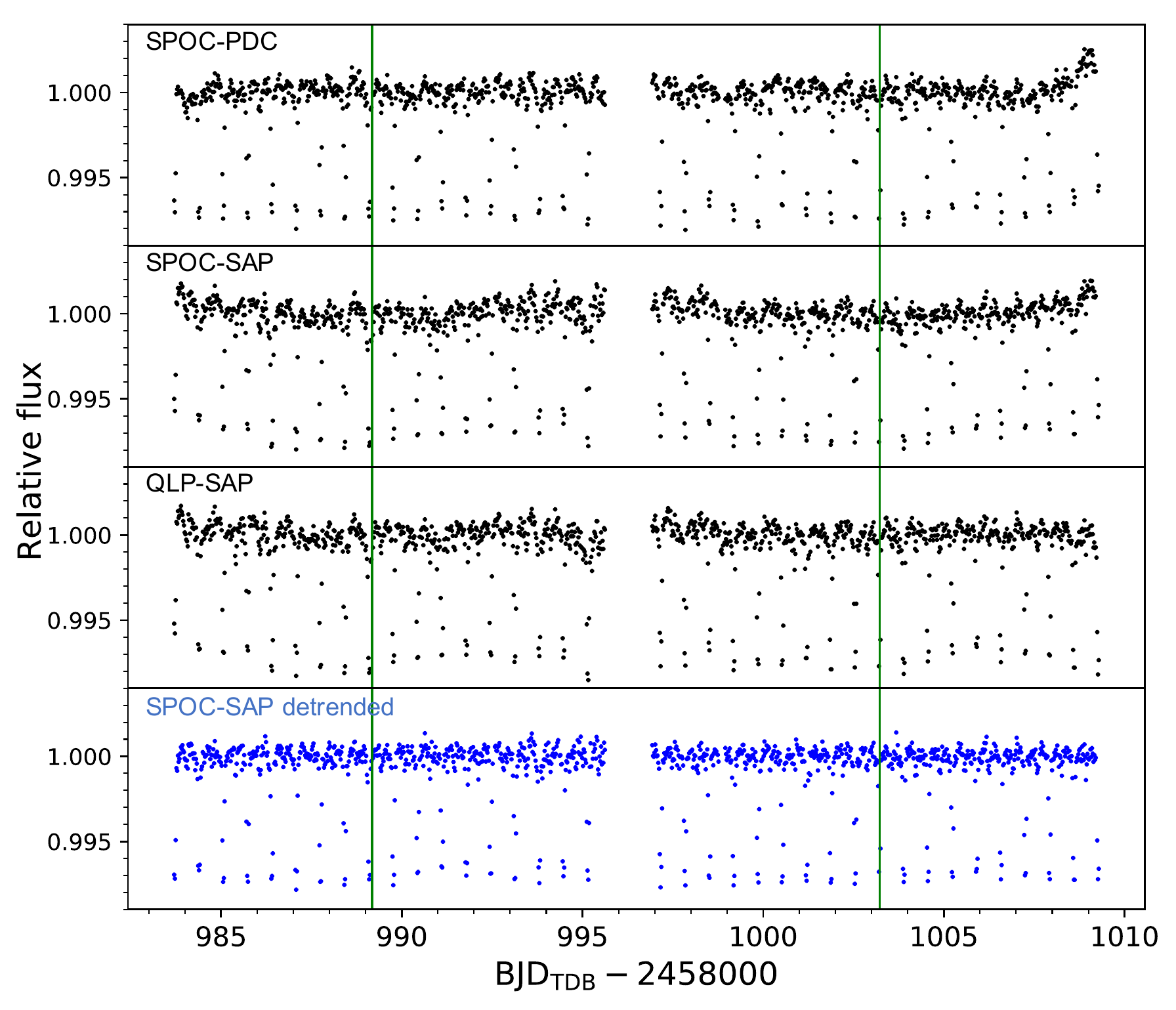}
\caption{Panels 1--3: the undetrended light curves of TOI-2109 extracted by the SPOC and QLP pipelines, prior to our fitting analysis. The vertical green lines indicate the momentum dumps, which divide the light curve into four segments. Careful inspection of the phase-folded light curves reveals a strong phase-curve signal, as well as additional photometric modulations due to systematics and stellar variability. Panel 4: the systematics-corrected SPOC-SAP light curve, which was used to produce the final light-curve fitting results.}
\label{fig:lcs}
\end{figure*}

\subsection{\tess\ Light Curves}
\label{sec:tess}

TOI-2109, listed in the \tess\ input catalog (TIC; \citealt{stassun18a}) as TIC 392476080, was observed by the \tess\ spacecraft from UT 2020 May 13 to 2020 June 8 during the Sector 25 campaign. Photometry of the target was obtained from the full frame images (FFIs), which have a cadence of 30 minutes. Initial light-curve extraction and analysis were carried out using the MIT Quick Look Pipeline (QLP; \citealt{huang2020a,huang2020b}). Subsequent vetting identified a transit-like signal in the target's light curve with a depth of roughly 7000 ppm and a duration of $\sim$1.8 hr that occurred every $\sim$0.67 days. The planet candidate was added to the list of \tess\ objects of interest (TOIs) as TOI-2109.01. Table~\ref{tab:info} lists astrometric and photometric information about the target provided in TIC Version 8.1 \citep{stassun19}.

Two different extractions of the TOI-2109 photometry are available on the Mikulski Archive for Space Telescopes\footnote{\url{https://mast.stsci.edu/}} (MAST). The first extraction was carried out using the Science Processing and Operations Center (SPOC) pipeline based at NASA Ames Research Center \citep{jenkins2016} as part of the \tess\ Light Curves from Full Frame Images (TESS-SPOC) High Level Science Products project \citep{caldwell2020}. The corresponding datafile contains two versions of the light curve: (1) the simple aperture photometry (SAP) light curve, which consists of the flux extracted from the optimized aperture on the FFI, and (2) the pre-search data conditioning (PDC) light curve, which was produced by the SPOC pipeline's detrending routine that utilizes the empirically determined co-trending basis vectors (CBVs) to remove instrumental systematics trends common to all sources on the detector \citep{smith2012,stumpe2012,stumpe2014}. The second data extraction available on MAST is the QLP-derived photometry; the SAP light curve contained therein was extracted using a different aperture than the SPOC light curves. 

\begin{deluxetable*}{lcccccc}[t]
\setlength{\tabcolsep}{5pt}
\tablewidth{0pc}
\renewcommand{\arraystretch}{1.0}
\tabletypesize{\footnotesize}
\tablecaption{
    TFOP Primary Transit Observations
    \label{tab:SG1obs}
}
\tablehead{\vspace{-0.2cm} \\
    \multicolumn{1}{l}{Date (UT)} &  
    \multicolumn{1}{l}{Telescope} & 
    \multicolumn{1}{l}{Filter} & 
    \multicolumn{1}{l}{Coverage} &
    \multicolumn{1}{c}{Exposure time (s)} &
    \multicolumn{1}{c}{Duration (hr)} &
    \multicolumn{1}{c}{Aperture radius (arcsec)}
}
\startdata
2020 Jul 31	& FLWO KeplerCam (1.2 m) & $z'$ & Partial   & 5     & 2.8   & 7.4 \\
2020 Jul 31	& ULMT (0.6 m) & $r'$	& Full      & 16    & 2.9   & 4.3\\
2020 Jul 31\tablenotemark{\scriptsize a}	& LCOGT McD (0.4 m)&	$g',i'$	& Full	&17, 25	&3.8	&6.8\\
2020 Aug 2	& MLO (0.36 m)&	$I$	& Full	&30	&3.9	&6.7\\
2020 Aug 4\tablenotemark{\scriptsize a}	&TCS MuSCAT2 (1.5 m)&	$g', r', i', z_s$	& Full	&3, 3, 3, 3	&2.5	&13.9\\
2020 Aug 24	&LCOGT SSO (1.0 m)&	$B, z_s$	& Full	&10, 24	&3.6	&15.6\\
2021 Apr 7	&WBRO (0.24 m)&	$R$	& Full	&100	&4.8	&8.7\\
2021 Apr 7	&LCOGT SAAO (1.0 m)&	$i'$	& Full	&16	&5.1	&9.0\\
2021 Apr 7	&GdP (0.4 m)&	$i'$	& Full	&20	&4.4	&6.6\\
2021 Apr 8	&LCOGT SSO (1.0 m)&	$i'$	& Partial	&16	&3.5	&7.4\\
2021 Apr 9	&LCOGT CTIO (1.0 m)&	$i'$	& Partial	&16	&3.5	&7.4\\
2021 Apr 9	&ULMT (0.6 m)&	$i'$	& Full	&32	&4.4	&4.3\\
2021 Apr 9	&FLWO KeplerCam (1.2 m)&	$z'$	& Full	&5	&5.2	&6.7\\
2021 Apr 11	&FLWO KeplerCam (1.2 m)&	$z'$	& Full	&5	&4.7	&6.7\\
2021 May 14\tablenotemark{\scriptsize a}	&TCS MuSCAT2 (1.5 m)&	$g', r', i', z_s$	& Full	& 25, 25, 50, 50	& 3.4	&10.9\\
2021 May 22\tablenotemark{\scriptsize a}	&TCS MuSCAT2 (1.5 m)&	$g', r', i', z_s$	& Full	& 10, 5, 10, 5	& 3.9	&10.9\\
2021 May 24 & FTN MuSCAT3 (2.0 m)& $g',r',i',z_s$ & Full & 20, 12, 15, 33 & 5.7 & 6.1 \\
2021 May 25\tablenotemark{\scriptsize a}	&TCS MuSCAT2 (1.5 m)&	$g', r', i', z_s$	& Full	& 5, 5, 10, 5	& 3.5	&10.9\\
2021 Jun 12	&TCS MuSCAT2 (1.5 m)&	$g', r', i', z_s$	& Full	& 5, 5, 10, 5	& 3.0	& 10.9 \\
2021 Jun 26 & FTN MuSCAT3 (2.0 m)& $g',r',i',z_s$ & Full & 20, 12, 15, 33 & 6.6 & 5.3 \\
\enddata
\vspace{+0.15cm}\textbf{Note.}
\vspace{-0.15cm}\tablenotetext{\textrm{a}}{These four MuSCAT2 observations and the McD light curve were affected by dome issues and clouds, respectively, and were not included in the final set of ground-based light curves analyzed in this paper.}
\vspace{-0.5cm}
\end{deluxetable*}

We used the \texttt{ExoTEP} pipeline \citep{benneke2019,wong2020hatp12} to analyze the \tess\ photometry. Prior to fitting the light curves, we removed all points in the time series with a NaN flux value or a nonzero quality flag (as indicated by the SPOC pipeline). We also applied a 16 point wide moving median filter to the transit-trimmed light curves and removed $5\sigma$ outliers. From an initial time series of 1231 points, these preprocessing steps removed 82 points (6.7\% of the data). Next, we divided the light curve into four segments that are separated by the scheduled momentum dumps (i.e., when the spacecraft thrusters were engaged to reset the onboard reaction wheels, leading to increased pointing jitter and poorer data quality) and data downlink interruptions, similar to what was done in previous analyses of \tess\ photometry \citep[e.g.,][]{wong2020wasp19,wong2020year1,wong2020kelt9,wong2021year2}. Figure~\ref{fig:lcs} shows the three versions of the TOI-2109 light curve: SPOC-PDC, SPOC-SAP, and QLP-SAP.

Aside from the transits, there are periodic brightness modulations that are commensurate with the orbital period (i.e., a phase-curve signal), which were initially discerned from careful inspection of the phase-folded QLP photometry produced as part of the initial candidate vetting process. In addition, there are long-term flux trends in the data attributable to uncorrected instrumental systematics. The time-correlated noise is particularly severe in the SPOC-PDC light curve, which also displays sharp flux ramps near the beginning and end of several data segments. 

While the SPOC detrending routine that produces the PDC photometry typically reduces instrumental systematics and improves time-correlated noise behavior, the presence of stellar variability and/or residual systematics features that are not shared by other sources on the detector can result in poorer data quality, as was previously reported, for example, in the \tess\ light curve of the active planet-hosting star WASP-19 \citep{wong2020wasp19}. In our \tess\ light-curve analysis, we only considered the SPOC-SAP and QLP-SAP light curves.


\subsection{Ground-based Primary Transit Light Curves}
\label{sec:groundphot}

We acquired ground-based time-series photometry of primary transits of TOI-2109 as part of the \tess\ Follow-up Observing Program\footnote{\url{http://tess.mit.edu/followup}} (TFOP) Sub Group 1 (SG1; seeing-limited time-series photometry) collaboration. The TFOP SG1 network includes both professional and amateur astronomers at more than a hundred observatories around the world. Observers choose targets to follow up using the {\tt TESS Transit Finder}, which is a customized version of the {\tt Tapir} software package \citep{jensen2013}. 

For TOI-2109, 20 full- and partial-transit observations were obtained between 2020 July 31 and 2021 June 26. The photometric data sets contributed by TFOP SG1 observers were uploaded to the ExoFOP-TESS repository.\footnote{\url{http://exofop.ipac.caltech.edu/tess/}} These observations are summarized in Table~\ref{tab:SG1obs}; detailed descriptions of the instruments and observing methodology are provided in the following subsections. 

Unless otherwise noted, the data reduction and photometric extraction of the TFOP SG1 observations were performed using the {\tt AstroImageJ} (AIJ) software package \citep{collins17}. The extraction aperture radii ranged from 4\farcs3 to 15\farcs6 across the data sets. The nearest Gaia DR2 star to TOI-2109 is a faint neighbor at a projected separation of 18\arcsec, well outside all of the apertures used for the ground-based photometry. In the case of the heavily defocused observations on 2020 August 4 and 24, some light from the neighboring star would have fallen within the aperture; however, given that the neighbor is 7 mag fainter than the target star, the dilution is negligible. These ground-based time-series observations excluded all stars in the vicinity of TOI-2109 (at separations larger than $\sim$1\arcsec) as the source of the transit signal. We describe our high-angular-resolution search for smaller-separation companion stars in Section~\ref{sec:imaging}.


\subsubsection{FLWO KeplerCam}
We captured a transit ingress and two full transits of TOI-2109b at the Fred Lawrence Whipple Observatory (FLWO) on Mt.\,Hopkins in Arizona, USA. We used the KeplerCam instrument on the 1.2 m robotic, queue-scheduled telescope, which features a Fairchild CCD 486 detector with a $23\farcm1 \times 23\farcm1$ field of view (FOV). The ingress was observed on UT 2020 July 31 (referred to hereafter as KeplerCam $z'$ \#1), while the full-transit observations were taken on UT 2021 April 9 and 11 (KeplerCam $z'$ \#2 and \#3). For all three visits, we obtained 5 s exposures in the Sloan $z'$-band filter with $2 \times 2$ pixel binning, resulting in a 0\farcs67 pixel scale. 

\subsubsection{ULMT}
We observed two full transits with the 0.61 m University of Louisville Manner Telescope (ULMT) at Mt.\,Lemmon in Arizona, USA, using an STX 16803 camera with 0\farcs39 pixel scale and a $26\arcmin \times 26\arcmin$ FOV. The UT 2020 July 31 observation utilized the Sloan $r'$-band filter, while the UT 2021 April 9 observation was taken in the Sloan $i'$ bandpass. The exposure times for the two visits were 16 and 32 s, respectively.

\subsubsection{LCOGT McD, SSO, SAAO, and CTIO}
We obtained five transit light curves with the Las Cumbres Observatory Global Telescope network (LCOGT; \citealt{brown13}). Four visits were obtained on 1.0 m network nodes and used Sinistro cameras that have 0\farcs39 pixels and a $26\arcmin \times 26\arcmin$ FOV. On the 0.4 m network node, we used the SBIG STX 6303 camera with a pixel scale of 0\farcs57 and an FOV of $29\farcm2 \times 19\farcm5$. Data were calibrated using the standard LCOGT {\tt BANZAI} pipeline \citep{mccully2018}. 

We captured a full transit of TOI-2109b on UT 2020 July 31 in Sloan $g'$ and $i'$ bands with the 0.4 m telescope at the McDonald Observatory (McD) in Texas, USA. These observations were affected by periods of significant cloud cover, with widely varying transparency throughout the 3.8 hr visit. We therefore did not include the light curve from this observation in our analysis.

With the 1.0 m telescope at the Siding Spring Observatory (SSO) in New South Wales, Australia, we observed a full transit with the Johnson $B$ and Pan-STARRS $z_s$ ($\lambda_{\mathrm{eff}} = 870$ nm) filters on UT 2020 August 24. The exposure times in the two bands were 10 and 24 s, respectively. Due to the defocused nature of the observations, a large 15\farcs6 extraction aperture was applied when obtaining the light curve. We also recorded an ingress with the same instrument in the $i'$ band on UT 2021 April 8 using 16 s exposures.

We observed a full transit in the $i'$ band on UT 2021 April 7 using the 1.0 m telescope at the South African Astronomical Observatory (SAAO) in Sutherland, South Africa. The exposure time was set at 16 s.

Finally, on UT 2021 April 9, we used the 1.0 m telescope at the Cerro Tololo Interamerican Observatory (CTIO) in Chile to record a partial $i'$-band transit with 16 s exposures. The 3.5 hr visit included the transit ingress and mid-transit, ending just before the beginning of egress.

\subsubsection{MLO}
We observed a full transit of TOI-2109b on UT 2021 August 31 with the 0.36 m telescope at the Maury Lewin Astronomical Observatory (MLO) in California, USA. The SBIG STF 8300M CCD has an FOV of $23\arcmin \times 17\arcmin$ and was operated with the Cousins $I$-band filter in the $2 \times 2$ binning mode, yielding a pixel scale of 0\farcs84. An exposure time of 30 s was used.

\subsubsection{TCS MuSCAT2}\label{muscat2}
A full transit was observed on UT 2020 August 4 simultaneously in the $g'$, $r'$, $i'$, and $z_s$ bands with the MuSCAT2 multicolor imager \citep{narita2019} installed on the 1.5 m Telescopio Carlos S{\'a}nchez (TCS) at Teide Observatory, Spain. MuSCAT2 is equipped with four $1024 \times 1024$ pixel CCDs, each having a $7\farcm4 \times 7\farcm4$ FOV with a pixel scale of 0\farcs44. The defocused exposures had a total integration time of 3 s.

Four additional multiband transit observations were carried out on UT 2021 May 14, May 22, May 25, and June 12. These subsequent visits used 
different exposure times across the four photometric bands. All of the MuSCAT2 data sets were passed through the dedicated MuSCAT2 photometry pipeline \citep{Parviainen:2019} for standard image calibration, photometric extraction, and instrumental systematics detrending.

The first four MuSCAT2 observations suffered to varying degrees from operational issues with the dome, which occasionally caused field occlusion and severe, variable vignetting in the four bands. The resultant instrumental systematics features led to strongly discrepant estimates of transit depth when compared to the values obtained from the other ground-based data sets. On the other hand, the last visit (2021 June 12) was not affected by dome issues throughout the duration of the observation. As such, only this final set of MuSCAT2 transit photometry was included in our fitting analysis. 

\subsubsection{WBRO}
We captured a full transit with the Cousins $R$-band filter on UT 2021 April 7 using the 0.24 m telescope at the Wild Boar Remote Observatory (WBRO) near Florence, Italy. The SBIG ST-8 XME camera has a pixel scale of 0\farcs79 and an FOV of $20\arcmin \times 14\arcmin$. With an exposure time of 100 s, the resultant light curve has the longest cadence among the ground-based observations.

\subsubsection{GdP}
Using the FLI 4710 camera mounted on the RCO 40 cm telescope at the Grand-Pra (GdP) Observatory in Switzerland, we observed a full transit of TOI-2109b on UT 2021 April 7. The FLI 4710 camera is a back-illuminated CCD with an E2V CCD47-10 sensor and a $11\farcm7 \times 11\farcm7$ FOV. Observations were taken with a 20 s exposure time through an $i'$-band filter without pixel binning, yielding a pixel scale of 0\farcs73. 

\subsubsection{FTN MuSCAT3}\label{muscat3}
On UT 2021 May 24 and June 26, we collected simultaneous time-series observations of two full transits in four photometric bands with the MuSCAT3 multicolor imager \citep{narita2020}. This instrument, which is operationally similar to the MuSCAT2 imager (Section~\ref{muscat2}), was recently installed on the 2.0 m Faulkes Telescope North (FTN) at Haleakala Observatory on Maui, Hawai'i and is operated by Las Cumbres Observatory. MuSCAT3 is equipped with four $2048 \times 2048$ pixel CCDs that provide a $9\farcm1 \times 9\farcm1$ FOV and a pixel scale of 0\farcs266.

The exposure times in the $g'$, $r'$, $i'$, and $z_s$ filters were 20, 12, 15, and 33 s, respectively. Data processing and aperture photometry were carried out using AIJ. The resultant light curves from the two visits, referred to hereafter as \#1 and \#2 respectively, have the highest signal-to-noise ratio (S/N, scaled to a 30 s exposure) of any photometric series collected for TOI-2109. Along with the MuSCAT2 transit light curves (Section~\ref{muscat2}), these high-precision photometric series, obtained roughly one year after the initial \tess\ observations, provide exquisite constraints on both the orbital ephemeris and the transit-shape parameters.

\subsection{Ground-based Secondary Eclipse Light Curve}
\label{sec:wirc}

On UT 2021 March 6, we observed a secondary eclipse of TOI-2109b in the $K_\mathrm{s}$ band ($\lambda_{\mathrm{eff}} = 2.13$ $\mu$m) with the Wide-field Infared Camera \citep[WIRC;][]{wilson2003} on the Hale 200\arcsec\ Telescope at Palomar Observatory, California, USA. The data were taken with a beam-shaping diffuser that increased our observing efficiency and improved guiding stability on this bright target \citep{stefansson2017,vissapragada2020}. In order to mitigate a known detector systematic at short exposure times, we initiated the observations with a four-point dither near the target to construct a background frame. We then collected 239 exposures, each consisting of 15 coadds of 0.92 s for a total per-image exposure time of 13.8 s.

The data were dark-subtracted, flat-fielded, and corrected for bad pixels using the same methodology as in \citet{vissapragada2020}. We scaled and subtracted the aforementioned background frame from each image to remove the full frame background structure. Next, we performed circular aperture photometry with the \texttt{photutils} package \citep{photutils} on TOI-2109 and two nearby comparison stars of similar brightness located within the FOV. We experimented with a range of photometric extraction apertures from 10 to 25 pixels in 1 pixel steps (with $0\farcs25\,\mathrm{pixel}^{-1}$), eventually selecting a 13 pixel ($3\farcs25$) aperture that minimized the per-point scatter of the fitted photometry. 


\begin{figure}
\includegraphics[width=\linewidth]{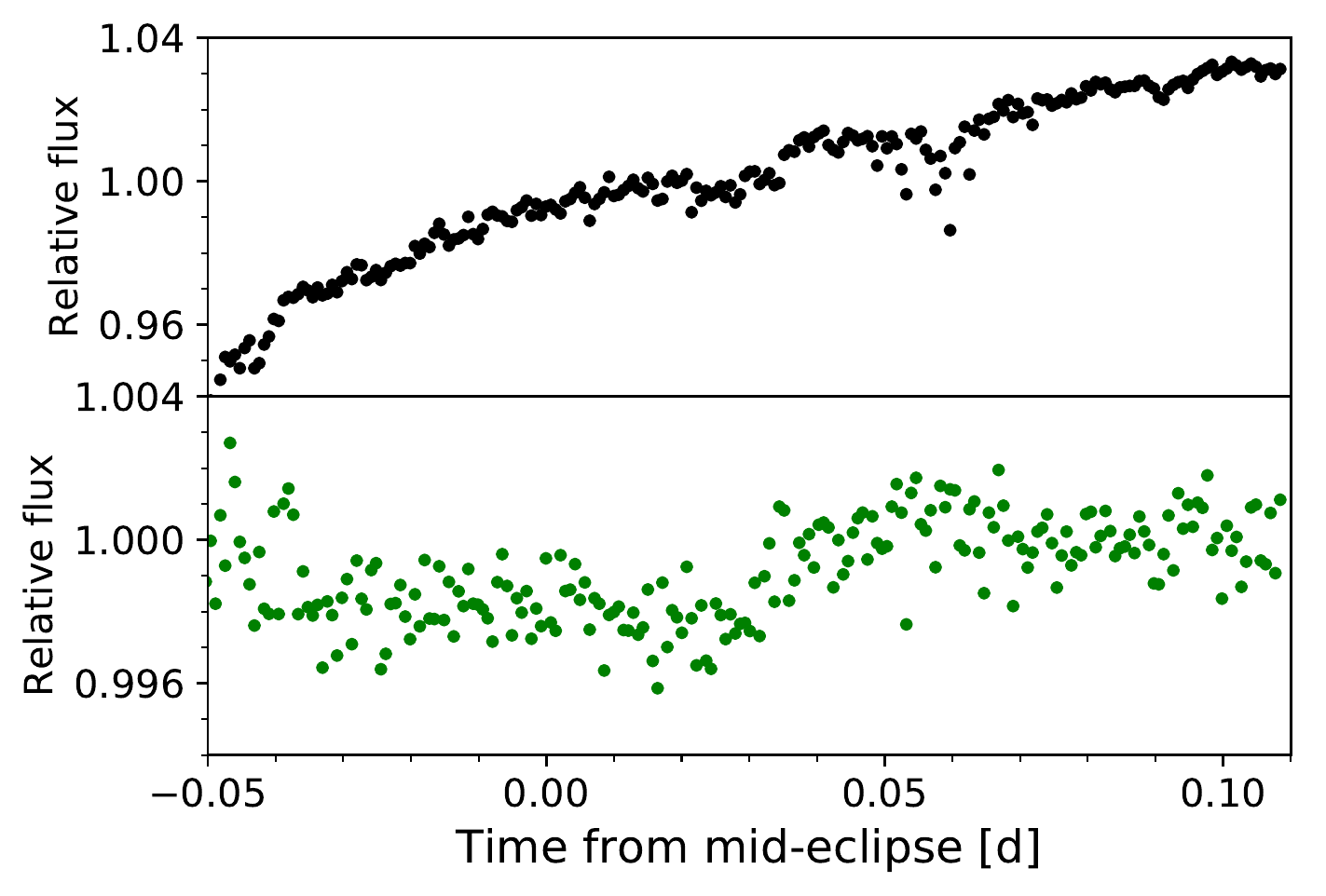}
\caption{The raw (top) and detrended (bottom) \Ks-band secondary eclipse light curve obtained with WIRC.}
\label{fig:wirc}
\end{figure}

The raw extracted photometry is shown in the top panel of Figure~\ref{fig:wirc}. The light curve begins just before the start of ingress, when the target reached an airmass above 2. While the sky was clear throughout the $\sim$3.8 hr observation, the sky background in the vicinity of TOI-2109 varied considerably, particularly during the last hour, resulting in more severe residual systematics. 



\subsection{High-angular-resolution Imaging}
\label{sec:imaging}
A bound or line-of-sight companion in close proximity to the target can create a false positive transit signal if it is an eclipsing binary. The so-called ``third-light” flux from a companion star can also yield an underestimated planetary radius if not accounted for in the transit model \citep{ciardi2015}. Likewise, the photometric contamination can lead to nondetections of small planets residing within the same exoplanetary system \citep{lester2021}. The discovery of binary systems containing close, bound companions, which exist around nearly half of FGK-type stars \citep{matson2018}, can help further our understanding of exoplanet formation, dynamics, and evolution \citep{howell2021}. In order to search for close-separation companions unresolved by \tess\ and other seeing-limited ground-based follow-up observations, we carried out high-resolution speckle imaging of TOI-2109.

\begin{figure}
\includegraphics[width=\linewidth]{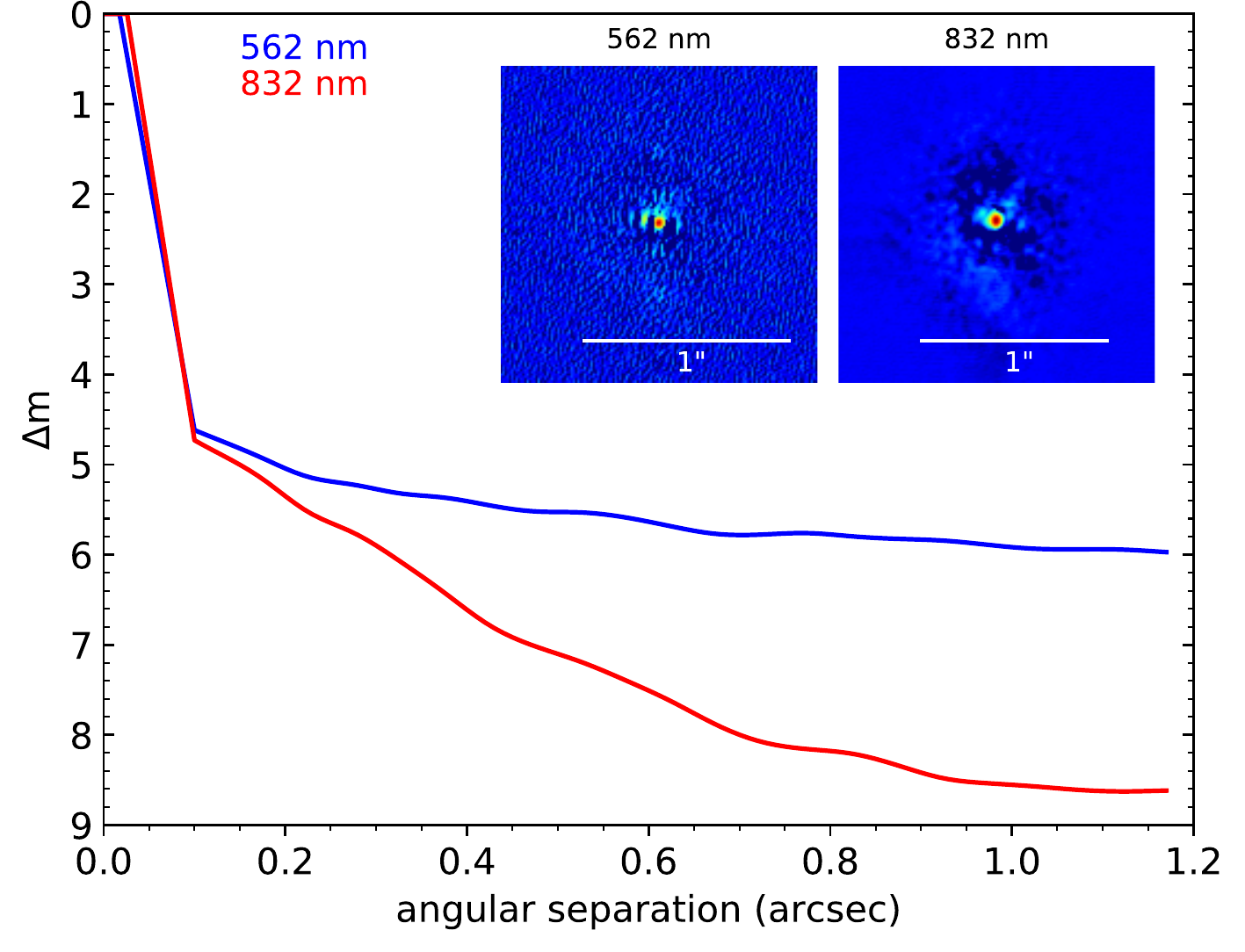}
\caption{Results of speckle-imaging observations of TOI-2109 at Gemini-North. The contrast curves provide $5\sigma$ upper limits (in units of magnitude difference) on the location of a nearby star as a function of angular separation (in arcsec). The blue and red curves correspond to the 562 and 832 nm bands of ‘Alopeke, respectively.}
\label{fig:speckle}
\end{figure}

The target was observed on UT 2020 September 17 using the ‘Alopeke speckle instrument on Gemini-North. ‘Alopeke provides simultaneous speckle imaging in two bands (562 and 832 nm), producing a reconstructed image with robust contrast limits on companion detections \citep[e.g.,][]{howell2016}. Five sets of $1000 \times 0.06$ s exposures were collected and passed through a standard Fourier analysis in our reduction pipeline \citep[see][]{howell2011}. Figure~\ref{fig:speckle} shows the resultant $5\sigma$ contrast curves and the reconstructed speckle images in both bands. We find TOI-2109 to be a single star with no companion brighter than 5--9 mag below the target star's brightness from the diffraction limit ($\sim$20 mas) out to 1.2\arcsec. At the distance of TOI-2109 ($d = 262$ pc; Table~\ref{tab:info}), these angular limits correspond to physical separations of 5 au to 314 au, respectively.

\subsection{High-resolution Spectroscopy}
\label{sec:spec}

\subsubsection{TRES}
\label{sec:tres}

\renewcommand{\arraystretch}{0.99}
\begin{deluxetable}{lccc}[t]
\tablewidth{0pc}
\tabletypesize{\scriptsize}
\tablecaption{
    TOI-2109 Radial Velocities
    \label{tab:rvs}
}
\tablehead{ \vspace{-0.1cm} \\ 
    \multicolumn{1}{l}{BJD$_{\mathrm{TDB}}$} &
    \multicolumn{1}{c}{$v$ (\kms)}    &
    \multicolumn{1}{c}{$\sigma_{v}$ (\kms)}   &
    \multicolumn{1}{c}{Instrument} \vspace{-0.1cm}
}
\startdata
 2{,}459{,}071.67553 & $-$24.70 & 0.41 & TRES \\
 2{,}459{,}080.72393 & $-$25.45 & 0.63 & TRES \\
 2{,}459{,}093.43976 & $-$27.20 & 0.58 & FIES \\
 2{,}459{,}095.43218 & $-$26.05 & 0.80 & FIES \\
 2{,}459{,}103.63381 & $-$26.96 & 0.97 & TRES \\
 2{,}459{,}105.37042$^{\dag}$ & $-$27.05 & 1.57 & FIES \\
 2{,}459{,}108.62802 & $-$26.02 & 0.79 & TRES \\
 2{,}459{,}109.64697$^{\dag}$ & $-$26.49 & 1.22 & TRES \\ 
 2{,}459{,}110.65230$^{\dag}$ & $-$24.68 & 2.30 & TRES \\
 2{,}459{,}111.67394$^{\dag}$ & $-$26.51 & 1.39 & TRES \\
 2{,}459{,}112.68604$^{\dag}$ & $-$26.67 & 1.77 & TRES \\ 
 2{,}459{,}114.60697$^{\dag}$ & $-$25.77 & 1.15 & TRES \\
 2{,}459{,}115.61703$^{\dag}$ & $-$26.87 & 1.02 & TRES \\ 
 2{,}459{,}116.60570 & $-$25.00 & 0.78 & TRES \\
 2{,}459{,}117.61174 & $-$26.26 & 0.63 & TRES \\
 2{,}459{,}119.36331 & $-$24.88 & 0.33 & FIES \\
 2{,}459{,}123.35113 & $-$25.10 & 0.33 & FIES \\
 2{,}459{,}133.34116 & $-$26.36 & 0.50 & FIES \\
 2{,}459{,}246.04456 & $-$25.91 & 0.29 & TRES \\
 2{,}459{,}264.97842 & $-$26.52 & 0.41 & TRES \\
 2{,}459{,}265.99880 & $-$25.05 & 0.35 & TRES \\
 2{,}459{,}269.01286 & $-$26.64 & 0.47 & TRES \\
 2{,}459{,}269.98900 & $-$24.44 & 0.25 & TRES \\
 2{,}459{,}270.96574 & $-$25.92 & 0.43 & TRES \\
 2{,}459{,}272.71821 & $-$24.42 & 0.44 & FIES \\
 2{,}459{,}277.02309 & $-$26.66 & 0.33 & TRES \\
 2{,}459{,}286.68469 & $-$24.85 & 0.92  & FIES \\
 2{,}459{,}293.66293 & $-$25.28 & 0.29  & FIES \\
 2{,}459{,}294.70777 & $-$24.83 & 0.34  & FIES \\
 2{,}459{,}297.68209 &  $-$25.19 & 0.39  & FIES \\
 2{,}459{,}313.83792 &  $-$25.39 &   0.42    & TRES \\
 2{,}459{,}313.85012$^{\ddag}$  &  $-$25.17 &   0.48    & TRES \\
 2{,}459{,}313.87085$^{\ddag}$  &  $-$25.42 &   0.66    & TRES \\
 2{,}459{,}313.88294$^{\ddag}$  &  $-$25.84 &   0.53    & TRES \\
 2{,}459{,}313.89533$^{\ddag}$  &  $-$25.81 &   0.37    & TRES \\
 2{,}459{,}313.90775$^{\ddag}$  &  $-$25.03 &   0.46    & TRES \\
 2{,}459{,}313.91975$^{\ddag}$  &  $-$25.84 &   0.40    & TRES \\
 2{,}459{,}313.93224$^{\ddag}$  &  $-$25.87 &   0.49    & TRES \\
 2{,}459{,}313.94506$^{\ddag}$  &  $-$26.70 &   0.38    & TRES \\
 2{,}459{,}313.95763  &  $-$25.45 &   0.37    & TRES \\
 2{,}459{,}313.96966  &  $-$27.31 &   0.40    & TRES \\
 2{,}459{,}313.98194  &  $-$25.93 &   0.35    & TRES \\
 2{,}459{,}313.99409  &  $-$26.66 &   0.41    & TRES \\
 2{,}459{,}314.00616  &  $-$26.85 &   0.45    & TRES \\
 2{,}459{,}315.83142  &  $-$25.13 &   0.54    & TRES \\
 2{,}459{,}315.84371  &  $-$25.33 &   0.40    & TRES \\
 2{,}459{,}315.85572  &  $-$24.90 &   0.33    & TRES \\
 2{,}459{,}315.86779$^{\ddag}$  &  $-$25.79 &   0.42    & TRES \\
 2{,}459{,}315.87981$^{\ddag}$  &  $-$25.56 &   0.46    & TRES \\
 2{,}459{,}315.89217$^{\ddag}$  &  $-$25.24 &   0.37    & TRES \\
 2{,}459{,}315.90416$^{\ddag}$  &  $-$26.05 &   0.30    & TRES \\
 2{,}459{,}315.91647$^{\ddag}$  &  $-$25.64 &   0.35    & TRES \\
 2{,}459{,}315.92848$^{\ddag}$  &  $-$25.76 &   0.26    & TRES \\
 2{,}459{,}315.94055$^{\ddag}$  &  $-$25.56 &   0.29    & TRES \\
 2{,}459{,}315.95263$^{\ddag}$  &  $-$25.83 &   0.57    & TRES \\
 2{,}459{,}315.96486$^{\ddag}$  &  $-$26.29 &   0.37    & TRES \\
 2{,}459{,}315.97680  &  $-$26.28 &   0.39    & TRES \\
 2{,}459{,}315.98895  &  $-$25.77 &   0.45    & TRES \\
\enddata
\textbf{Note.}
\vspace{-0.2cm}\tablenotetext{}{RVs marked with $\dag$ have uncertainties $>$1 \kms, while entries marked with $\ddag$ were obtained during the primary transit. All of the marked RVs were excluded from the final RV orbit analysis.}
\vspace{-0.3cm}
\end{deluxetable}

We obtained 19 individual observations of TOI-2109 using the Tillinghast Reflector Echelle Spectrograph (TRES) on the 1.5 m telescope at FLWO. TRES is a fiber-fed echelle spectrograph with a spectral resolving power of $R \approx 44,000$ over the wavelength range 3850--9100 \AA. The exposure times were set between 540 and 1800 s, and the S/N per spectral resolution element at the peak of the Mg b order near 519 nm ranged from 24 to 80 across the 19 spectra. Wavelength calibration was achieved through ThAr hollow-cathode lamp exposures that bracketed each on-target observation. 

In addition to the individual observations, we also collected two spectroscopic transits of TOI-2109b to measure its orbital obliquity and help eliminate additional false positive scenarios. During the transit, the planet successively blocks different parts of the rotating stellar surface. Spectroscopically, the rotationally broadened stellar absorption lines exhibit variations due to occultation by the transiting planet, resulting in an apparent velocity shift \citep{mclaughlin1924,rossiter1924,gaudi2007} and a line-profile variation \citep{colliercameron2010}. Modeling the path of the transit in the spectra (i.e., the Doppler shadow; Section~\ref{sec:jointfit}) yields the projected obliquity of the planet's orbital plane and confirms that the planet is indeed orbiting the designated host star, not an unseen background star. 

The two spectroscopic transit observations occurred on UT 2021 April 9 and 11. Each night's observation spanned the entire transit, with at least one hour of baseline on either side of ingress and egress. A total of 28 individual spectra were obtained with a 900 s exposure time, bracketed by ThAr lamp exposures for wavelength calibration. 

The stellar spectra from both the individual observations and the Doppler spectroscopic transits were extracted as per \citet{buchhave2010}. Radial velocities (RVs) were derived by modeling the line-broadening profiles of the spectra, which were constructed via a least-squares deconvolution of each spectrum against a nonrotating synthetic template \citep{donati1997}. The template was generated using ATLAS9 model atmospheres \citep{castelli2003}, with stellar parameters matching that of TOI-2109. We fit broadening models to the line profiles following the method outlined in \citet{gray2005}, incorporating the effects of rotational broadening, macroturbulent broadening, instrumental broadening, and a radial velocity shift. The resulting velocities are listed in Table~\ref{tab:rvs}. By accounting for the rotational broadening of the stellar line profiles, our least-squares deconvolution analysis simultaneously produced an estimate of the host star's sky-projected rotational velocity: $\vsini = 81.9\pm1.7$ \kms.

\subsubsection{FIES}
\label{sec:fies}

From UT 2020 August 31 to 2021 March 24, we collected 11 spectra of TOI-2109 using the Fiber-fed Echelle Spectrograph (FIES; \citealt{telting2014}) installed on the 2.56 m Nordic Optical Telescope (NOT) at the Roque de los Muchachos Observatory in La Palma, Spain. The high-resolution mode of FIES provides a spectral resolution of up to $R \approx 67,000$ across the wavelength range 3760--8840 \AA. The observations utilized an exposure time of 1800 s. The S/N per resolution element varied from 24 to 62 at the peak of the Mg b order. To construct the wavelength solution, a pair of ThAr calibration spectra were taken before and after each science observation. Optimal spectral extraction was carried out using the methods described in \citet{buchhave2010}, and the RVs were derived using the same procedure as for the TRES data (Section~\ref{sec:tres}).
The 11 FIES RVs are provided in Table~\ref{tab:rvs}.

\section{Stellar Characterization}
\label{sec:star}

To obtain an initial set of basic stellar parameters for the host star TOI-2109, we used the Spectral Parameter Classification (SPC) tool \citep[e.g.,][]{buchhave2012}. Given the relatively low S/N of each TRES spectrum, we combined all of the spectra obtained outside of the primary transit. With SPC, this combined spectrum was cross-correlated against a grid of synthetic stellar spectra based on ATLAS9 model atmospheres \citep{castelli2003}. The parameters that we allowed to vary freely were the stellar effective temperature \teff, \loggstar, \feh, and \vsini. We retrieved $\teff = 6808 \pm 50$ K, $\loggstar = 4.12 \pm 0.10$, $\feh = 0.082 \pm 0.080$ dex, and $\vsini = 85.2 \pm 0.5$ \kms. We note that the \vsini\ derived using SPC differs by about 2$\sigma$ from the value obtained from the least-squares deconvolution analysis of the same TRES spectra (Section~\ref{sec:tres}). The latter technique directly accounts for the effects of both rotational broadening and macroturbulence on the stellar spectra and therefore provides a more dependable estimate of \vsini; we use that value ($81.9 \pm 1.7$ \kms) when modeling the spectroscopic transits in Section~\ref{sec:jointfit}.

\begin{deluxetable}{llc}[t]
\tablewidth{0pc}
\setlength{\tabcolsep}{6pt}
\renewcommand{\arraystretch}{1.0}
\tabletypesize{\footnotesize}
\tablecaption{
    Stellar Parameters of TOI-2109 from the \texttt{EXOFASTv2} SED Analysis
    \label{tab:stellar}
}
\tablehead{
\vspace{-0.2cm}\\
    \multicolumn{1}{l}{Parameter} &
    \multicolumn{1}{c}{Description}    &
    \multicolumn{1}{c}{Value}   
}
\startdata
$M_*$ &Mass (\msun) &$1.447^{+0.075}_{-0.078}$\\
$R_*$ &Radius (\rsun) &$1.698^{+0.062}_{-0.057}$\\
$L_*$ &Luminosity (\lsun) &$4.71^{+0.38}_{-0.27}$\\
$\rho_*$ &Density (cgs)&$0.417^{+0.056}_{-0.053}$\\
$\loggstar $ &Surface gravity (cgs) &$4.139^{+0.041}_{-0.046}$\\
$T_{\rm eff}$ &Effective temperature (K) &$6530^{+160}_{-150}$\\
$\feh$ &Metallicity\tablenotemark{\scriptsize a} (dex) &$0.068^{+0.070}_{-0.062}$\\
$\feh_{0}$ &Initial metallicity\tablenotemark{\scriptsize b} (dex) &$0.212^{+0.067}_{-0.072}$\\
Age &Age (Gyr)&$1.77^{+0.88}_{-0.68}$\\
$A_V$ &$V$-band extinction (mag) &$0.087^{+0.086}_{-0.062}$\\
$\sigma_{\mathrm{SED}}$ &SED photometry error scaling  &$0.86^{+0.29}_{-0.19}$\\
$\varpi$ &Parallax\tablenotemark{\scriptsize a} (mas) &$3.817^{+0.049}_{-0.047}$\\
$d$ &Distance (pc) &$262.0 \pm 3.3$\\
\enddata
\textbf{Notes.}
\vspace{-0.15cm}\tablenotetext{\textrm{a}}{These parameters were constrained by priors derived from the SPC modeling of the TRES spectra and Gaia data.}
\vspace{-0.15cm}\tablenotetext{\textrm{b}}{The initial metallicity of the host star when it was formed.}
\vspace{-0.8cm}
\end{deluxetable}

To expand our characterization of TOI-2109, we modeled the spectral energy distribution (SED) using the publicly available exoplanet fitting suite \texttt{EXOFASTv2} \citep{eastman2013,eastman2019,eastman2017}. We fit the broadband photometric measurements in the $B_T$, $V_T$, Gaia ($G$, $G_{\mathrm{BP}}$, $G_{\mathrm{RP}}$), $J$, $H$, $K$, and $W1$--$W3$ bandpasses, which are listed in Table~\ref{tab:info}. Gaussian priors were placed on the stellar metallicity ($0.082 \pm 0.080$ dex, as derived from the SPC analysis of the TRES spectra) and parallax ($3.818 \pm 0.047$ mas; from Gaia DR2, corrected for the systematic offset reported by \citealt{lindegren2018}). Our analysis used MESA Isochrones and Stellar Tracks (MIST) stellar evolution models \citep{paxton2011,paxton2013,paxton2015,choi2016,dotter2016} to constrain the stellar parameters. We included an upper limit on the line-of-sight $V$-band extinction from \citet{schlegel1998} and \citet{schlafly2011}, as well as systematic floors on the broadband photometric errors \citep{stassun2016}. We used the \texttt{EXOFASTv2} default lower limit of 3\% on the systematic error on the bolometric flux, which is consistent with the spread seen from various techniques used to calculate that quantity \citep{zinn2019}.

The full results of the SED fit are provided in Table~\ref{tab:stellar}. We find that TOI-2109 is a mid--late F-type star with an effective temperature of $\teff = 6530^{+160}_{-150}$ K and roughly solar metallicity ($\feh = 0.068^{+0.070}_{-0.062}$ dex). The stellar radius and mass are $1.698^{+0.062}_{-0.057}$ \rsun\ and $1.447^{+0.075}_{-0.078}$ \msun, respectively. The star lies on the main sequence, with $\loggstar = 4.139^{+0.041}_{-0.046}$ and a weakly constrained age of $1.77^{+0.88}_{-0.68}$ Gyr. In addition, the measured \vsini\ from the TRES spectra and the $R_*$ from the SED fit imply a stellar rotation period of $P_{\rm rot}/\sin i_* = 1.05 \pm 0.04$ days, which is consistent with one of the stellar periodicity signals detected in the \tess\ light curve (see Section~\ref{sec:activity}).

\section{Data Analysis}
\label{sec:ana}

\subsection{\tess\ Light-curve Fit}
\label{sec:tessfit}

The \tess\ light curves of TOI-2109 (Figure~\ref{fig:lcs}) show photometric variability that is synchronous with the orbital period of the planet. We used a full-orbit phase-curve model to fit the light curves. To address the instrumental systematics present in the photometry, we experimented with several different methods for detrending the light curves. Significant periodic brightness modulations attributed to stellar variability were also detected in the \tess\ photometry and included in our light-curve model.

\subsubsection{Full-orbit Phase-curve Model}
\label{sec:phasemodel}

Following previous phase-curve analyses of \tess\ data \citep[e.g.,][]{shporer2019,wong2020wasp19,wong2020year1,wong2020kelt9,wong2021year2}, we used a simple sinusoidal phase-curve model that treats the stellar and planetary fluxes separately:
\begin{align}
\label{astro}F(t) &= \frac{F_{*}(t)\lambda_t(t) + F_{p}(t)\lambda_e(t)}{1 + \bar{f_p}},\\
\label{planet}F_{p}(t) &= \bar{f_{p}} - A_{\mathrm{atm}} \cos(\phi + \psi),\\
\label{star}F_{*}(t) &= 1 - A_{\mathrm{ellip}}\cos(2\phi) + A_{\mathrm{Dopp}}\sin(\phi).
\end{align}
The orbital phase $\phi$ is given by $2\pi(t - T_c)$, where $T_c$ is the mid-transit time. The transit and eclipse light curves are represented by $\lambda_t$ and $\lambda_e$. The planetary flux $F_{p}$, which is expected to be dominated by thermal emission from the atmosphere, is defined relative to the average flux level $\bar{f_{p}}$, with the semiamplitude of the atmospheric brightness modulation represented by $A_{\mathrm{atm}}$. The parameter $\psi$ denotes a shift in the planetary phase curve, which may arise from an offset dayside hotspot due to superrotating equatorial winds or inhomogeneous clouds. In this parameterization, the secondary eclipse depth is $D_d = \bar{f_{p}} - A_{\mathrm{atm}}\cos(\pi + \psi)$, and the nightside flux is $D_n = \bar{f_{p}} - A_{\mathrm{atm}}\cos{\psi}$.

The stellar flux $F_{*}$ contains terms corresponding to two separate physical processes \citep[e.g.,][]{faigler2011,faigler2015,shporer2017}: (1) the tidal response of the stellar surface to the gravitational pull of the orbiting companion, typically referred to as ellipsoidal distortion ($A_{\mathrm{ellip}}$), and (2) the modulation in the band-integrated flux due primarily to the periodic Doppler shifting of the stellar spectrum, i.e., Doppler boosting ($A_{\mathrm{Dopp}}$). Here the sign convention for the various phase-curve amplitudes was chosen so as to produce positive values under normal circumstances. 

In an unconstrained fit, $A_{\mathrm{Dopp}}$ and $\psi$ are degenerate, as a phase shift in the planetary phase-curve signal can be absorbed by the coefficient of the $\sin\phi$ term (i.e., where the Doppler-boosting signal lies). Therefore, when fitting the \tess\ light curves alone, we did not consider any phase offset and simply fit for the total harmonic power at the sine of the orbital frequency, which we denote as $A_1$. It follows that the values for $A_1$ that we obtained from our dedicated \tess\ phase-curve analysis contain contributions from both Doppler boosting on the host star and a phase offset in the planet's atmospheric brightness modulation. In our final global analysis of all available data sets (Section~\ref{sec:jointfit}), we leveraged the constraint on planet mass provided by the RV measurements to self-consistently model the Doppler-boosting signal and disentangle the phase offset.

\begin{figure*}[t]
\includegraphics[width=\linewidth]{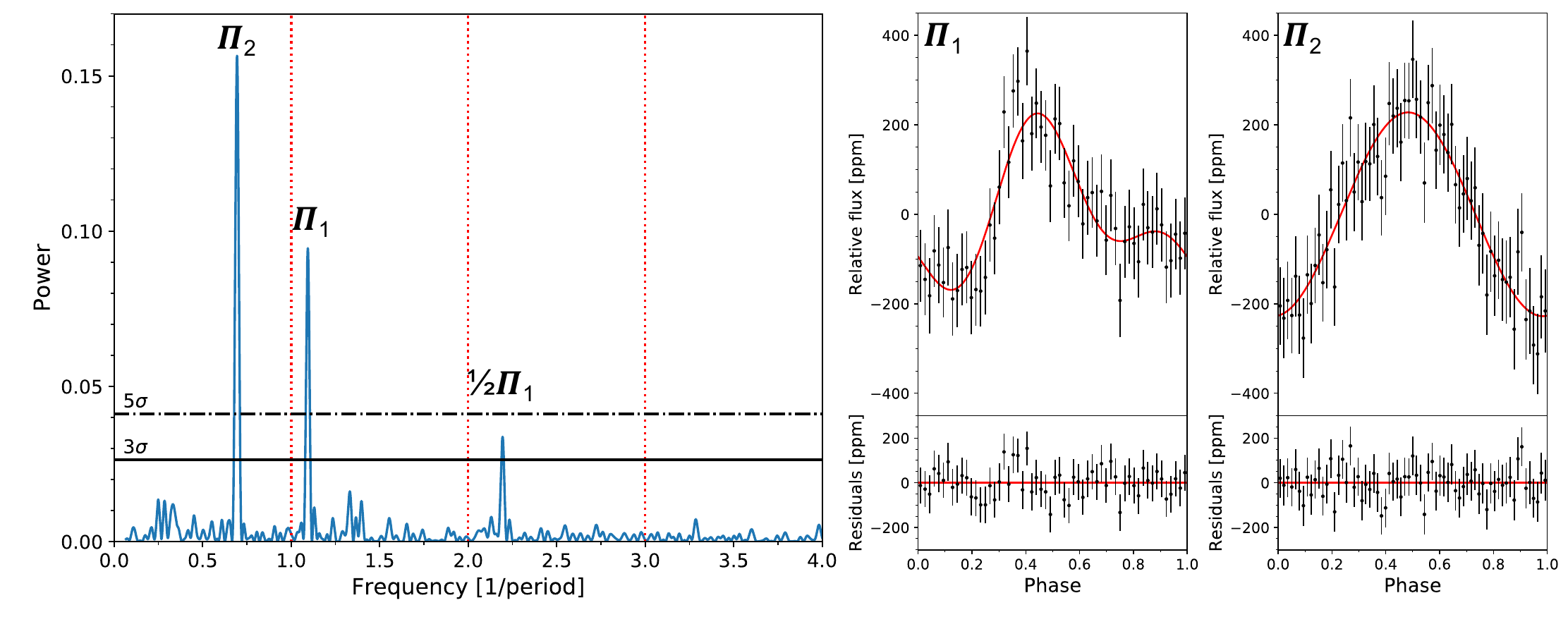}
\caption{The left plot shows the Lomb--Scargle periodogram of the residuals from the fit to the SPOC-SAP light curve when not accounting for any additional astrophysical signals besides the orbital phase curve. The frequencies are given as a multiple of the orbital frequency. The vertical red lines indicate the location of the fundamental and first two harmonics of the planet's orbital frequency; no residual synchronous signals are present after the fit. The horizontal lines denote statistical significance levels. The three salient peaks in the periodogram are labeled for the two stellar variability periods, $\Pi_1$ and $\Pi_2$, as well as the first harmonic of the modulation at $\Pi_1$. The two plots on the right show the best-fit stellar variability signals at the two characteristic periods (red curves), along with the systematics-corrected light curve after removing the astrophysical phase-curve model and the eclipses (black data points). The photometry is phase-folded to the corresponding periods and binned at 15 and 20 minute intervals, respectively. The residuals are plotted in the bottom panels.}
\label{fig:act}
\end{figure*}

We mention in passing that the ellipsoidal distortion component of the star's phase-curve modulation contains additional higher-order terms at other harmonics of the cosine \citep[e.g.,][]{morris1985,morris1993}. The model in Equation~\eqref{star} only includes the first-order term at the first harmonic of the cosine. However, the second-order term, which is at the second harmonic (i.e., $\cos{3\phi}$), has a theoretically predicted amplitude that is at least an order of magnitude smaller than the leading-order term. In the context of our light-curve fits, this makes the expected second-order ellipsoidal distortion amplitude smaller than the characteristic uncertainties on the phase-curve amplitudes. We also note that previous analyses of Kepler phase curves revealed anomalously large second-harmonic phase-curve signals on HAT-P-7 and KOI-13, which were attributed to the large spin-orbit misalignments in both of those systems \citep[e.g.,][]{esteves2013,esteves2015}. In contrast, the TOI-2109 system is well aligned (see Section~\ref{sec:jointresults}), and we therefore do not expect an additional contribution to the photometric modulation at the second harmonic. Indeed, when fitting the \tess\ light curve using only the leading-order term (i.e., $\cos{2\phi}$), we did not find any periodicity in the residuals at the second harmonic of the orbital phase (see Figure~\ref{fig:act}). Therefore, we ignored all higher-order terms of the ellipsoidal distortion in the final analysis.

In the \texttt{ExoTEP} pipeline, both the transit and secondary eclipse light curves are modeled using \texttt{batman} \citep{batman}. For all of our \tess\ light-curve fits, we allowed the mid-transit time $T_c$, orbital period $P$, radius ratio $R_p/R_*$, impact parameter $b$, and scaled orbital semimajor axis $a/R_*$ to vary freely. Due to the low 30 minute cadence of the \tess\ data, the ingress and egress are not well resolved in the light curve. As such, we did not allow the quadratic limb-darkening coefficients $u_1$ and $u_2$ to vary freely, but instead placed Gaussian priors. The mean values were set to the coefficients tabulated in \citet{claret2018} for the nearest available combination of stellar parameters ($u_1 = 0.33$, $u_2 = 0.22$), and the width of each Gaussian was conservatively set at 0.05.

\subsubsection{Systematics Detrending}
\label{sec:systematics}

The SPOC-SAP light curve is not corrected for instrumental systematics. We downloaded the CBVs\footnote{\url{http://archive.stsci.edu/tess/bulk_downloads}} for the specific \tess\ camera and detector on which the target was located (Camera 1, CCD 4) and carried out a customized detrending procedure. The systematics were modeled as a linear combination of the CBVs $\nu_k(t)$:
\begin{equation}\label{cbv}
S_{\mathrm{CBV}}(t) = c_{0} + \sum\limits_{k=1}^{8} c_k \nu_k(t).
\end{equation}
Eight CBVs were determined by the SPOC pipeline in Sector 25 and included in the downloaded light-curve files.

We fit the SPOC-SAP light curve to the combined phase-curve and systematics model using two approaches. For the first approach, we included all eight CBVs in the detrending model (CBV-full), while for the second approach we only included the CBVs with significant coefficients in the fit (CBV-opt). Each of the four data segments was fit separately, with the optimal combination of CBVs determined for each segment using the Bayesian information criterion (BIC). The sets of CBVs included in the CBV-opt fits are (3,7), (2,3,7), (2,3,6,7), and (1,2,3,6) for segments 1--4, respectively.

The QLP-SAP light curve was generated using a different aperture than the SPOC pipeline. Therefore, the SPOC-generated CBVs are not applicable, and we instead utilized a standard polynomial in time to model the long-term systematics in each data segment:
\begin{equation}\label{poly}
S_{\mathrm{poly}}(t) = \sum\limits_{k=0}^{N} a_k (t - t_{0})^k.
\end{equation}
Here, $t_{0}$ is the first timestamp of the segment, and $N$ is the order of the detrending polynomial. When determining the optimal polynomial order for each data segment, we considered both the BIC and the Akaike information criterion \citep[AIC;][]{akaike1974}; the AIC penalizes the addition of free parameters less severely than the BIC, resulting in higher-order polynomials. With simultaneous stellar variability modeling (Section~\ref{sec:activity}), the optimal orders for the four segments when considering the BIC are 1, 3, 2, and 2, while the AIC prefers 10, 8, 10, and 5, respectively.

For both SPOC-SAP and QLP-SAP light-curve fits, the best-fit systematics model for each segment was removed from the photometry prior to the joint fits of all four segments. The joint fits did not include any additional systematics modeling.

\subsubsection{Stellar Variability}
\label{sec:activity}

Close inspection of the residuals from the SPOC-SAP light-curve fits revealed additional short-term time-correlated brightness variations. Figure~\ref{fig:act} shows the Lomb--Scargle periodogram of the residuals as a function of the planet's orbital frequency. While the phase-curve model has fit away all photometric modulation that is synchronous with the orbit, there are two prominent peaks corresponding to periods of roughly $\Pi_1 = 0.61$ days and $\Pi_2 = 0.97$ days; additionally, a smaller peak is located at twice the frequency of the 0.61 day period, i.e., a variation at the first harmonic. These signals indicate that two distinct stellar variability frequencies are present in the data. The 0.97 day signal lies close to the host star's rotation period as implied by the measured \vsini\ and $R_*$ ($P_{\mathrm{rot}}/\sin i_* = 1.05 \pm 0.04$ days; see Section~\ref{sec:star}). 

We checked the Lomb--Scargle periodograms for the QLP-SAP light curves of all targets within $4'$ of TOI-2109: TIC 284467967 ($T = 13.2$ mag), TIC 284467994 ($T = 13.5$ mag), TIC 392476048 ($T = 11.6$ mag), and TIC 392476087 ($T = 11.6$ mag). None of them shows any periodicity near $\Pi_1$ or $\Pi_2$. Therefore, we assumed that the stellar variability signal belongs to the target host star.

We modeled the two variability signals using generalized sinusoids at the characteristic periods $\Pi_1$ and $\Pi_2$:
\begin{align}
    \Psi_1(t)&=1 + \alpha_1\sin(\phi_1) + \beta_1\cos(\phi_1) \notag\\
            &\quad\quad  + \alpha'_1\sin(2\phi_1) + \beta'_1\cos(2\phi_1),\\
    \Psi_2(t)&=1 + \alpha_2\sin(\phi_2) + \beta_2\cos(\phi_2).
\end{align}
The zero-points of the corresponding variability phases are set at the reference time $T_\mathrm{ref} =  2{,}458{,}997$ $\mathrm{BJD}_{\mathrm{TDB}}$, which is the integer Julian date closest to the median of the \tess\ time series: $\phi_1 = 2\pi(t - T_\mathrm{ref})/\Pi_1$ and $\phi_2 = 2\pi(t - T_\mathrm{ref})/\Pi_2$. The coefficients marked with primes are the amplitudes of the first harmonic terms at the $\sim$0.61 day period.

To simultaneously retrieve the parameter values from the full-orbit phase-curve and stellar variability models, we multiplied $\Psi_1(t)$ and $\Psi_2(t)$ by $F_{*}(t)$ in Equation~\eqref{astro}. The amplitudes and variability periods were allowed to vary freely in the fits. For completeness, we also present the results from the initial SPOC-SAP light-curve fits that did not account for stellar variability. The stellar variability was included in the full light-curve model when fitting the QLP-SAP photometry; QLP-SAP light-curve fits without stellar variability modeling necessitated very high orders ($>$15) in the systematics detrending polynomials for every data segment, making the analysis untenable.

\subsubsection{Fitting and Model Selection}
\label{sec:select}

\texttt{ExoTEP} uses the affine-invariant Markov Chain Monte Carlo (MCMC) routine \texttt{emcee} \citep{emcee} to compute the posterior distributions of all fit parameters. Each fit consisted of two steps. In the first MCMC run, we included a per-point uncertainty parameter $\sigma$, which was allowed to vary freely to ensure that the resultant best-fit model has a reduced $\chi^2$ of one. The value of $\sigma$ represents the scatter in the light curve and includes the contributions from both photon noise (i.e., white noise) and time-correlated noise (i.e., red noise) at the 30 minute cadence of the observations.

\begin{figure}[t]
\includegraphics[width=\linewidth]{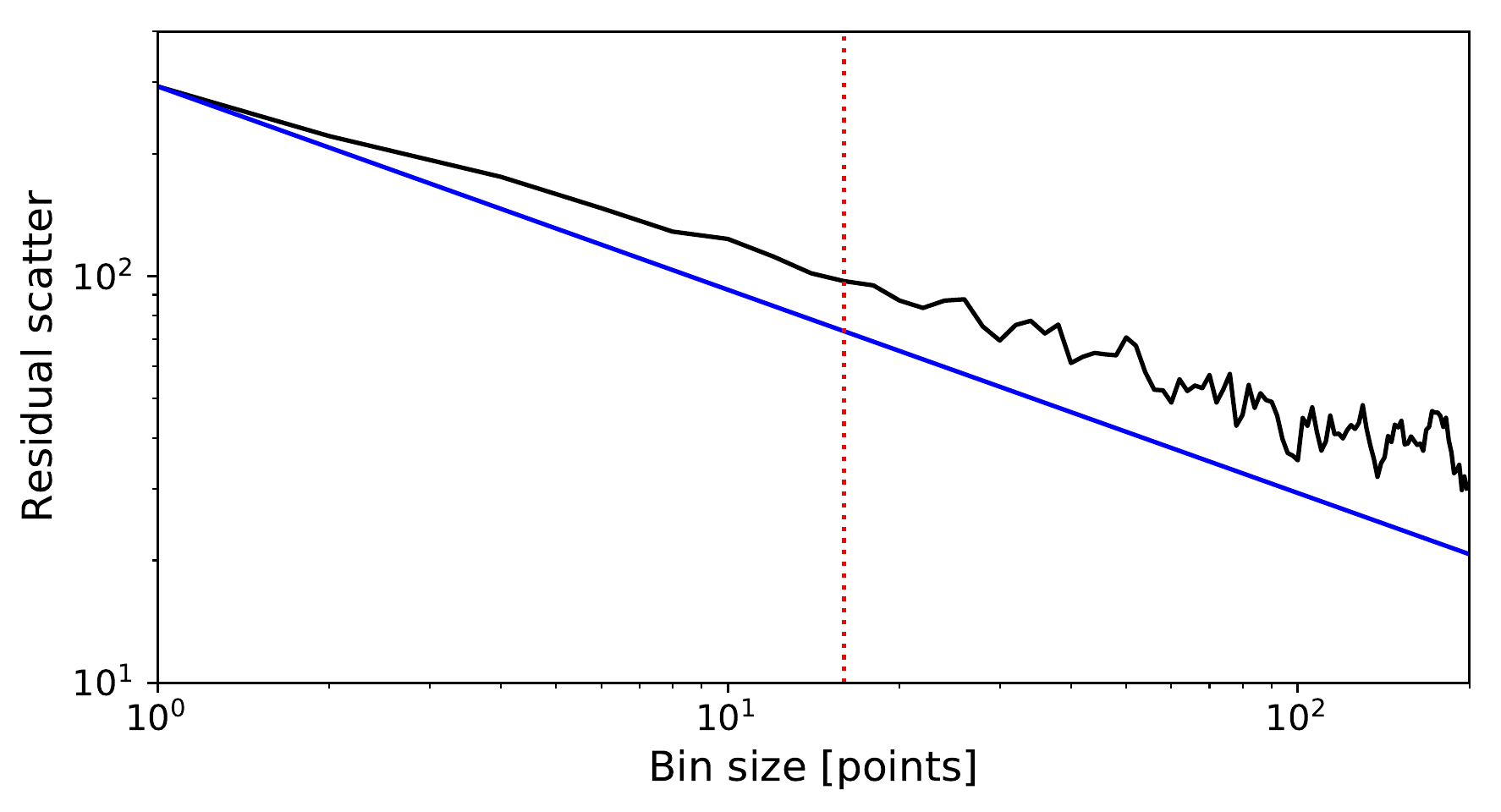}
\caption{Plot of the binned residual scatter from the SPOC-SAP light-curve fit using the full CBV detrending method, for various bin sizes (black curve). The expected $1/\sqrt{n}$ scaling for pure white noise is shown by the blue line. The residual scatter shows a positive deviation from the white noise trend, indicating the presence of significant time-correlated noise at hour-long timescales. The vertical red dashed line denotes a bin size of 16 (i.e., 8 hr).}
\label{fig:rms}
\end{figure}

To address the effect of red noise at longer timescales on the \tess\ light-curve fit results, we computed the standard deviation of the residuals, binned at various intervals, and compared the resultant values to the expected $1/\sqrt{n}$ scaling for pure white noise (e.g., \citealt{pont2006}; see \citealt{wong2020year1} for details on the specific implementation described here). In all cases, the binned residual scatter showed a positive deviation from the white noise trend (see Figure~\ref{fig:rms} for an example plot). This indicates the presence of significant time-correlated noise at timescales longer than a few hours, which is particularly relevant for our phase-curve analysis, because the characteristic flux modulations from the atmospheric brightness variation and ellipsoidal distortion occur on those timescales.

We accounted for the contribution of additional red noise in our fits by computing the average fractional deviation $\beta$ between the binned residual scatter and the white noise trend for bin sizes up to 16 (i.e., 8 hr, or roughly half the orbital period) and multiplying this ratio by the fitted per-point uncertainty $\sigma$ from the first MCMC run. Typical values of $\beta$ for fits that included stellar variability modeling ranged from 1.15 to 1.25. We then fixed the flux uncertainty to this inflated value $\sigma_{r} \equiv \beta\sigma$ and ran the MCMC analysis a second time. The resultant posteriors are broader, reflecting the added contribution of red noise to the overall photometric uncertainty.

We carried out fits to the full \tess\ light curves from both SPOC and QLP using various systematics detrending techniques (SPOC: CBV-full and CBV-opt; QLP: Poly-BIC and Poly-AIC). Table~\ref{tab:comparison} lists the results from these fits, along with the corresponding fitted scatter levels $\sigma$ and red-noise-inflated per-point uncertainties $\sigma_{r}$. Unsurprisingly, the SPOC-SAP fits that did not account for stellar variability yielded significantly higher scatter at both the native 30 minute cadence and longer timescales. Meanwhile, the noise level in the residuals from the QLP-SAP fits is systematically higher than the residual scatter from the SPOC-SAP fits.

\begin{deluxetable*}{lcccccc}[t]
\setlength{\tabcolsep}{2.5pt}
\tablewidth{0pt}
\tabletypesize{\scriptsize}
\tablecaption{
    Comparison of SPOC and QLP Light-curve Fits with Different Fitting and Detrending Methods 
    \label{tab:comparison}
}
\tablehead{\vspace{-0.1cm} \\ & \multicolumn{1}{c}{\textbf{SPOC-SAP}} & \multicolumn{1}{c}{SPOC-SAP} & \multicolumn{1}{c}{SPOC-SAP} & \multicolumn{1}{c}{SPOC-SAP} & \multicolumn{1}{c}{QLP-SAP} & \multicolumn{1}{c}{QLP-SAP} \\ & \multicolumn{1}{c}{\textbf{with Variability}} & \multicolumn{1}{c}{with Variability} & \multicolumn{1}{c}{without Variability} & \multicolumn{1}{c}{without Variability} & \multicolumn{1}{c}{with Variability} & \multicolumn{1}{c}{with Variability} \vspace{-0.2cm}\\
    \multicolumn{1}{l}{Parameter} \vspace{-0.1cm}&
    \colhead{\textbf{CBV-full}}                     &
    \colhead{CBV-opt}  &
    \colhead{CBV-full}                     &
    \colhead{CBV-opt}                     &
    \colhead{Poly-BIC}  &
    \colhead{Poly-AIC}            
}
\startdata
\multicolumn{2}{l}{Transit and Orbital Parameters} & & & & & \\
$R_p/R_*$   & $\mathbf{0.0789 \pm 0.0020}$ & $0.0790 \pm 0.0027$  & $0.0765_{-0.0012}^{+0.0030}$ & $0.0768_{-0.0012}^{+0.0025}$ & $0.0789_{-0.0018}^{+0.0028}$ & $0.0795 \pm 0.0024$ \\
$T_{c}$ \tablenotemark{\scriptsize a} & $\mathbf{997.16653 \pm 0.00016}$ & $997.16652 \pm 0.00017$ & $997.16653 \pm 0.00024$ & $997.16650 \pm 0.00024$ & $997.16636 \pm 0.00019$ & $997.16642 \pm 0.00017$ \\
$P$ (days)  & $\mathbf{0.672469 \pm 0.000015}$ & $0.672469 \pm 0.000015$ & $0.672470 \pm 0.000022$ & $0.672468 \pm 0.000020$ & $0.672478 \pm 0.000018$ & $0.672483 \pm 0.000014$ \\
$b$       & $\mathbf{0.61_{-0.18}^{+0.10}}$ &   $0.62_{-0.24}^{+0.13}$ & $0.29_{-0.24}^{+0.35}$ & $0.37_{-0.25}^{+0.24}$ & $0.53_{-0.20}^{+0.17}$ & $0.59_{-0.23}^{+0.13}$ \\
$a/R_*$     & $\mathbf{2.60_{-0.24}^{+0.30}}$ &   $2.57_{-0.30}^{+0.39}$ & $3.04_{-0.51}^{+0.12}$ & $2.96_{-0.38}^{+0.19}$ & $2.73_{-0.37}^{+0.27}$ & $2.62_{-0.31}^{+0.34}$ \\
$u_1$\tablenotemark{\scriptsize b} & $\mathbf{0.34 \pm 0.04}$ & $0.34 \pm 0.05$ & $0.34 \pm 0.05$ & $0.34 \pm 0.05$ & $0.34 \pm 0.05$ & $0.33 \pm 0.05$ \\
$u_2$\tablenotemark{\scriptsize b} & $\mathbf{0.24 \pm 0.05}$   & $0.22 \pm 0.04$ & $0.23 \pm 0.04$ & $0.23 \pm 0.05$ & $0.23 \pm 0.05$ & $0.23 \pm 0.05$ \\
\multicolumn{2}{l}{Stellar Variability Parameters} & & & & & \\
$\Pi_1$ (days) & $\mathbf{0.61393 \pm 0.00051}$ & $0.61389 \pm 0.00048$ & $\dots$ & $\dots$ & $0.61450 \pm 0.00058$ & $0.61437 \pm 0.00051$ \\
$\alpha_1$ (ppm) & $\mathbf{-11 \pm 15}$ & $-15 \pm 15$ & $\dots$ & $\dots$ & $-7 \pm 18$ & $-6 \pm 17$ \\
$\beta_1$ (ppm) & $\mathbf{-149 \pm 14}$ & $-158 \pm 15$ & $\dots$ & $\dots$ & $-180 \pm 21$ & $-169 \pm 17$ \\
$\alpha'_1$ (ppm) & $\mathbf{-71 \pm 15}$ & $-65 \pm 17$ & $\dots$ & $\dots$ & $-62 \pm 22$ & $-65 \pm 18$ \\
$\beta'_1$ (ppm) & $\mathbf{64 \pm 15}$ & $61 \pm 17$ & $\dots$ & $\dots$ & $88 \pm 18$ & $95 \pm 18$ \\
$\Pi_2$ (days) & $\mathbf{0.9675 \pm 0.0013}$ & $0.9676 \pm 0.0014$ & $\dots$ & $\dots$ & $0.9690 \pm 0.0013$ & $0.9688 \pm 0.0013$ \\
$\alpha_2$ (ppm) & $\mathbf{25 \pm 17}$ & $21 \pm 15$ & $\dots$ & $\dots$ & $44 \pm 24$ & $29 \pm 18$ \\
$\beta_2$ (ppm) & $\mathbf{-222 \pm 14}$ & $-213 \pm 15$ & $\dots$ & $\dots$ & $-262 \pm 20$ & $-248 \pm 16$ \\
\multicolumn{2}{l}{Phase-curve Parameters} & & & & & \\
$\bar{f_p}$ (ppm)  & $\mathbf{367 \pm 39}$  &  $363 \pm 42$ & $387 \pm 61$ & $371 \pm 69$ & $384 \pm 52$ & $380 \pm 44$ \\
$A_{\mathrm{atm}}$ (ppm) & $\mathbf{360 \pm 22}$  & $357 \pm 21$ & $366 \pm 27$ & $367 \pm 31$ & $401 \pm 29$ & $396 \pm 25$ \\
$A_{\mathrm{ellip}}$ (ppm)   & $\mathbf{240 \pm 19}$ & $243 \pm 23$ & $223 \pm 27$ & $231 \pm 28$ & $245 \pm 25$ & $249 \pm 23$ \\
$A_{1}$ (ppm)\tablenotemark{\scriptsize c} & $\mathbf{31 \pm 16}$ & $40 \pm 14$ & $25 \pm 24$ & $20 \pm 23$ & $25 \pm 19$ & $31 \pm 16$ \\
\multicolumn{2}{l}{Derived Parameters} & & & & &  \\
$D_{d}$ (ppm)\tablenotemark{\scriptsize d}  & $\mathbf{726 \pm 46}$ & $720 \pm 47$ & $751 \pm 71$ & $736 \pm 81$ & $785 \pm 56$ & $778 \pm 47$\\
$D_{n}$ (ppm)\tablenotemark{\scriptsize d}  & $\mathbf{8 \pm 44}$ & $3 \pm 46$ & $23 \pm 63$  & $9 \pm 70$ & $-18 \pm 61$ & $-16 \pm 54$ \\
$i$ (deg)   & $\mathbf{76.5^{+5.1}_{-4.0}}$ & $76.2^{+6.6}_{-5.2}$ & $84.6^{+4.5}_{-9.1}$ & $82.8^{+5.0}_{-6.7}$ & $78.8^{+4.9}_{-6.1}$ & $77.0^{+6.1}_{-5.2}$\\
\multicolumn{2}{l}{Fit-quality Metrics} & & & & &  \\
$\sigma$ (ppm)\tablenotemark{\scriptsize e} & $\mathbf{295}$ & 293 & 349 & 351 & 367 & 356\\
$\sigma_{r}$ (ppm)\tablenotemark{\scriptsize e} & $\mathbf{359}$ & 363 & 547 & 549 & 472 & 409 \\
\enddata
\textbf{Notes.}
\vspace{-0.25cm}\tablenotetext{\textrm{a}}{$\mathrm{BJD}_{\mathrm{TDB}}- 2{,}458{,}000$.}
\vspace{-0.25cm}\tablenotetext{\textrm{b}}{Limb-darkening coefficients were constrained by priors: $u_1 = 0.33 \pm 0.05$, $u_2 = 0.22 \pm 0.05$.}
\vspace{-0.25cm}\tablenotetext{\textrm{c}}{$A_1$ is the total harmonic power in the photometry at the sine of the orbital phase, which includes the Doppler-boosting signal from the host star and any phase shift in the planet's atmospheric brightness modulation.}
\vspace{-0.25cm}\tablenotetext{\textrm{d}}{Dayside flux (secondary eclipse depth) and nightside brightness of the planet, derived from the fitted average planetary flux $\bar{f_p}$ and atmospheric brightness modulation amplitude $A_{\mathrm{atm}}$.}
\vspace{-0.25cm}\tablenotetext{\textrm{e}}{$\sigma$: scatter in the residuals from the best-fit phase-curve model; $\sigma_{r}$: per-point uncertainty, inflated to account for red noise.}
\vspace{-0.8cm}
\end{deluxetable*}

When comparing the values from the six listed sets of results, we report a high level of mutual consistency. Most notably, the results from the SPOC-SAP fits that did or did not account for stellar variability agree with each other at much better than the $1\sigma$ level, which indicates that our treatment of stellar variability does not have any significant effect on the phase-curve results, aside from improving the time-correlated noise and reducing parameter uncertainties. Likewise, comparisons of fits that utilized full CBV detrending vs. optimized CBV detrending show full statistical consistency, as do the QLP-SAP light-curve fits with Poly-BIC vs. Poly-AIC detrending.

Looking at the results of the SPOC-SAP and QLP-SAP light-curve fits (with stellar variability modeling) side by side, we find that the transit-shape parameters and orbital ephemeris are mutually consistent to well within $1\sigma$. Similarly, most of the phase-curve parameters and derived quantities such as secondary eclipse depth $D_{d}$ and nightside flux $D_{n}$ agree with one another at better than the $1\sigma$ level. The exception is the atmospheric brightness modulation amplitude $A_{\mathrm{atm}}$, for which the QLP-SAP photometry prefers a value that is up to $1.2\sigma$ larger than the corresponding measurements derived from the SPOC-SAP light curve. Meanwhile, the stellar variability parameters from the SPOC-SAP and QLP-SAP fits are broadly consistent, with no deviations larger than $2\sigma$. All in all, we find that the astrophysical parameter values of interest are highly robust to the specific choice of photometric extraction, systematics detrending methodology, and stellar variability modeling.

The versions of the SPOC-SAP light curve corrected using the full and optimized CBV detrending methods yielded very similar fit quality with respect to both residual scatter and time-correlated noise. We selected the former for the main results of this paper, on account of the marginally better red noise level. 

\begin{figure}[t!]
\includegraphics[width=\linewidth]{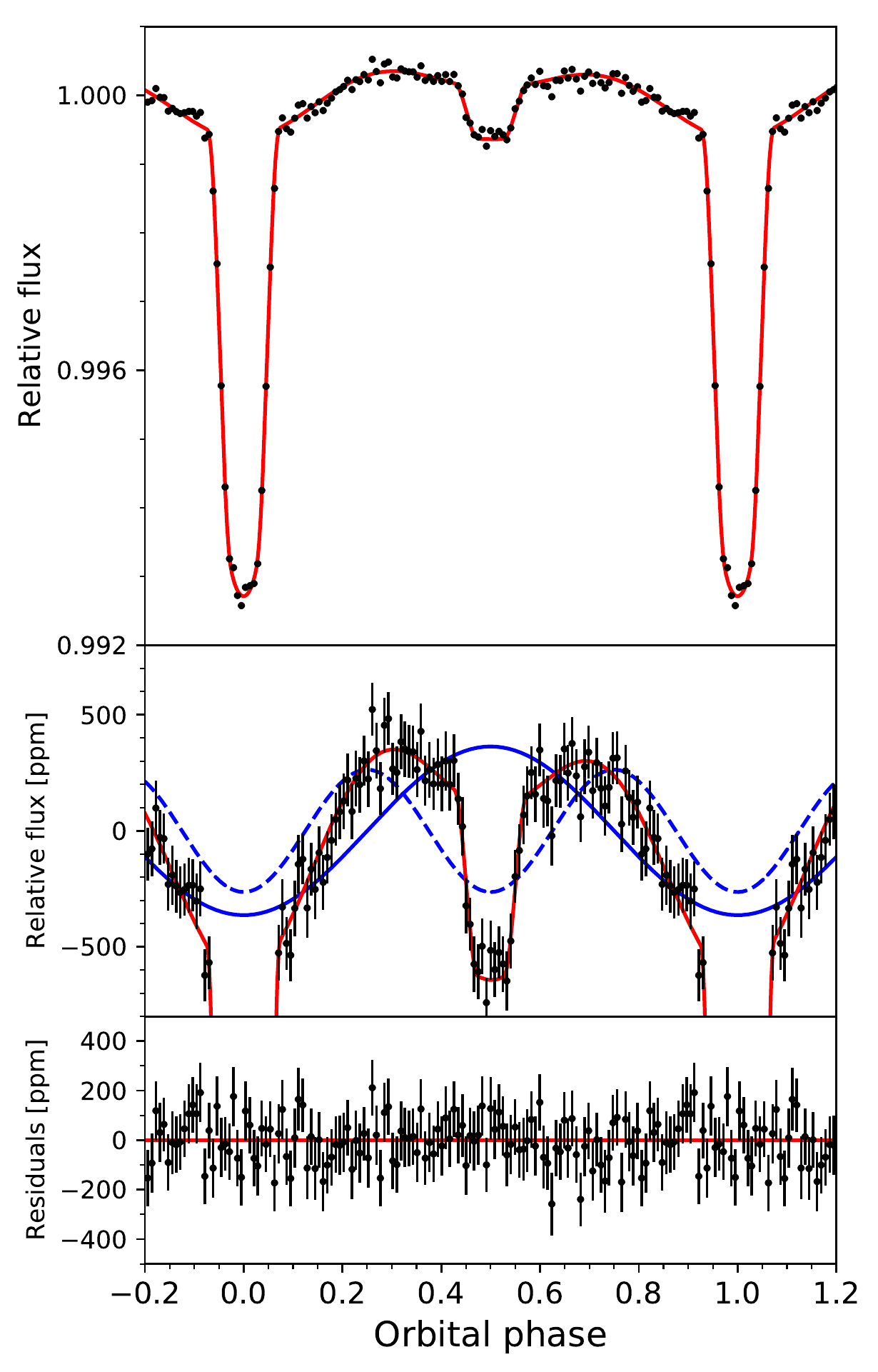}
\caption{Top panel: the phase-folded \tess\ SPOC-SAP light curve of TOI-2109, binned at 8 minute intervals and corrected for instrumental systematics using the full CBV detrending method. The stellar variability modulations have also been divided out from the light curve. The red curve is the best-fit light-curve model from the final joint fit (Table~\ref{tab:jointfit}). Middle panel: a zoomed-in view of the phase-curve modulation and secondary eclipse. The solid and dashed blue lines show the high-S/N atmospheric brightness modulation and ellipsoidal distortion components of the phase curve individually. Bottom panel: the corresponding residuals from the best-fit model.}
\label{fig:phasecurve}
\end{figure}

\subsubsection{Results}
\label{sec:results}

The phase-folded SPOC-SAP light curve, corrected for systematics using the full CBV detrending model and with the best-fit stellar variability signals removed, is shown in Figure~\ref{fig:phasecurve}. In addition to the high-S/N secondary eclipse with a depth of $726 \pm 46$ ppm, we retrieved significant ($>$$10\sigma$) phase-curve amplitudes corresponding to the atmospheric brightness modulation of the planet and the ellipsoidal distortion of the star. The nightside flux is consistent with zero. We also obtained a marginal phase-curve amplitude at the sine of the orbital frequency ($31 \pm 16$ ppm).

The transit-shape parameters $b$ and $a/R_*$ are not well constrained by the \tess\ light curve alone, due to the low cadence of the observations. The best-fit values indicate an orbit that is moderately inclined from edge-on. The precision of these values is substantially improved when including the ground-based light curves and spectroscopic transit observations in the final joint fit (Section~\ref{sec:jointfit}). The two fitted stellar variability periods are $\Pi_1 = 0.61393 \pm 0.00051$ days and $\Pi_2 = 0.9675 \pm 0.0013$ days. Figure~\ref{fig:act} shows the \tess\ light curve phase-folded on the two different variability periods, with the best-fit full-orbit phase-curve model removed.

The extremely small planet--star separation makes any significant orbital eccentricity highly unlikely. Nevertheless, to probe for possible deviations from a circular orbit, we carried out a fit that included $e\cos\omega$ and $e\sin\omega$ as free parameters. The constraints on eccentricity here are primarily driven by the relative timing of the secondary eclipse. We obtained $e\cos\omega = -0.0005 \pm 0.0029$ and $e\sin\omega = 0.003 \pm 0.045$, corresponding to a formal $2\sigma$ upper limit of $e < 0.07$. Therefore, we conclude that the orbit of TOI-2109b is indeed consistent with circular.

\subsection{Ground-based Light-curve Fits}
\label{sec:groundfit}

The raw light curves obtained from the various ground-based observations described in Section~\ref{sec:groundphot} were affected by instrumental systematics and observing conditions, such as airmass and sky background level. To detrend these systematics, we considered all possible linear combinations of relevant quantities, including the measured centroid position of the target ($x$, $y$), width of the target's point-spread function $\Delta_{\mathrm{PSF}}$, airmass AM, sky background level in the vicinity of the target $F_{\mathrm{sky}}$, and the total flux of nearby companion stars $F_{\mathrm{comp}}$ used to derive the differential photometry. For every data set, an additional baseline offset $c_0$ was used to properly normalize the light curve. We also experimented with modeling a linear trend in time $t$. In the case of the WIRC \Ks-band light curve of the secondary eclipse, the extracted fluxes from the two selected companion stars $F_1$ and $F_2$ were included as additional detrending vectors (see Section~\ref{sec:wirc}). The optimized combinations of detrending vectors were determined through minimizing the BIC for each data set.

\begin{deluxetable}{lcccc}[t!]
\setlength{\tabcolsep}{15pt}
\tablewidth{0pc}
\renewcommand{\arraystretch}{0.9}
\tabletypesize{\footnotesize}
\tablecaption{
    Priors on Limb-darkening Coefficients
    \label{tab:priors}
}
\tablehead{\vspace{-0.2cm}\\
    \multicolumn{1}{l}{Parameter} & & & &
    \multicolumn{1}{c}{Value}   
}
\startdata
$u_{1,\mathrm{TESS}}$ & & & & $0.33 \pm 0.05$\\
$u_{2,\mathrm{TESS}}$ & & & & $0.22 \pm 0.05$\\
$u_{1,B}$ & & & & $0.56 \pm 0.05$\\
$u_{2,B}$ & & &  & $0.25 \pm 0.05$\\
$u_{1,g'}$ & & & & $0.54 \pm 0.05$\\
$u_{2,g'}$ & & &  & $0.22 \pm 0.05$\\
$u_{1,r'}$ & & & & $0.40 \pm 0.05$\\
$u_{2,r'}$ & & &  & $0.23 \pm 0.05$\\
$u_{1,R}$ & & & & $0.39 \pm 0.05$\\
$u_{2,R}$ & & &  & $0.22 \pm 0.05$\\
$u_{1,i'}$ & & & & $0.33 \pm 0.05$\\
$u_{2,i'}$ & & &  & $0.21 \pm 0.05$\\
$u_{1,I}$ & & & & $0.31 \pm 0.05$\\
$u_{2,I}$ & & &  & $0.20 \pm 0.05$\\
$u_{1,z'/z_s}$  & & & & $0.26 \pm 0.05$\\
$u_{2,z'/z_s}$  & & & & $0.21 \pm 0.05$\\
\enddata
\vspace{-2cm}
\end{deluxetable}

In our fits, we only considered ground-based light curves with full transit coverage, due to the possibility of significant biases to the transit timing and transit-shape parameters when modeling partial light curves. Each transit data set was fit using \texttt{ExoTEP}. The $B$-, $r'$-, $R$-, $i'$-, $I$-, and $z'$/$z_s$-band quadratic limb-darkening coefficients were constrained by priors derived from the tabulated values in \citet{claretsloan}, with $1\sigma$ Gaussian widths uniformly set to 0.05. The full set of limb-darkening priors used in our analysis is given in Table~\ref{tab:priors}.

\begin{deluxetable*}{lcccc}[t!]
\setlength{\tabcolsep}{20pt}
\tablewidth{0pc}
\renewcommand{\arraystretch}{1.0}
\tabletypesize{\footnotesize}
\tablecaption{
    Individual Ground-based Transit Fit Results
    \label{tab:groundbased}
}
\tablehead{\vspace{-0.25cm} \\
    \multicolumn{1}{l}{Data Set} &  \multicolumn{1}{c}{UT Date} & \multicolumn{1}{c}{Detrending Vectors\tablenotemark{\scriptsize a}} & \multicolumn{1}{c}{$R_p/R_*$} &
    \multicolumn{1}{c}{$\sigma$ (ppm)}   
}
\startdata
MuSCAT2 $g'$ & 2021 Jun 12 & $\dots$ & $0.0799 \pm 0.0012$ & 2555 \\
MuSCAT2 $r'$ & 2021 Jun 12 & $\dots$ & $0.0834 \pm 0.0011$ & 2597 \\
MuSCAT2 $i'$ & 2021 Jun 12 & $\dots$ & $0.0817 \pm 0.0012$ & 2113\\
MuSCAT2 $z_s$ & 2021 Jun 12 & $\dots$ & $0.0798 \pm 0.0010$ & 2914 \\
\hline
MuSCAT3 $g'$ \#1 & 2021 May 24 & $\dots$ & $0.0818 \pm 0.0007$ & 810 \\
MuSCAT3 $r'$ \#1 & 2021 May 24 & $t$, AM, $F_{\mathrm{comp}}$, $\Delta_{\mathrm{PSF}}$ & $0.0821 \pm 0.0011$ & 811 \\
MuSCAT3 $i'$ \#1 & 2021 May 24 & $t$, AM, $F_{\mathrm{comp}}$, $\Delta_{\mathrm{PSF}}$ & $0.0827 \pm 0.0012$ & 914 \\
MuSCAT3 $z_s$ \#1 & 2021 May 24 & $F_{\mathrm{comp}}$, $\Delta_{\mathrm{PSF}}$ & $0.0811 \pm 0.0008$ & 666 \\
\hline
MuSCAT3 $g'$ \#2 & 2021 Jun 26 & $F_{\mathrm{sky}}$ & $0.0815 \pm 0.0008$ & 785 \\
MuSCAT3 $r'$ \#2 & 2021 Jun 26 & $F_{\mathrm{sky}}$ & $0.0809 \pm 0.0006$ & 808 \\
MuSCAT3 $i'$ \#2 & 2021 Jun 26 & AM, $F_{\mathrm{sky}}$ &  $0.0829 \pm 0.0008$ & 942 \\
MuSCAT3 $z_s$ \#2 & 2021 Jun 26 & $\Delta_{\mathrm{PSF}}$, $F_{\mathrm{sky}}$ & $0.0813 \pm 0.0007$ & 635 \\
\hline
SSO $B$ & 2020 Aug 24 & AM, $F_{\mathrm{comp}}$, $\Delta_{\mathrm{PSF}}$, $F_{\mathrm{sky}}$ & $0.0777 \pm 0.0075$  & 3711 \\
ULMT $r'$ & 2020 Jul 31 & $x$, $t$, $\Delta_{\mathrm{PSF}}$ & $0.0794 \pm 0.0022$ & 2580  \\
WBRO $R$ & 2021 Apr 7 & $x$, $t$ & $0.0795 \pm 0.0025$ & 1631 \\
SAAO $i'$ & 2021 Apr 7 & AM, $F_{\mathrm{comp}}$ & $0.0810 \pm 0.0025$ & 2500 \\
GdP $i'$ & 2021 Apr 7 & AM, $F_{\mathrm{comp}}$, $\Delta_{\mathrm{PSF}}$ &  $0.0811 \pm 0.0018$ & 2239 \\
ULMT $i'$ & 2021 Apr 9 & $F_{\mathrm{comp}}$ & $0.0820 \pm 0.0021$ & 2123 \\
MLO $I$ & 2020 Aug 2 & $t$, AM, $F_{\mathrm{comp}}$, $\Delta_{\mathrm{PSF}}$, $F_{\mathrm{sky}}$ & $0.0860 \pm 0.0066$ & 4260 \\
KeplerCam $z'$ \#2 & 2021 Apr 9 & AM, $F_{\mathrm{comp}}$, $\Delta_{\mathrm{PSF}}$, $F_{\mathrm{sky}}$ & $0.0837 \pm 0.0018$ & 2610 \\
KeplerCam $z'$ \#3 & 2021 Apr 11 & $\dots$ & $0.0831 \pm 0.0018$ & 3291 \\
SSO $z_s$ & 2020 Aug 24 & $\dots$ & $0.0811 \pm 0.0033$ & 2257 \\
\enddata
\vspace{+0.15cm}\textbf{Note.}
\vspace{-0.15cm}\tablenotetext{\textrm{a}}{The optimized sets of detrending vectors used in the light-curve fits. Refer to the text for the definition of variables.}
\vspace{-0.5cm}
\end{deluxetable*}

The transit and secondary eclipse depths measured from a light-curve fit can be systematically affected by the unmodeled photometric variability of the system during the eclipsing event. This variability includes the planetary phase curve and modulations in the stellar brightness, which shift the out-of-eclipse baseline. To minimize the possibility of biases in our ground-based transit light-curve fits while simultaneously preserving a sufficient out-of-transit baseline, we excluded all data points that lie more than $0.1 \times P$ from the mid-transit time. 

To leverage the high precision of the simultaneous multiband MuSCAT2 and MuSCAT3 photometry, we initiated our ground-based light-curve analysis by jointly fitting each set of transit light curves. The MuSCAT2 transit fit yielded tight constraints on the mid-transit time ($T_c =  2{,}459{,}378.45955 \pm 0.00017$ BJD$_{\mathrm{TDB}}$) and transit-shape parameters ($b = 0.743 \pm 0.020$, $a/R_* = 2.274 \pm 0.057$). The first set of MuSCAT3 light curves provided even more precise measurements: $T_c =  2{,}459{,}358.957820 \pm 0.000086$ BJD$_{\mathrm{TDB}}$, $b = 0.730 \pm 0.012$, $a/R_* = 2.321 \pm 0.031$. The second set of MuSCAT3 transit photometry had the best photometric precision of all, yielding $T_c =  2{,}459{,}391.90867 \pm 0.00010$ BJD$_{\mathrm{TDB}}$, $b = 0.766 \pm 0.011$, and $a/R_* = 2.220 \pm 0.032$.

\begin{figure}[t!]
\includegraphics[width=\linewidth]{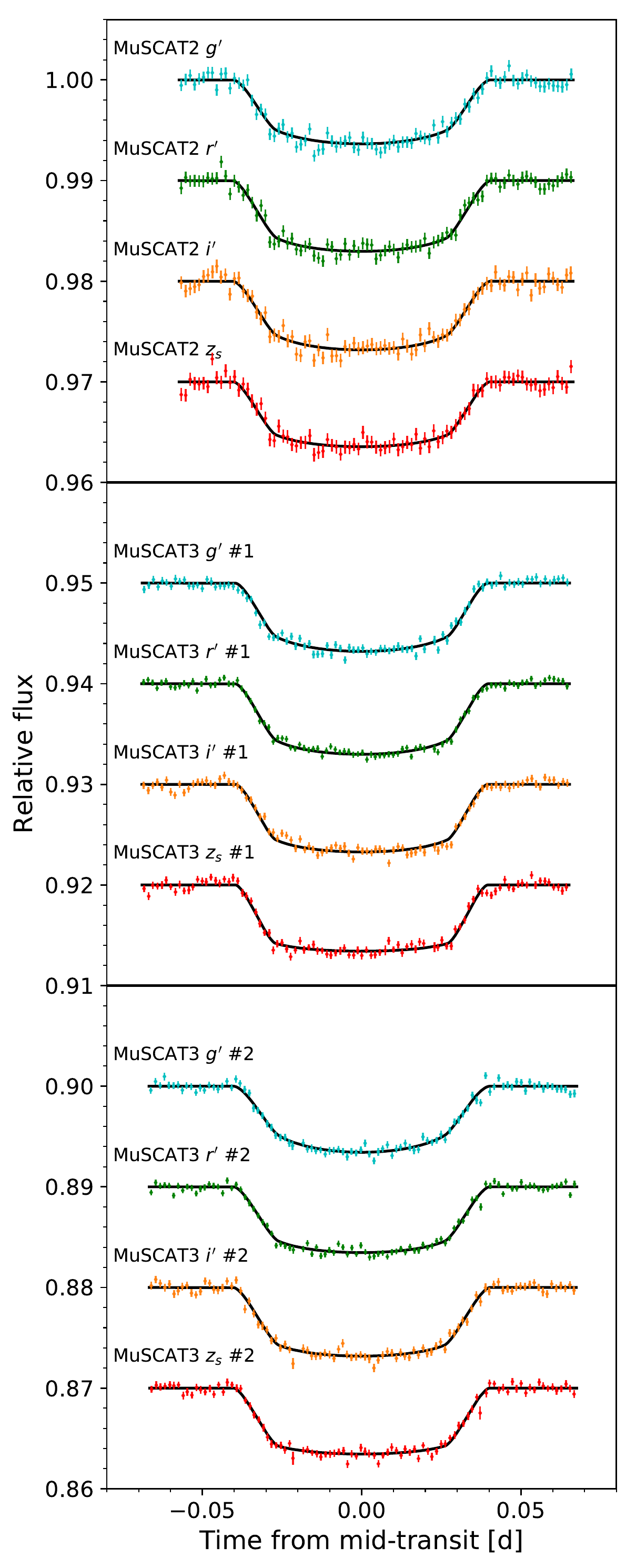}
\caption{The systematics-corrected MuSCAT2 and MuSCAT3 transit light curves, binned at 2 minute intervals. The black curves are the best-fit transit models, as derived from our independent joint fits of each set of light curves.}
\label{fig:muscattransits}
\vspace{+0.1cm}
\end{figure}

The individual MuSCAT2 and MuSCAT3 $R_p/R_*$ values are listed at the top of Table~\ref{tab:groundbased}. The detrended transit light curves are plotted in Figure~\ref{fig:muscattransits}. Notably, the MuSCAT3 \#1 transit depths show a high level of achromaticity, with all four measurements lying within $1.1\sigma$ of each other; meanwhile, the MuSCAT2  and MuSCAT3 \#2 transit depths show slightly larger variance ($2.4\sigma$ and $2.0\sigma$, respectively). The broad consistency in transit depths across the various photometric bands serves as supporting evidence against the false positive scenario of a blended eclipsing binary. 

\begin{figure}[t!]
\includegraphics[width=\linewidth]{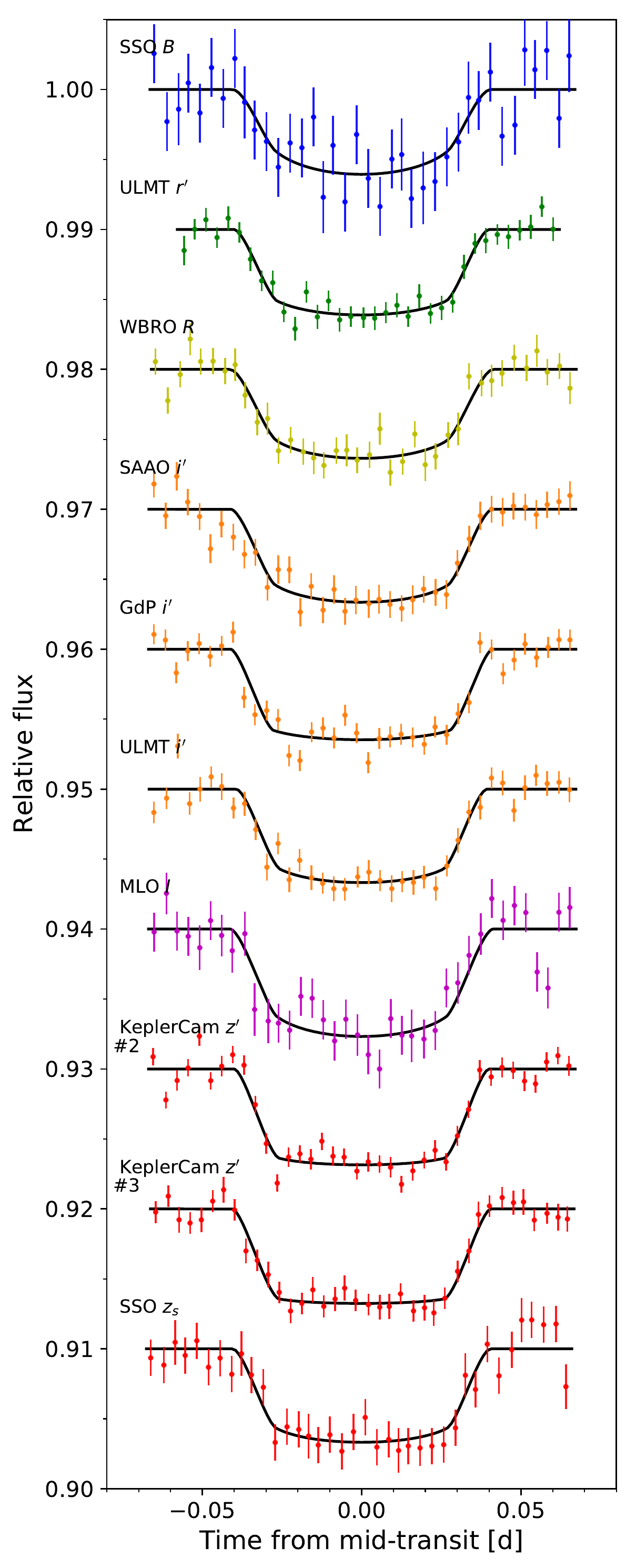}
\caption{Same as Figure~\ref{fig:muscattransits}, but for all other ground-based transit light curves considered in our analysis. The time series are binned at 5 minute intervals.}
\label{fig:transits}
\vspace{+0.6cm}
\end{figure}

The estimates of $b$ and $a/R_*$ from the MuSCAT2 and MuSCAT3 light-curve fits, which mutually agree at the $2.2\sigma$ level, have uncertainties that are almost an order of magnitude smaller than the corresponding errors derived from the \tess\ light curve alone (Table~\ref{tab:comparison}), highlighting the power of these high-cadence, high-S/N light curves in resolving the detailed transit geometry of the TOI-2109 system. For the remaining ground-based transit light curves, we used the most precise set of $b$ and $a/R_*$ values (from the MuSCAT3 \#2 fit) as Gaussian priors. Priors on $T_c$ were derived by interpolating the \tess, MuSCAT2, and MuSCAT3 timing measurements, assuming a linear orbital ephemeris. The full results of our individual ground-based transit light-curve fits are shown in Table~\ref{tab:groundbased}. In addition to the measured transit depth, the optimized detrending vector set and best-fit uniform per-point scatter $\sigma$ are provided for each light curve. The entries from non-MuSCAT facilities are sorted by bandpass. The corresponding systematics-corrected transit light curves are shown in Figure~\ref{fig:transits}. 

For the \Ks-band secondary eclipse fit, we utilized the same priors on orbital ephemeris and transit-shape parameters as in the ground-based transit light-curve fits; additionally, we fixed $R_p/R_*$ to the weighted average of the individual MuSCAT3 \#2 depths. Due to the relatively large planet--star flux contrast ratio in the \Ks band and the sizable temporal baseline, we expect some variation in the out-of-eclipse flux due to the planet's atmospheric brightness modulation. This phase-curve signal can bias the measured eclipse depth if not accounted for in the analysis \citep[e.g.,][]{bell2019}. We therefore included an additional quadratic function in time to remove any curvature from the planet's phase curve that is present in the out-of-eclipse data.

We measured an eclipse depth of $2027 \pm 88$ ppm and obtained a best-fit per-point uncertainty of 937 ppm. For the optimal set of detrending vectors, we utilized the two companion star fluxes $F_1$ and $F_2$, the airmass AM, and the sky background level $F_{\mathrm{sky}}$. The bottom panel of Figure~\ref{fig:wirc} shows the detrended light curve.

\subsection{Radial Velocity Fit}
\label{sec:rvfit}
To obtain a preliminary mass measurement of TOI-2109b, we used the \texttt{radvel} package \citep{radvel} and modeled the planet's orbital RV signal assuming a circular orbit and no additional planets in the system. The assumption of $e = 0$ was motivated by the results of our \tess\ light-curve fits (Section~\ref{sec:results}). The full transit duration is roughly 0.15 in orbital phase, and we excluded all RV measurements that were obtained within 0.08 in orbital phase of the mid-transit time; these RVs are denoted in Table~\ref{tab:rvs} by the superscripts. Due to the host star's rapid rotation, the precision of the RV measurements is quite poor, with some RV uncertainties as high as 1.5 \kms\ or more. Nevertheless, the phase-folded RVs show good phase coverage across the two quadratures, and the RVs calculated from our spectroscopic transit observations provide high-S/N data points near the primary transit. For the final results presented in this paper, we removed all RV measurements with uncertainties greater than 1 \kms.

In this stand-alone analysis of the RV signal, we placed Gaussian priors on the mid-transit time $T_c$ and orbital period $P$, using the median and $1\sigma$ values from the \tess\ phase-curve fit results (Table~\ref{tab:comparison}). In addition to the orbital ephemeris parameters and the RV semiamplitude $K_p$, we fit for the systemic RV offset and jitter of each instrument, $\gamma_{i}$ and $\sigma_{\mathrm{jit},i}$. The parameter space was sampled using the default MCMC routine within \texttt{radvel}. 

\begin{figure}[t]
\includegraphics[width = \linewidth]{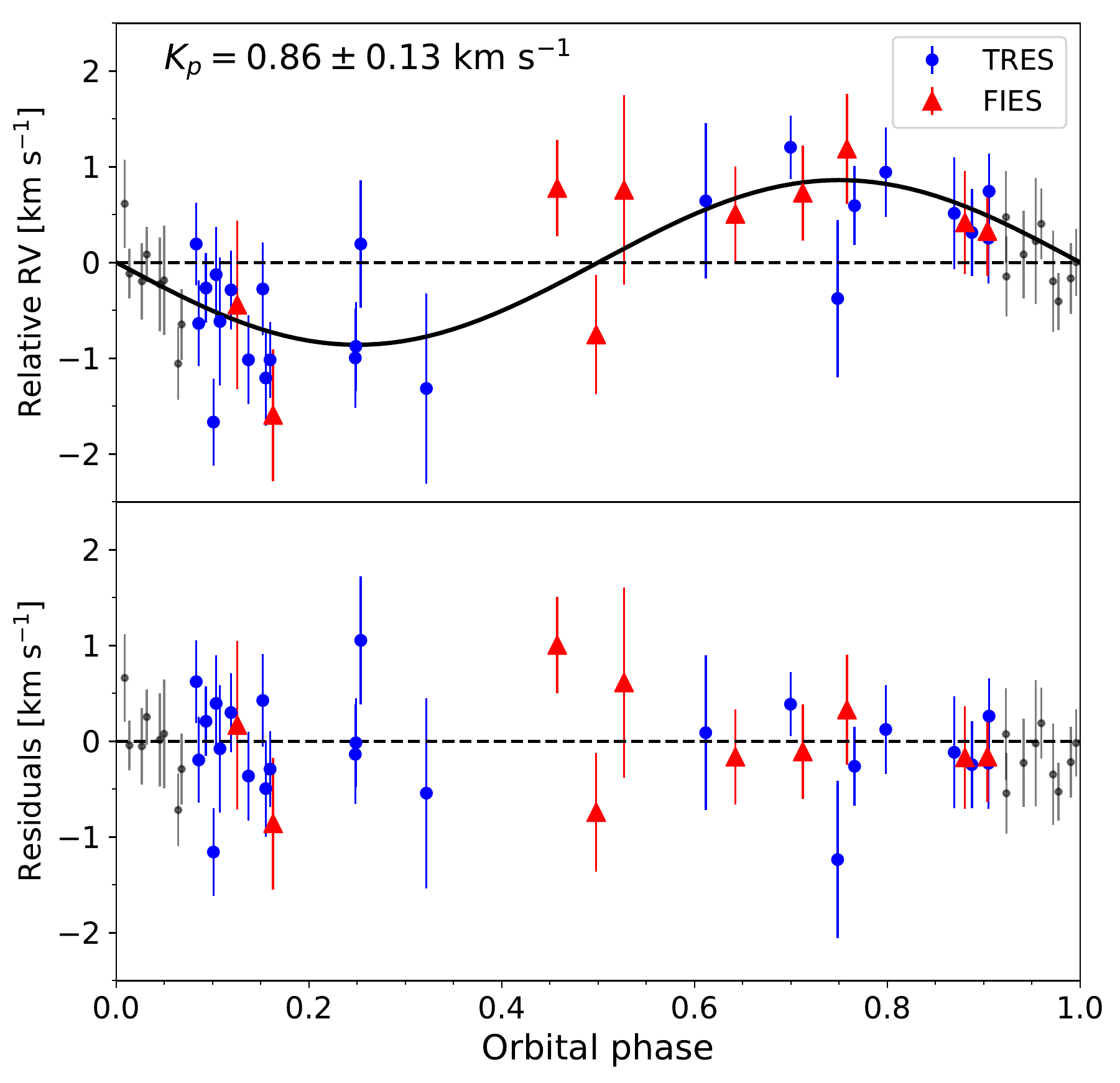}
\caption{Top panel: phase-folded RV measurements from TRES (blue points) and FIES (red triangles), with the best-fit systemic velocities removed; data points during transit are shown in gray and were excluded from the RV fit. RVs with uncertainties greater than 1 \kms\ were not considered in the final fit and are not included in the plot. The solid black curve indicates the best-fit planetary RV signal from the joint fit (Table~\ref{tab:jointfit}), which has a measured semiamplitude of $K_p = 0.86 \pm 0.13$ \kms. Bottom panel: the corresponding residuals from the best-fit model.}
\label{fig:rvfit}
\vspace{-0.3cm}
\end{figure}

We obtained a $7.1\sigma$ detection of the planetary RV signal, with a semiamplitude of $850 \pm 120$ \ms. Including the RVs with uncertainties greater than 1 \kms\ yielded a $K_p$ value that agrees with the previously listed amplitude at better than the $0.1\sigma$ level. No significant long-term RV trends were measured when allowing for linear and quadratic temporal terms in the RV model. Using the stellar mass determined from our SED fitting ($M_* = 1.447^{+0.075}_{-0.078}$ \msun; Table~\ref{tab:stellar}) and the orbital parameters from the \tess\ phase-curve fit, we derived a planet mass of $M_p = 4.77 \pm 0.70$ \mjup. The phase-folded and offset-corrected RVs are plotted in Figure~\ref{fig:rvfit}.

Visual inspection of the residuals for the RVs taken during transit does not reveal any significant deviations. Such deviations can arise due to the Rossiter--McLaughlin (RM) effect, wherein the planet occults regions of the stellar disk with different rotational velocities and creates systematic aberrations in the stellar line shapes and resultant RVs \citep{rossiter1924,mclaughlin1924}. The maximum value of the RV anomaly due to the RM effect is given by \citep[e.g.,][]{gaudi2007,albrecht2011}
\begin{equation}\label{rm}
K_{\mathrm{RM}} \sim \left(\frac{R_{p}}{R_{*}}\right)^2 \vsini(\sqrt{1 - b^2} + b\tan\lambda)\cos\lambda,
\end{equation}
where $\lambda$ is the sky-projected obliquity of the planet's orbit. Using the median values of $R_p/R_*$, $b$, and $\lambda$ from our global analysis of the \tess\ photometry, ground-based light curves, RVs, and spectroscopic transit observations (Section~\ref{sec:jointfit}; Table~\ref{tab:jointfit}), we find $K_{\mathrm{RM}} = 0.37$ \kms. This value is smaller than the average uncertainty of the RV measurements obtained during the planetary transit. We also note that the estimate in Equation~\eqref{rm} neglects the effects of limb and gravity darkening, which reduce the maximum RM anomaly, particularly for systems that are close to aligned. Therefore, we posit that the precision of the RVs, which is severely affected by the fast stellar rotation, is insufficient for securing the detection of an RM signal.

\subsection{Joint Photometric and Spectroscopic Fit}
\label{sec:jointfit}

To obtain the final results from our analysis of the TOI-2109 system, we carried out a joint fit of all available data sets --- \tess\ photometry, ground-based transit and secondary eclipse light curves, RV measurements, and spectroscopic transits --- to simultaneously measure all of the astrophysical quantities of interest. Given the mutual consistency between the individually measured planet--star radius ratios from the \tess\ and ground-based transit fits (Tables~\ref{tab:comparison} and \ref{tab:groundbased}), we defined a single $R_p/R_*$ parameter for all data sets.

By fitting the RV measurements jointly with the \tess\ light curve, we were able to separate the Doppler-boosting signal from the total phase-curve modulation at the sine of the orbital frequency and retrieve the phase offset $\psi$ in the planet's atmospheric brightness modulation (see discussion in Section~\ref{sec:phasemodel}). Both the RV semiamplitude $K_{\mathrm{RV}}$ and the Doppler-boosting semiamplitude $A_{\mathrm{Dopp}}$ depend on the planet mass $M_{p}$ \citep[e.g.,][]{loeb2003,perryman2011,shporer2017}: 
\begin{align}
A_{\mathrm{Dopp}}  &= \left\lbrack\frac{2\pi G} {Pc^{3}}\frac{M_p^3\sin^3 i}{(M_* + M_p)^2}\right\rbrack^{1/3}  \left\langle \frac{xe^{x}}{e^{x} - 1}\right\rangle_{\mathrm{TESS}}, \label{dopp} \\
K_{\mathrm{RV}} &= \left(\frac{2\pi G}{P}\right)^{1/3}\frac{M_p\sin i}{(M_* + M_p)^{2/3}}.\label{krv}
\end{align}
In Equation~\eqref{dopp}, $x \equiv hc/k\lambda T_{\mathrm{eff}}$, and the term in the angled brackets is integrated with respect to wavelength $\lambda$ across the \tess\ bandpass, weighted by the transmission function of the instrument. The expression in Equation~\eqref{krv} assumes zero orbital eccentricity.

Instead of fitting for $A_{\mathrm{Dopp}}$ and $K_{\mathrm{RV}}$ independently, we included the planet mass $M_p$, stellar mass $M_s$, and stellar effective temperature $T_{\mathrm{eff}}$ as free parameters and self-consistently modeled both the RV trend and the Doppler-boosting signal. The stellar mass and effective temperature were constrained by Gaussian priors based on the results of the SED modeling (Table~\ref{tab:stellar}). All other orbital ephemeris, transit-shape, limb-darkening, phase-curve, stellar variability, and RV parameters were treated in an identical manner to the corresponding analyses of individual data sets (Sections~\ref{sec:tessfit}--\ref{sec:rvfit}).

The ellipsoidal distortion amplitude is also dependent on $M_p$. However, unlike in the case of Doppler boosting, the physical processes driving the stellar tidal response are strongly contingent upon the detailed characteristics of the stellar interior and atmosphere. Secondary effects from the stellar rotation and the interaction between the external tidal force and pulsation modes can often lead to significant discrepancies between the predicted behavior and the measured amplitude (see, for example, \citealt{burkart2012} and \citealt{wong2020koi964}). Given this caveat, we did not model the ellipsoidal distortion using the planet mass in our joint fit, but instead kept the ellipsoidal distortion amplitude $A_{\mathrm{ellip}}$ as an independent fit parameter. A comparison of the predicted and measured amplitudes is provided in Section~\ref{sec:tidal}.

\begin{figure}[t!]
\includegraphics[width=\linewidth]{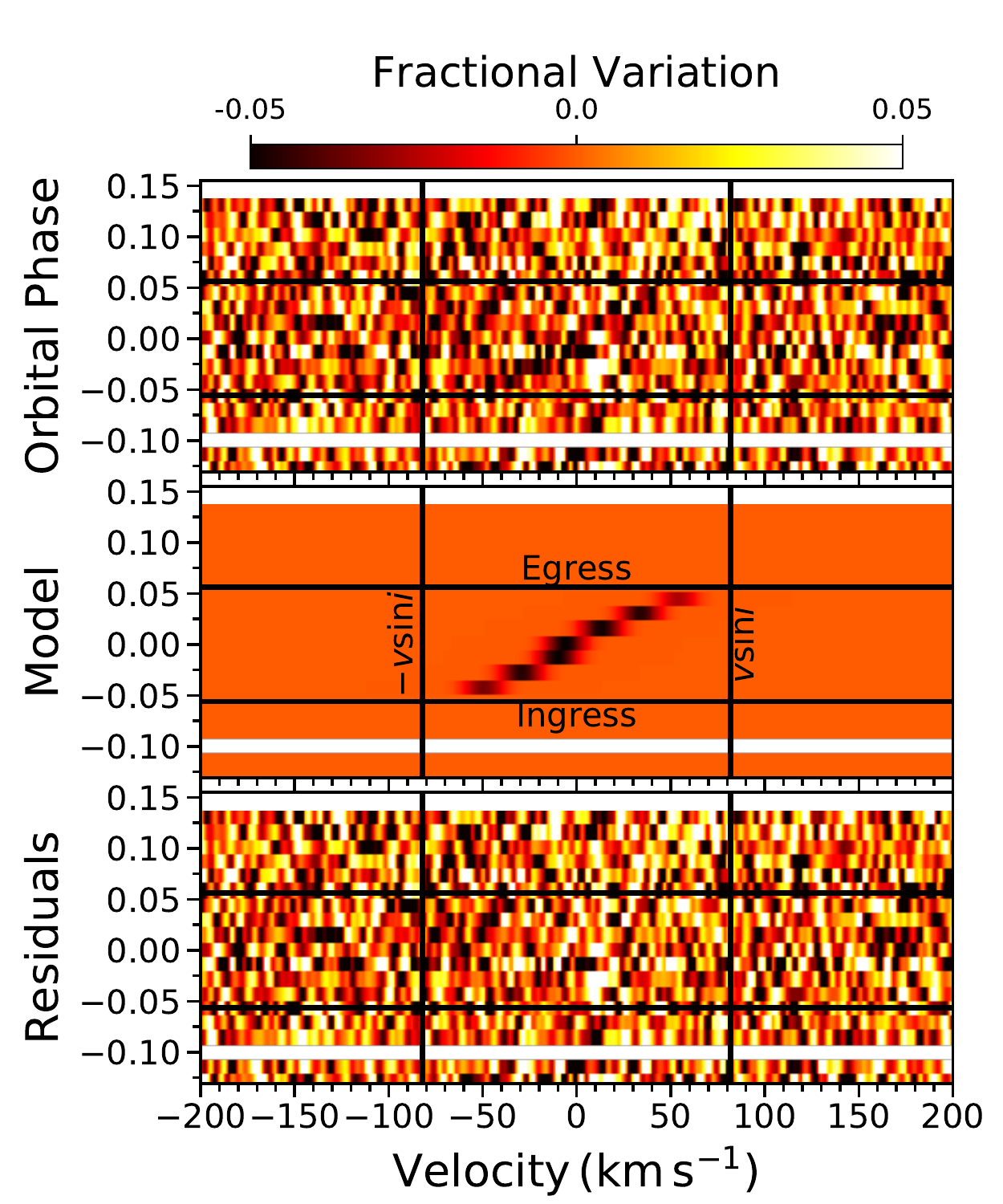}
\caption{The detection of the Doppler transit of TOI-2109b. The top panel shows the line-broadening profiles from both nights of TRES observations, with the average out-of-transit profile subtracted. The best-fit model is plotted in the middle panel, with the corresponding residuals provided in the bottom panel.}
\label{fig:dt}
\vspace{-0.2cm}
\end{figure}

When modeling the spectroscopic transit observations, we followed the methodology of \citet{zhou2016}. Prior to fitting, the average line-broadening profile derived from the out-of-transit spectra (Sections~\ref{sec:tres}) was subtracted from the full set of line-broadening profiles obtained during the two spectroscopic transits. The resultant differential rotational profiles from both nights of observation are displayed in the top panel of Figure~\ref{fig:dt}. The Doppler shadow of the transiting planet was modeled as a Gaussian centered at $v_{p}(t)$ --- the projected rotational velocity of the region occulted by the planet; the width contains contributions from the spectral resolution of the instrument and a nonrotational broadening component $v_{\mathrm{nonrot}}$, which covers both instrumental broadening and macroturbulence in the stellar atmosphere. The area of the signal at each timestamp is equivalent to $1 - \lambda_{t}(t)$, where $\lambda_{t}(t)$ is the transit light curve evaluated at time $t$. 

The location and orientation of the Doppler transit signal depend on the sky-projected stellar rotational velocity \vsini, sky-projected orbital obliquity $\lambda$, and impact parameter $b$. These parameters, in addition to the nonrotational broadening component $v_{\mathrm{nonrot}}$, were included as free parameters in the joint fit, with \vsini\ constrained by a Gaussian prior based on the TRES-derived measurement: $\vsini = 81.9 \pm 1.7$ \kms\ (Section~\ref{sec:tres}). 

The total number of free parameters in the joint fit is 44. The photometric light curves were corrected for instrumental systematics prior to fitting, with the uniform per-point flux uncertainties fixed to their respective best-fit values from the individual analyses (see Tables~\ref{tab:comparison} and \ref{tab:groundbased}). For the \tess\ light curve, the red-noise-inflated per-point uncertainty $\sigma_{r}$ was used.

\subsubsection{Results}\label{sec:jointresults}

\begin{figure}[t!]
\includegraphics[width=\linewidth]{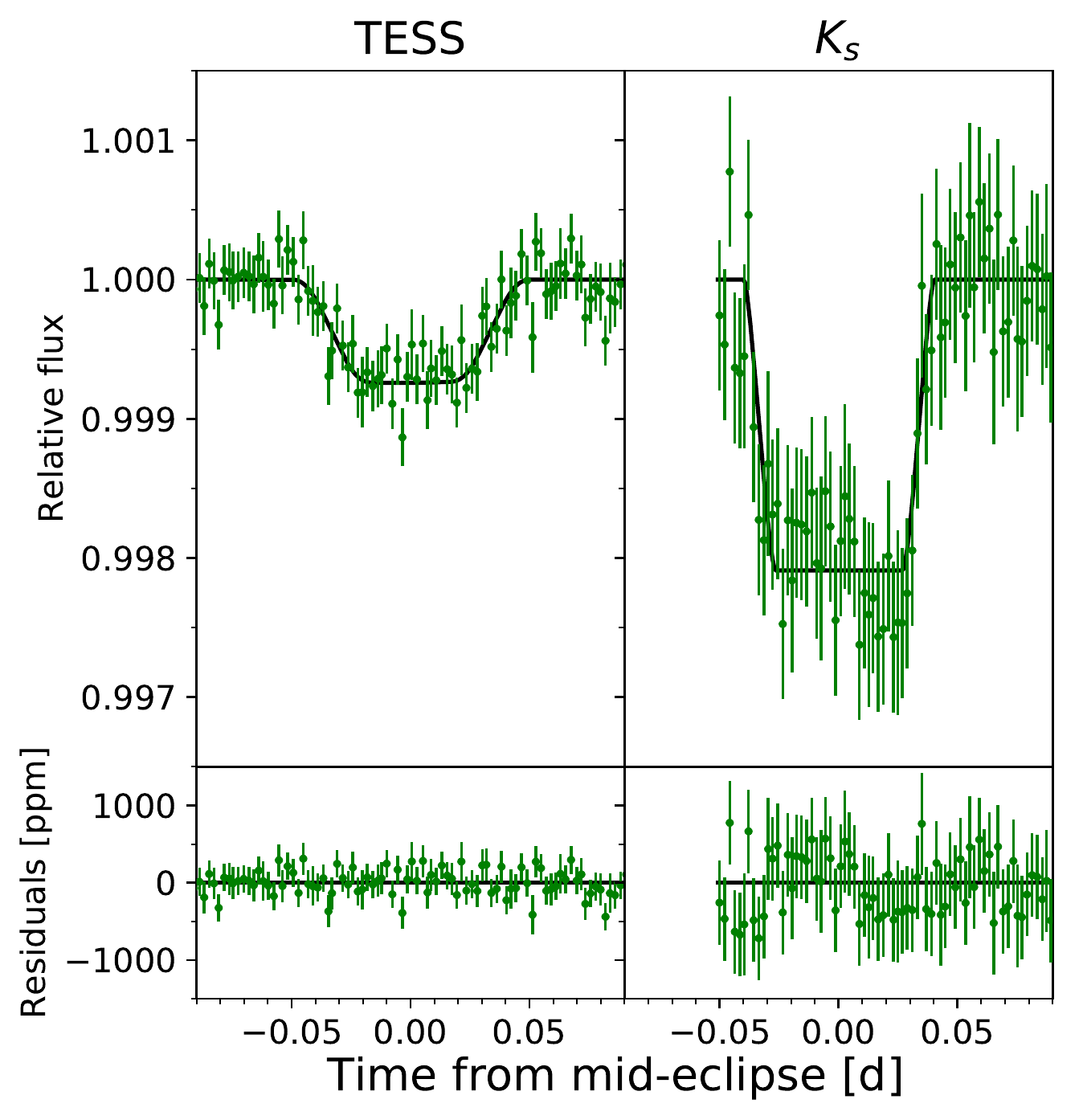}
\caption{The systematics-corrected \tess- and \Ks-band secondary eclipse light curves, binned at 3 minute intervals. The best-fit phase-curve and stellar variability models have been removed from the \tess\ photometry.}
\label{fig:occults}
\end{figure}

The results of our joint MCMC fit are shown in Table~\ref{tab:jointfit}; the values of relevant derived parameters are also provided. The best-fit full-orbit phase-curve, \tess- and \Ks-band secondary eclipse, RV, and Doppler transit models are displayed in Figures~\ref{fig:phasecurve}, \ref{fig:rvfit}, \ref{fig:dt}, and \ref{fig:occults}, alongside the corresponding residuals. 

TOI-2109b has an orbital period of $0.67247414 \pm 0.00000028$ days and a radius of $1.347 \pm 0.047$ \rjup. The addition of the ground-based transits in the joint fit greatly enhanced the precision of the orbital ephemeris and transit-shape parameters when compared to the \tess-only fit results in Table~\ref{tab:comparison}. Using the measured impact parameter and scaled semimajor axis --- $b = 0.7481 \pm 0.0073$ and $a/R_* = 2.268 \pm 0.021$ --- we derived an orbital inclination of $i = 70\overset{\circ}{.}74 \pm 0\overset{\circ}{.}37$, indicating a moderately inclined viewing geometry.

\clearpage


\startlongtable
\begin{deluxetable*}{lcc}
\setlength{\tabcolsep}{12pt}
\renewcommand{\arraystretch}{1.0}
\tabletypesize{\footnotesize}
\tablecaption{Results from the Joint Fit to the \tess\ Photometry, Ground-based Light Curves, Radial Velocities, and Spectroscopic Transit Observations}
\label{tab:jointfit}
\tablehead{\vspace{-0.2cm} \\
\multicolumn{1}{c}{Parameter} & \multicolumn{1}{c}{Description} & \multicolumn{1}{c}{Value}
}
\startdata
\hline 
\multicolumn{3}{c}{Fitted Parameters} \smallskip \\ 
\hline
$R_{p}/R_{*}$ & Planet--star radius ratio & $0.08155 \pm 0.00022$ \\
$T_{c}$ & Mid-transit time (BJD$_{\mathrm{TDB}}$) & $ 2{,}459{,}378.459370 \pm 0.000059$  \\ 
$P$ & Orbital period (days) & $0.67247414 \pm 0.00000028$  \\
$b$ & Impact parameter & $0.7481 \pm 0.0073$  \\ 
$a/R_{*}$ & Scaled semimajor axis & $2.268 \pm 0.021$ \\ 
$\bar{f}_{p}$ & Average \tess-band relative planetary flux (ppm) & $370 \pm 41$\\
$D_{d,K}$ & \Ks-band secondary eclipse depth (ppm) & $2012 \pm 80$\\
$A_{\mathrm{atm}}$ & Atmospheric brightness modulation semiamplitude (ppm) & $362 \pm 19$\\
$\psi$ & Phase offset of the atmospheric brightness modulation (deg) & $4.0 \pm 2.3$ \\
$A_{\mathrm{ellip}}$  & Ellipsoidal distortion semiamplitude (ppm) & $245 \pm 19$\\
$M_{p}$ & Planet mass (\mjup) & $5.02 \pm 0.75$\\
$M_{*}$  \tablenotemark{\scriptsize b} & Stellar mass (\msun) & $1.453 \pm 0.074$ \\
$\teff$  \tablenotemark{\scriptsize b} & Stellar effective temperature (K) & $6540 \pm 160$ \\
$\gamma_{\mathrm{TRES}}$ & Radial velocity offset for TRES (\kms) & $-25.64 \pm 0.11$\\
$\gamma_{\mathrm{FIES}}$ & Radial velocity offset for FIES (\kms) & $-25.61 \pm 0.21$\\
$\sigma_{\mathrm{jit,TRES}}$ & Radial velocity jitter for TRES (\kms) & $0.22 \pm 0.15$ \\
$\sigma_{\mathrm{jit,FIES}}$ & Radial velocity jitter for FIES (\kms) & $0.37 \pm 0.24$\\
$\vsini$ \tablenotemark{\scriptsize b} & Sky-projected stellar rotational velocity (\kms)& $81.2 \pm 1.6$ \\
$\lambda$ & Sky-projected obliquity (deg)& $1.7 \pm 1.7$ \\
$v_{\mathrm{nonrot}}$& Nonrotational broadening component (\kms)& $7.5 \pm 1.6$\\
$\Pi_{1}$ & First stellar variability period (days) & $0.61395 \pm 0.00055$ \\
$\alpha_{1}$  & Sine semiamplitude at $\Pi_1$ (ppm) & $-9 \pm 15$\\
$\beta_{1}$  & Cosine semiamplitude at $\Pi_1$ (ppm) & $-153 \pm 16$\\
$\alpha'_{1}$  & Sine semiamplitude at $1/2 \times \Pi_1$ (ppm) & $-68 \pm 16$\\
$\beta'_{1}$  & Cosine semiamplitude at $1/2 \times \Pi_1$ (ppm) & $62 \pm 15$ \\
$\Pi_{2}$ & Second stellar variability period (days) & $0.9674 \pm 0.0013$\\
$\alpha_{2}$  & Sine semiamplitude at $\Pi_2$ (ppm) & $24 \pm 16$ \\
$\beta_{2}$  & Cosine semiamplitude at $\Pi_2$ (ppm) & $-222 \pm 15$ \\
$u_{1,\mathrm{TESS}}$ \tablenotemark{\scriptsize a} & \tess-band quadratic limb-darkening coefficient & $0.29 \pm 0.03$\\
$u_{2,\mathrm{TESS}}$ \tablenotemark{\scriptsize a} & \tess-band quadratic limb-darkening coefficient & $0.20 \pm 0.04$\\
$u_{1,B}$ \tablenotemark{\scriptsize a} & $B$-band quadratic limb-darkening coefficient & $0.56 \pm 0.05$\\
$u_{2,B}$ \tablenotemark{\scriptsize a} & $B$-band quadratic limb-darkening coefficient & $0.25 \pm 0.05$\\
$u_{1,g'}$ \tablenotemark{\scriptsize a} & $g'$-band quadratic limb-darkening coefficient & $0.53 \pm 0.03$\\
$u_{2,g'}$ \tablenotemark{\scriptsize a} & $g'$-band quadratic limb-darkening coefficient & $0.18 \pm 0.04$\\
$u_{1,r'}$ \tablenotemark{\scriptsize a} & $r'$-band quadratic limb-darkening coefficient & $0.34 \pm 0.03$\\
$u_{2,r'}$ \tablenotemark{\scriptsize a} & $r'$-band quadratic limb-darkening coefficient & $0.23 \pm 0.04$\\
$u_{1,R}$ \tablenotemark{\scriptsize a} & $R$-band quadratic limb-darkening coefficient & $0.40 \pm 0.05$\\
$u_{2,R}$ \tablenotemark{\scriptsize a} & $R$-band quadratic limb-darkening coefficient & $0.22 \pm 0.05$\\
$u_{1,i'}$ \tablenotemark{\scriptsize a} & $i'$-band quadratic limb-darkening coefficient & $0.32 \pm 0.03$\\
$u_{2,i'}$ \tablenotemark{\scriptsize a} & $i'$-band quadratic limb-darkening coefficient & $0.23 \pm 0.04$\\
$u_{1,I}$ \tablenotemark{\scriptsize a} & $I$-band quadratic limb-darkening coefficient & $0.30 \pm 0.05$\\
$u_{2,I}$ \tablenotemark{\scriptsize a} & $I$-band quadratic limb-darkening coefficient & $0.20 \pm 0.05$\\
$u_{1,z'/z_s}$ \tablenotemark{\scriptsize a} & $z'/z_s$-band quadratic limb-darkening coefficient & $0.29 \pm 0.03$\\
$u_{2,z'/z_s}$ \tablenotemark{\scriptsize a} & $z'/z_s$-band quadratic limb-darkening coefficient & $0.19 \pm 0.04$\\
\hline 
\multicolumn{3}{c}{Derived Parameters} \smallskip \\
\hline
$R_{p}$ & Planet radius (\rjup) & $1.347 \pm 0.047$ \\
$a$  & Semimajor axis (au) & $0.01791 \pm 0.00065$ \\
$i$  & Orbital inclination (deg) & $70.74 \pm 0.37$\\
$q$ & Planet--star mass ratio & $0.00331 \pm 0.00052$ \\
$K_p$ & Radial velocity semiamplitude (\kms) & $0.86 \pm 0.13$ \\
$\loggpl$ & Planet surface gravity (cgs) & $3.836 \pm 0.071$ \\
$A_{\mathrm{Dopp}}$ & Doppler-boosting semiamplitude (ppm) & $8.6 \pm 1.3$ \\
$\left(R_p/R_*\right)^2$ & Transit depth (ppm) & $6651 \pm 36$ \\
$D_{d,\mathrm{TESS}}$ & \tess-band secondary eclipse depth (ppm) & $731 \pm 46$ \\
$D_{n,\mathrm{TESS}}$ & \tess-band nightside flux (ppm) & $9 \pm 43$ \\
$T_{\mathrm{irr}}$ \tablenotemark{\scriptsize c} & Irradiation temperature (K) & $4340 \pm 100$ \\
$T_{\mathrm{eq}}$ \tablenotemark{\scriptsize c} & Dayside-redistribution equilibrium temperature (K) & $3646 \pm 88$ \\
$T_{b,\mathrm{day}}$ \tablenotemark{\scriptsize c} & Dayside brightness temperature (K) & $3631 \pm 69$ \\
$T_{b,\mathrm{night}}$ \tablenotemark{\scriptsize c} & Nightside brightness temperature (K) & $<$2500 \\
\enddata
\vspace{+0.15cm}\textbf{Notes.}
\vspace{-0.15cm}\tablenotetext{\textrm{a}}{These parameters were constrained by priors based on the tabulated limb-darkening coefficients from \citet{claret2017}. See Table~\ref{tab:priors}.}
\vspace{-0.15cm}\tablenotetext{\textrm{b}}{These parameters were constrained by priors based on modeling of the host star's SED and TRES spectra: $M_* = 1.447 \pm 0.077$ \msun, $T_{\mathrm{eff}} = 6530 \pm 160$ K, and $\vsini = 81.9 \pm 1.7$ \kms.}
\vspace{-0.15cm}\tablenotetext{\textrm{c}}{The irradiation temperature is defined as $T_{\mathrm{irr}} \equiv T_{\mathrm{eff}}\sqrt{R_{*}/a}$. The dayside-redistribution equilibrium temperature assumes zero Bond albedo, a uniform dayside temperature, and $T = 0$ K on the nightside. The dayside brightness temperature is derived from a joint blackbody fit to the \tess- and \Ks-band secondary eclipse depths, assuming zero geometric albedo. For the nightside temperature, the $2\sigma$ upper limit is given. See Section~\ref{sec:atm}.}
\end{deluxetable*}

A notable result is that the projected spin--orbit misalignment is consistent with zero. The tight constraints on the transit shape and the mid-transit time at the epoch of the spectroscopic transit observations yielded an extremely precise measurement of the projected orbital obliquity: $\lambda = 1\overset{\circ}{.}7 \pm 1\overset{\circ}{.}7$). The faint Doppler shadow of the planet is discernible in the line-broadening profiles (Figure~\ref{fig:dt}), which definitively confirms that the planet is orbiting the host star and excludes all blended scenarios. The low obliquity of TOI-2109b stands in stark contrast to the sizable subset of ultrahot Jupiters with large obliquities, including HAT-P-7b ($182\overset{\circ}{.}5 \pm 9\overset{\circ}{.}4$; \citealt{narita2009,winn2009}), KELT-9b ($-88^{\circ} \pm 15^{\circ}$; \citealt{ahlers2020}), WASP-33b ($-108\overset{\circ}{.}8 \pm 1\overset{\circ}{.}0$; \citealt{colliercameron2010}), and WASP-121b ($-102\overset{\circ}{.}2 \pm 5\overset{\circ}{.}4$; \citealt{wasp121}). Meanwhile, other ultrahot Jupiters with near-zero obliquities include HATS-70b ($8\overset{\circ}{.}9 \pm 5\overset{\circ}{.}1$; \citealt{zhou2019}), WASP-18b ($4^{\circ} \pm 5^{\circ}$; \citealt{triaud2010}), and WASP-19b ($1\overset{\circ}{.}0 \pm 1\overset{\circ}{.}2$; \citealt{southworth2016}).

From the joint fit, we obtained $M_{p} = 5.02 \pm 0.75$ \mjup, which is consistent with the mass estimate we obtained from the dedicated RV fit in Section~\ref{sec:rvfit} ($4.77 \pm 0.70$ \mjup); the derived planet surface gravity is $\loggpl = 3.836 \pm 0.071$ (in cgs units). When combined with the orbital parameters via Equation~\eqref{dopp}, the planet mass corresponds to a predicted Doppler-boosting amplitude of $A_{\mathrm{Dopp}} = 8.6 \pm 1.3$ ppm.

From the \tess\ light curve, we measured a high-S/N secondary eclipse depth of $731 \pm 46$ ppm and a nightside flux consistent with zero. The semiamplitude of the planet's atmospheric brightness modulation is $A_{\mathrm{atm}} = 362 \pm 19$ ppm, with a marginal eastward offset of $\psi = 4\overset{\circ}{.}0 \pm 2\overset{\circ}{.}3$. The amplitude of the stellar ellipsoidal distortion signal is $A_{\mathrm{ellip}} = 245 \pm 19$ ppm. The two robustly detected phase-curve components are plotted separately in the middle panel of Figure~\ref{fig:phasecurve}. All of the phase-curve parameter values from our joint fit are statistically identical to the results we obtained from fitting the \tess\ light curve alone. Meanwhile, the \Ks-band secondary eclipse depth is $D_{d,K} = 2012 \pm 80$ ppm. 

The two fitted stellar variability periods are $\Pi_1 = 0.61395 \pm 0.00055$ days and $\Pi_2 = 0.9674 \pm 0.0013$ days. Combining the sine and cosine coefficients at each period into a single value, we find that the variability amplitudes at the fundamentals are $154 \pm 16$ ppm and $224 \pm 15$ ppm, respectively. The total first harmonic amplitude at $\Pi_1$ is $93 \pm 15$ ppm. The best-fit stellar variability signals are shown in the right panels of Figure~\ref{fig:act}, phase-folded on the corresponding periods.

Given the rapid stellar rotation, we checked the MuSCAT3 $g'$-band transit light curves for asymmetry that could arise due to the stellar oblateness and gravity darkening \citep[e.g.,][]{barnes2011}. These light curves were chosen on account of their high S/N and the more pronounced gravity darkening of the star at bluer wavelengths. A measurement of such asymmetry can yield the full three-dimensional stellar rotation axis and the true obliquity of the system. Following the methodology of \citet{ahlers2020}, we used the formalism of \citet{espinosa2011} to model the gravity-darkened transit, with the stellar inclination $i_*$ and rotation period $P_{\mathrm{rot}}$ as input parameters. A prior on $P_{\mathrm{rot}}$ based on the measured \vsini\ was applied. The near-zero projected obliquity of the system makes significant asymmetries unlikely, with the effects of gravity darkening becoming largely degenerate with the impact parameter and limb-darkening coefficients. We indeed did not detect any statistically significant transit asymmetry in the light-curve modeling, and as such no constraints on the stellar inclination were obtained from the gravity-darkening analysis.

The detection of periodic stellar variability presents another method for estimating the stellar inclination. From the measured sky-projected rotational velocity \vsini\ and the stellar radius $R_{*}$, a given variability period $\Pi$ can be converted to stellar inclination via $i_{*} = \sin^{-1}(\Pi\vsini/2\pi R_*)$. Plugging in $\Pi_1$ and $\Pi_2$, we obtained two independent stellar inclination estimates: $i_{*,1} = 36^{\circ} \pm 2^{\circ}$ and $i_{*,2} = 67^{\circ} \pm 6^{\circ}$. Given the roughly zero sky-projected obliquity and the measured orbital inclination of $i = 70\overset{\circ}{.}74 \pm 0\overset{\circ}{.}37$, we consider the second scenario, which corresponds to full spin--orbit alignment, to be more plausible. However, it should be acknowledged that the observed stellar variability need not be a manifestation of the star's rotation (e.g., from starspots) and may instead stem from intrinsic pulsation modes. In the latter case, the variability period is entirely unrelated to the measured sky-projected rotational velocity.

Assuming $i_{*,2} = 67^{\circ} \pm 6^{\circ}$, we computed a stellar oblateness of $1 - R_{*,\mathrm{pole}}/R_{*,\mathrm{eq}} \simeq 0.02$. When accounting for the viewing geometry, the mean radius of the sky-projected disk is $R_{*,\mathrm{mean}} \simeq 1.716$ \rsun, which differs from the $R_{*}$ value derived from the SED fitting analysis by just $\sim$1\% and is well within the $1\sigma$ confidence region. Therefore, the stellar oblateness is not expected to significantly bias the derived $R_p$ and $a$ values listed in Table~\ref{tab:jointfit}.

\begin{figure*}[t]
\includegraphics[width = \linewidth]{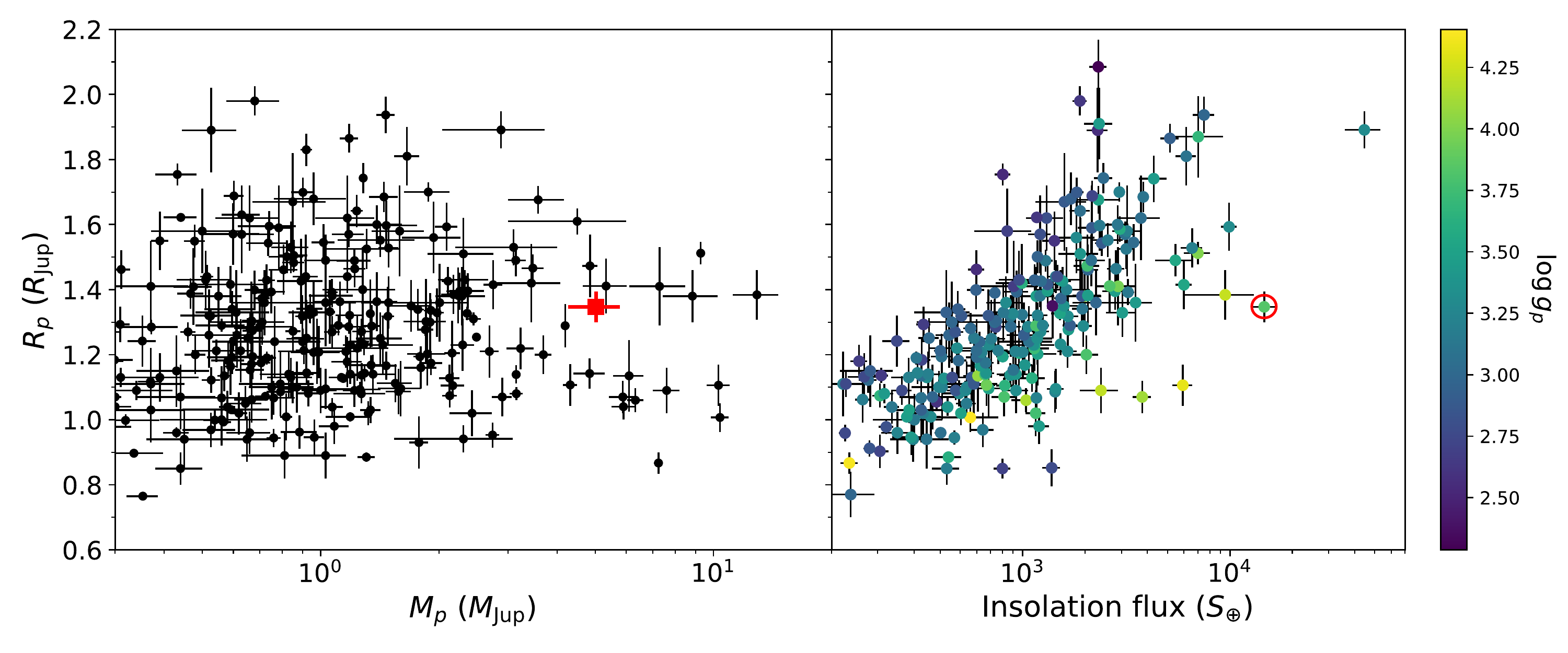}
\caption{Left: plot of planet radius vs. planet mass for hot Jupiters ($P < 5$ days, $R_p > 0.75$ \rjup, $M_p < 15$ \mjup). Only objects with relative radius and mass uncertainties less than 10\% and 50\%, respectively, are included. Right: an analogous plot of planet radius vs. dayside insolation flux (in units of Earth insolation). The color of the points is scaled to the planetary surface gravity \loggpl. TOI-2109b is marked in red.}
\label{fig:context}
\end{figure*}

\section{Discussion}
\label{sec:dis}

In the previous sections, we used space- and ground-based imaging and spectroscopy to confirm and characterize the TOI-2109 system. TOI-2109b has the shortest orbital period of any gas-giant exoplanet hitherto discovered. The massive 5 \mjup\ planet lies on a 16 hr orbit with near-zero sky-projected obliquity around a rapidly rotating F-type host star.

To place TOI-2109b in context, we compiled a list of hot Jupiters from the NASA Exoplanet Archive with $P < 5$ days, $R_p > 0.75~\rjup$, and $M_p < 15~\mjup$. To narrow the list to only well-characterized planets, we excluded objects with relative radius and mass uncertainties larger than 10\% and 50\%, respectively. We also included two recently discovered ultrahot Jupiters from \tess: TOI-1431b \citep{toi1431} and TOI-1518b \citep{toi1518}. 

In Figure~\ref{fig:context}, we plot the planet radius of the full sample as a function of planet mass and the calculated insolation flux. TOI-2109b is among the most massive known transiting planets, and it experiences the second most intense stellar irradiation environment behind KELT-9b. As with many other high-mass hot Jupiters, it is situated outside of the well-known radius--irradiation trend by virtue of its high surface gravity ($\loggpl = 3.836 \pm 0.071$). The extreme levels of irradiation and the strong tidal interaction with the host star have important implications for the planet's atmosphere, orbital evolution, and prospects for atmospheric characterization in the near future, which we discuss in the following.

\subsection{The Atmosphere of TOI-2109b}\label{sec:atm}

We obtained secondary eclipse observations of TOI-2109b in the \tess\ and \Ks\ bandpasses. The corresponding depths measured from the joint fit are $731 \pm 46$ and $2012 \pm 80$ ppm, respectively. Planets on ultrashort orbits are expected to be tidally locked, and so the depth of the secondary eclipse corresponds to the total brightness of the fixed dayside hemisphere relative to the stellar flux. When considering only thermal emission from the planet's atmosphere, we can express the relative planetary flux as 
\begin{equation}
\label{depth} D  = \left(\frac{R_{p}}{R_{*}}\right)^{2}\frac{\int B_{p,\lambda}(T_b)\tau(\lambda) d\lambda}{\int B_{*,\lambda}(T_{\mathrm{eff}})\tau(\lambda) d\lambda},
\end{equation}
where $B_{p,\lambda}$ and $B_{*,\lambda}$ are the emission spectra of the planet and star, respectively, and $\tau(\lambda)$ is the transmission function of the bandpass. The variable $T_{b}$ denotes the brightness temperature of the planet in the bandpass.

To calculate the dayside brightness temperature of TOI-2109b, we assumed that the planet's emission spectrum is that of a blackbody. For the stellar flux, we utilized PHOENIX models \citep{husser2013} and calculated the band-integrated flux (i.e., the denominator in the previous equation) for a grid of models spanning the measured stellar parameters for TOI-2109 (Table~\ref{tab:stellar}). Then, following the technique described in \citet{wong2020kelt9}, we fit for a polynomial function in (\teff, \loggstar, \feh) that interpolates the grid points, enabling smooth sampling of the integrated stellar flux. The posterior distribution of the planet's brightness temperature was calculated using \texttt{emcee}, with Gaussian priors on $D$ and $R_{p}/R_*$, and stellar parameters from the joint fit and the SED analysis.

From the \tess-band secondary eclipse depth, we derived a dayside brightness temperature of $T_{b,\mathrm{TESS}} = 3729 \pm 82$ K. A similar calculation with the \Ks-band secondary eclipse depth yielded $T_{b,K} = 3518 \pm 81$ K. Given the broad consistency between the \tess- and \Ks-band brightness temperatures, we also carried out a joint fit to both secondary eclipse depths and obtained a dayside brightness temperature of $T_{b,\mathrm{day}} = 3631 \pm 69$ K. Only KELT-9b has a higher measured dayside temperature: $4566 \pm 138$ K \citep{bell2021}. Through an analogous calculation, we converted the measured \tess-band nightside flux to an upper limit on the nightside brightness temperature. Due to the highly nonlinear relationship between the planet--star contrast ratio and the brightness temperature in the optical, we were only able to place a broad constraint: $T_{b,\mathrm{night}} < 2500$ K ($2\sigma$).

The previous discussion ignored the possibility of reflected starlight across the dayside hemisphere. A nonzero geometric albedo $A_g$ contributes an amount equal to $A_g \times \left(R_{p}/a\right)^2$ to the relative planetary flux, which serves to lower the dayside brightness temperature needed to match the \tess-band flux. While the relatively broad constraints on the dayside brightness temperature from the \Ks-band secondary eclipse alone and the marginally higher \tess-band dayside brightness temperature formally allow for a wide range of optical geometric albedos (up to $\sim$0.2 at $1\sigma$), we consider the scenario of significant dayside reflectivity unlikely. The extremely high temperatures across the dayside of TOI-2109b preclude the formation of condensate clouds or hazes, as all major condensate species are expected to be in the vapor phase, or even dissociated \citep[e.g.,][]{wakeford2017,lothringer2018}. Indeed, higher-precision optical and near-infrared secondary eclipse measurements of other ultrahot Jupiters have revealed very low geometric albedos consistent with zero \cite[see, for example,][and references therein]{bell2019,shporer2019,wong2020year1}. Moreover, an apparent excess in planetary flux at optical wavelengths may indicate deviations in the dayside emission spectrum from a simple blackbody, which can be caused by high-temperature opacity sources such as H$^{-}$ \citep[e.g.,][]{arcangeli2018}. We address this possibility with detailed emission spectrum modeling of TOI-2109b in Section~\ref{sec:future}.

We can compare the measured dayside brightness temperature to the equilibrium temperature \teq\ \citep{cowanagol}:
\begin{equation}\label{teq}
\teq = \teff\sqrt{\frac{R_{*}}{a}}\xi(1 - A_{\rm B})^{1/4}.\\
\end{equation}
The first two terms are referred to as the irradiation temperature $T_{\mathrm{irr}}\equiv \teff\sqrt{R_{*}/a}$; using the parameters for the TOI-2109 system, we obtained $T_{\mathrm{irr}} = 4340 \pm 100$ K. The variable $A_{\rm B}$ represents the Bond albedo, while $\xi$ is a factor that reflects the efficiency of day--night heat recirculation. For the dayside-redistribution equilibrium temperature, we assumed zero Bond albedo, a uniform dayside temperature, and no heat transport to the nightside atmosphere, which correspond to $A_{\rm B} = 0$ and $\xi = (1/2)^{1/4}$. This equilibrium temperature ($\teq = 3646 \pm 88$ K) is consistent with the measured dayside brightness temperature ($T_{\mathrm{day}} = 3631 \pm 69$ K). Meanwhile, the maximum theoretical temperature, which assumes no heat circulation across the dayside atmosphere ($\xi = (2/3)^{1/4}$) is $T_{\mathrm{max}} \sim 3900$ K, significantly hotter than $T_{\mathrm{day}}$.

\begin{figure}
\includegraphics[width=\linewidth]{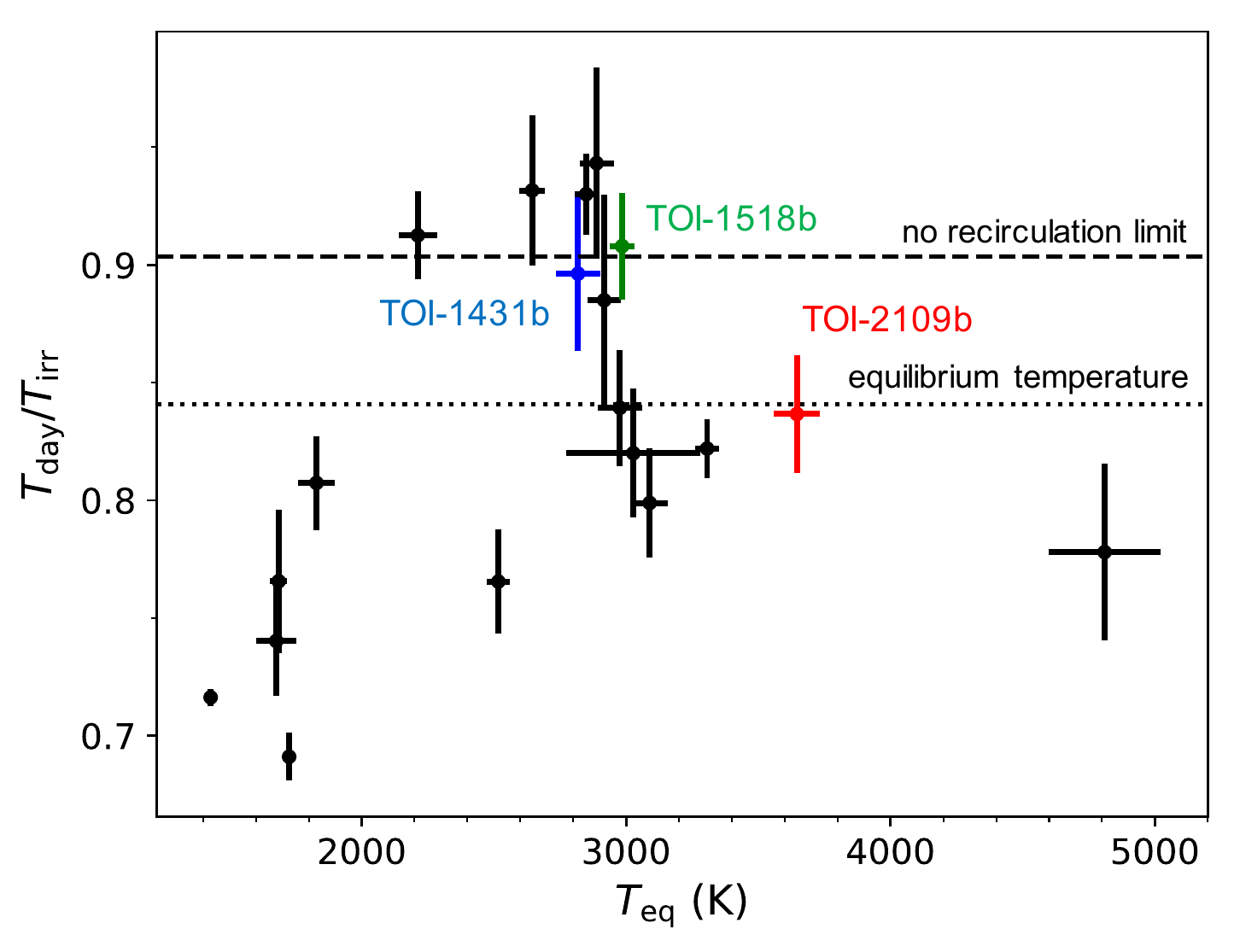}
\caption{Plot of the ratio between the measured dayside and irradiation temperatures for hot and ultrahot Jupiters, as a function of dayside-redistribution equilibrium temperature $T_{\mathrm{eq}}$. The black points are derived from the comprehensive analysis of Spitzer 4.5 $\mu$m phase curves by \citet{bell2021}. In addition to TOI-2109b, we have included the values for two recently published ultrahot Jupiters from \tess: TOI-1431b \citep{toi1431} and TOI-1518b \citep{toi1518}. The horizontal lines denote two fiducial scenarios: the maximum theoretical dayside temperature (no heat recirculation, $A_{\rm B} = 0$) and the dayside-redistribution equilibrium temperature (uniform dayside temperature, with $A_{\rm B} = 0$ and $T = 0$ K on the nightside).}
\label{fig:tday}
\end{figure}

We now place the dayside thermal energy budget of TOI-2109b in the context of other hot Jupiters. \citet{bell2021} carried out a uniform analysis of all available Spitzer 4.5 $\mu$m phase curves and derived $T_{\mathrm{day}}$ for 16 planets based on their secondary eclipse depths. In Figure~\ref{fig:tday}, we plot the ratio $T_{\mathrm{day}}/T_{\mathrm{irr}}$ as a function of \teq\ for these targets, alongside the corresponding values for TOI-2109b and the two recently discovered ultrahot Jupiters from \tess\: TOI-1431b and TOI-1518b. The horizontal lines denote two benchmark scenarios: the dayside-redistribution equilibrium temperature, with a uniform dayside temperature and $T = 0$ K on the nightside (i.e.,  $\xi = (1/2)^{1/4}$), and the maximum theoretical dayside temperature (i.e., the no-recirculation limit,  $\xi = (2/3)^{1/4}$). 

The first notable observation is that TOI-2109b lies intermediate between the dense cluster of well-characterized gas giants with $\teq \sim 3000$ K and the extreme endmember KELT-9b. It follows that intensive study of TOI-2109b's atmosphere may help connect the unique properties of KELT-9b to the broader underlying thermophysical processes driving the atmospheric dynamics and chemistry of hot gas giants. Next, we find a number of planets with $T_{\mathrm{day}} < \teq$, particularly at the ends of the temperature range shown. Such a scenario could be indicative of a nonzero Bond albedo and/or a significant amount of day--night heat transport. When examining the scatter plot for overall trends, we see that the $T_{\mathrm{day}}/T_{\mathrm{irr}}$ value of TOI-2109b lies in between the average value of the cluster of gas giants with $\teq \sim 3000$ K, which are consistent with extremely low levels of day--night heat recirculation, and the lower value of KELT-9b. This behavior suggests a tenuous decreasing trend that is predicted by recent atmospheric modeling of ultrahot Jupiters, which has demonstrated that the dissociation of molecular hydrogen on the hot dayside and its recombination on the cooler nightside can efficiently transport energy across the planet's surface, resulting in lower dayside temperatures and reduced day--night temperature contrasts in the hottest exoplanets \citep[e.g.,][]{bell2018,komacek2018,parmentier2018,tan2019,roth2021}. 

Past models simulating the atmospheric circulation of ultrahot Jupiters similar to TOI-2109b have shown that significant dayside hotspot offsets ($\sim$20$^{\circ}$) are expected under conditions of weak atmospheric drag and short planetary rotation period \citep{tan2019}. While the lack of a large eastward phase offset in the measured visible phase curves of cooler gas giants may be attributable to the contribution of advected reflective clouds from the nightside crossing over the western terminator, the extremely high dayside temperature of TOI-2109b renders the possibility of partial cloud cover highly unlikely. Instead, magnetohydrodynamic forces on the dayside atmosphere may be enhancing the atmospheric drag and reducing the magnitude of the hotspot offset \citep[e.g.,][]{rogers2014}. Observational tests of the detailed atmospheric circulation can be achieved through spectroscopic thermal phase curves, which we discuss briefly in Section~\ref{sec:future}.

\subsection{Planet--Star Tidal Interaction}\label{sec:tidal}
We measured a strong ellipsoidal distortion signal in the \tess\ light curve at almost $13\sigma$ significance. Having obtained the amplitude of this modulation independent of the constraint on planet mass from the RVs, we can now compare the measured value to the theoretical prediction. The expected ellipsoidal distortion amplitude at the first harmonic of the cosine of the orbital phase is related to the planet--star mass ratio $q$ via the following expression \citep[e.g.,][]{morris1985,morris1993,shporer2017}:
\begin{equation}\label{ellip}
A_{\mathrm{ellip}} = \alpha_{\mathrm{ellip}}q\left(\frac{R_{*}}{a}\right)^{3}\sin^2 i.
\end{equation}
The prefactor $\alpha_{\mathrm{ellip}}$ is a term that depends on the linear limb- and gravity-darkening coefficients $u$ and $g$ (see, for example, \citealt{morris1985} and \citealt{shporer2017}). We defined Gaussian priors centered on the values tabulated in \citet{claret2017}: $u = 0.4497$ and $g = 0.2273$; we used 0.05 for the width of the prior on $u$, while we selected 0.10 for $g$, given the steeper correlation between $g$ and \teff\ for F-type stars. After carrying out Monte Carlo sampling of the parameter distributions, we arrived at a predicted ellipsoidal distortion semiamplitude of $281 \pm 52$ ppm. 

This prediction is statistically consistent with the measured value from our joint fit ($A_{\mathrm{ellip}} = 245 \pm 19$ ppm) and shows that the classical theory of stellar tidal deformation accurately describes the gravitational response of TOI-2109 to its orbiting companion. This agreement holds despite the rapid rotation of the host star, which is not accounted for in the theoretical formalism underpinning Equation~\eqref{ellip}.

The ultrashort orbit of TOI-2109b raises the possibility that the planet may be close to being broken apart by tidal forces. The minimum distance from the star at which an orbiting companion can reside without being catastrophically disrupted is called the Roche limit. For gaseous planets, this quantity can be converted to a minimum orbital period  $P_{\mathrm{min}} \simeq 0.40 \mathrm{d}/\sqrt{\rho_{p}}$, where $\rho_p$ is the bulk density of the planet in units of \gcmc\ \citep{rappaport2013}. For the TOI-2109 system, we calculated $P_{\mathrm{min}} \simeq 0.12$ days, which is significantly shorter than the measured orbital period ($\sim$0.67 days). Therefore, we conclude that the high planet mass ensures that TOI-2109b is not threatened by tidal disruption, despite its extremely close orbit. For comparison, the roughly Jupiter-mass planets WASP-12b, WASP-19b, and WASP-121b all have orbital periods that are within three times their respective $P_{\mathrm{min}}$ limits.

Even when a planet is not close to the Roche limit, the strong tidal forces can lead to significant deformation of the planet's atmosphere and even mass loss through tidal stripping. A commonly used metric for evaluating the likelihood of atmospheric mass loss is the Roche lobe filling factor $r \equiv R_p/R_{\mathrm{Roche}}$, where $R_{\mathrm{Roche}}$ is the radius of the effective spherical surface around the planet within which material is gravitationally bound to the planet. This ratio can be approximated as \citep{eggleton1983}
\begin{equation}\label{roche}
r = \left(\frac{R_p}{R_*}\right)\left(\frac{a}{R_*}\right)^{-1}\frac{0.6q^{2/3} + \ln(1 + q^{1/3})}{0.49q^{2/3}}.
\end{equation}
For TOI-2109b, we used the values of $q$, $a/R_*$, and $R_p/R_*$ from the joint fit and obtained $r = 0.50 \pm 0.03$. 

The planet radius used in the previous calculation corresponds to the radial distance to which the atmosphere is optically thick at visible wavelengths. A low-density exosphere might still extend far beyond the planet's surface. For comparison, the two notable Jupiter-mass planets experiencing detectable atmospheric mass loss through Roche lobe overflow---WASP-12b \citep[e.g.,][]{haswell2012,fossati2013,bell2019} and KELT-9b \citep[e.g.,][]{yan2018,wyttenbach2020}---have $r \sim 0.8$ and $r \sim 0.5$, respectively. Meanwhile, outflow of metastable helium has been reported for WASP-107b \citep{spake2018,allart2019}, a hot sub-Saturn with a Roche lobe filling factor of just $r \sim 0.3$. The specific conditions necessary for large-scale atmospheric mass loss and the corresponding properties of the outflowing gas are still poorly understood. The combination of high surface gravity and high atmospheric temperature makes TOI-2109b a particularly compelling target for detecting signatures of gaseous outflow and evaluating models of tide- and irradiation-driven atmospheric mass loss.

\subsection{Orbital Decay}\label{sec:decay}

The strong gravitational interaction and tidal dissipation in ultrashort-period systems can cause the orbital period to measurably decay on year-long or decade-long timescales. The predicted rate of orbital decay depends on the host star's modified tidal quality factor $Q'_{*}$ \citep{goldreich1966,rasio1996,sasselov2003}:
\begin{equation}\label{decay}
\dot{P} = -\frac{27\pi q}{2Q'_{*}}\left(\frac{R_{*}}{a}\right)^5.
\end{equation}
Studies of tidal dissipation in late-type main-sequence stellar binaries and planet-hosting stars indicate that typical values of $Q'_{*}$ for F-type stars lie in the range $10^5$--$10^7$ \citep[e.g.,][]{ogilvie2007,lanza2011}. Using values of $q$ and $a/R_*$ from our joint fit (Table~\ref{tab:jointfit}), we found orbital decay rates that span $\sim$10--740 ms~yr$^{-1}$. 

For comparison, the most recently published measurement of orbital decay rate for WASP-12b is $32.5 \pm 1.6$ ms~yr$^{-1}$ \citep{turner2021}; that system is the only incontrovertible case of tidal orbital decay hitherto discovered. The inferred stellar tidal quality factor for the late-F-type host star WASP-12 is $Q'_{*} = 1.39 \pm 0.15 \times 10^5$, which is consistent with the lower end of the previously cited range. 

\begin{figure}
\includegraphics[width=\linewidth]{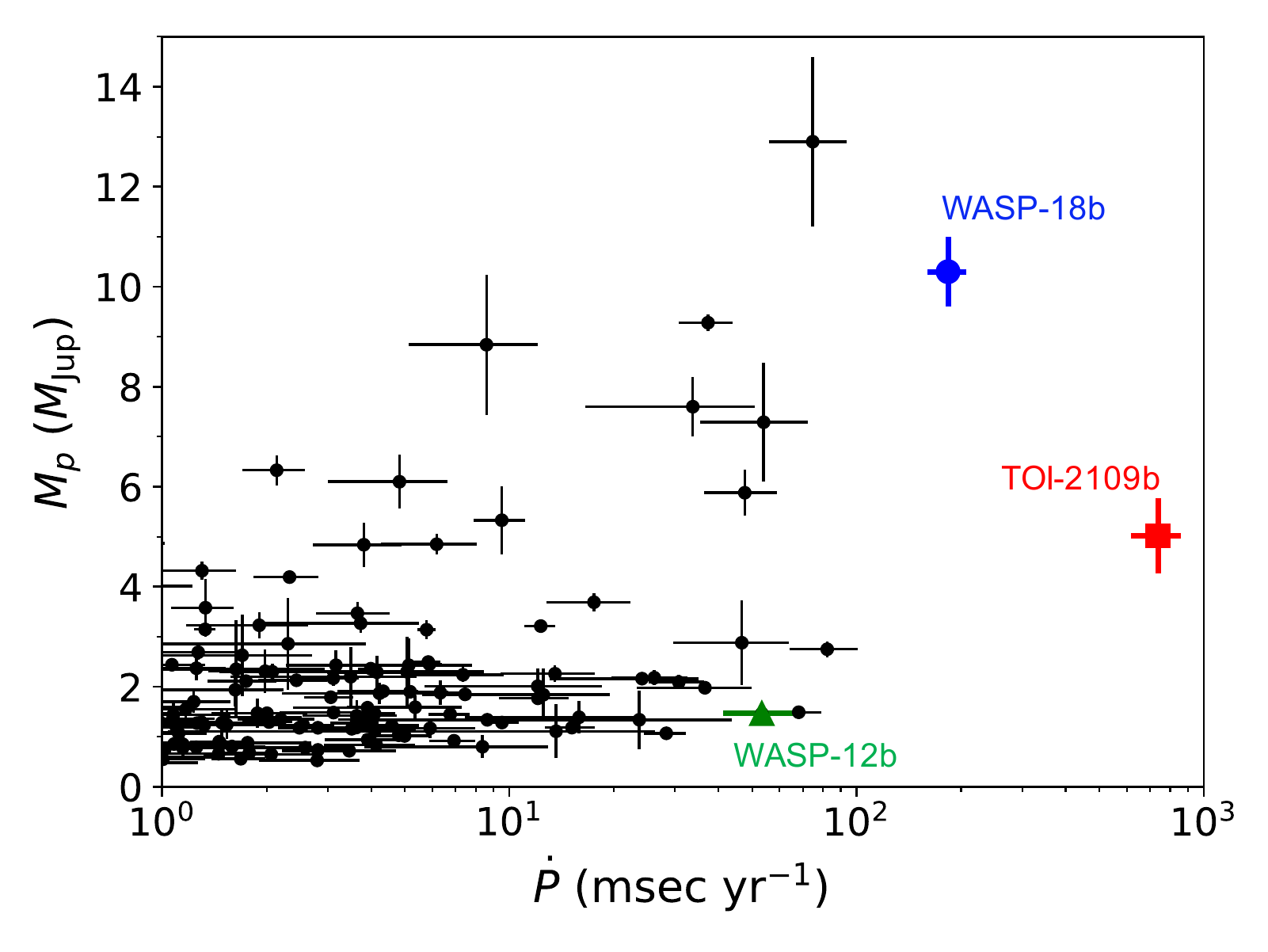}
\caption{Compilation of predicted tidal orbital decay rates for hot Jupiters, calculated from Equation~\eqref{decay} assuming $Q'_{*} = 10^5$. TOI-2109b is shown with the red square, while WASP-12b and WASP-18b are indicated by the green triangle and blue circle, respectively. For any given stellar tidal quality factor, TOI-2109b is expected to have by far the fastest orbital decay rate of any planetary mass companion.}
\label{fig:decay}
\end{figure}

Figure~\ref{fig:decay} shows the predicted orbital decay rate for all known hot Jupiters, assuming $Q'_{*} = 10^5$. For a given $Q'_{*}$, TOI-2109b has by far the fastest expected orbital decay rate among planetary mass companions, making this system the most promising candidate for probing orbital evolution and constraining the stellar tidal quality factor in the coming years. In second place is the supermassive gas giant WASP-18b, which also orbits an F-type host star. Curiously, no statistically significant variation in orbital period has been detected for WASP-18b, despite its high mass and the availability of high-precision transit-timing measurements spanning more than a decade \citep{patra2020}. If $Q'_{*}$ of TOI-2109 is comparable to that of WASP-12, and the cadence and precision of follow-up transit timings are similar to what was obtained for WASP-12b, then we may expect to detect a nonlinear ephemeris within 2--3 yr. \tess\ is slated to reobserve this system during the extended mission in Sector 52 (currently scheduled for 2022 May 18--June 13), at which point there should be sufficient time baseline to detect orbital decay rates at the upper end of the aforementioned range.

In addition to long-term transit monitoring, collecting further secondary eclipse timings will be crucial for distinguishing between tidal orbital decay and apsidal precession \citep[e.g.,][]{yee2020}: in the former case, the time interval between consecutive events decreases in tandem for both transits and secondary eclipses, while in the latter case, the directions of the timing deviations proceed in contrary motion. Likewise, continued RV monitoring is necessary to check for line-of-sight systemic acceleration due to the presence of a long-period bound companion (i.e., the R{\o}mer effect).

\subsection{Future Prospects}\label{sec:future}

The brightness of the host star ($V = 10.3$ mag, $K = 9.1$ mag) and the highly irradiated dayside atmosphere make TOI-2109b very amenable to intensive atmospheric characterization with current and near-future facilities. The rapid stellar rotation and associated rotational broadening of the stellar lines are conducive to ground-based high-dispersion spectroscopy, which has been used extensively in recent years to detect individual atomic, ionic, and molecular species in the atmospheres of ultrahot Jupiters \citep[e.g.,][]{hoeijmakers2018,casasayas2019}. Emission spectroscopy, both in eclipse and across the full orbital phase, is particularly promising. While the high surface gravity of TOI-2109b ($\loggpl = 3.836 \pm 0.071$) severely depresses the atmospheric scale height and therefore attenuates any spectral features observed in transmission relative to lower-mass hot Jupiters, the prospects for emission studies are not affected by the surface gravity. 

\begin{figure}
\includegraphics[width=\linewidth]{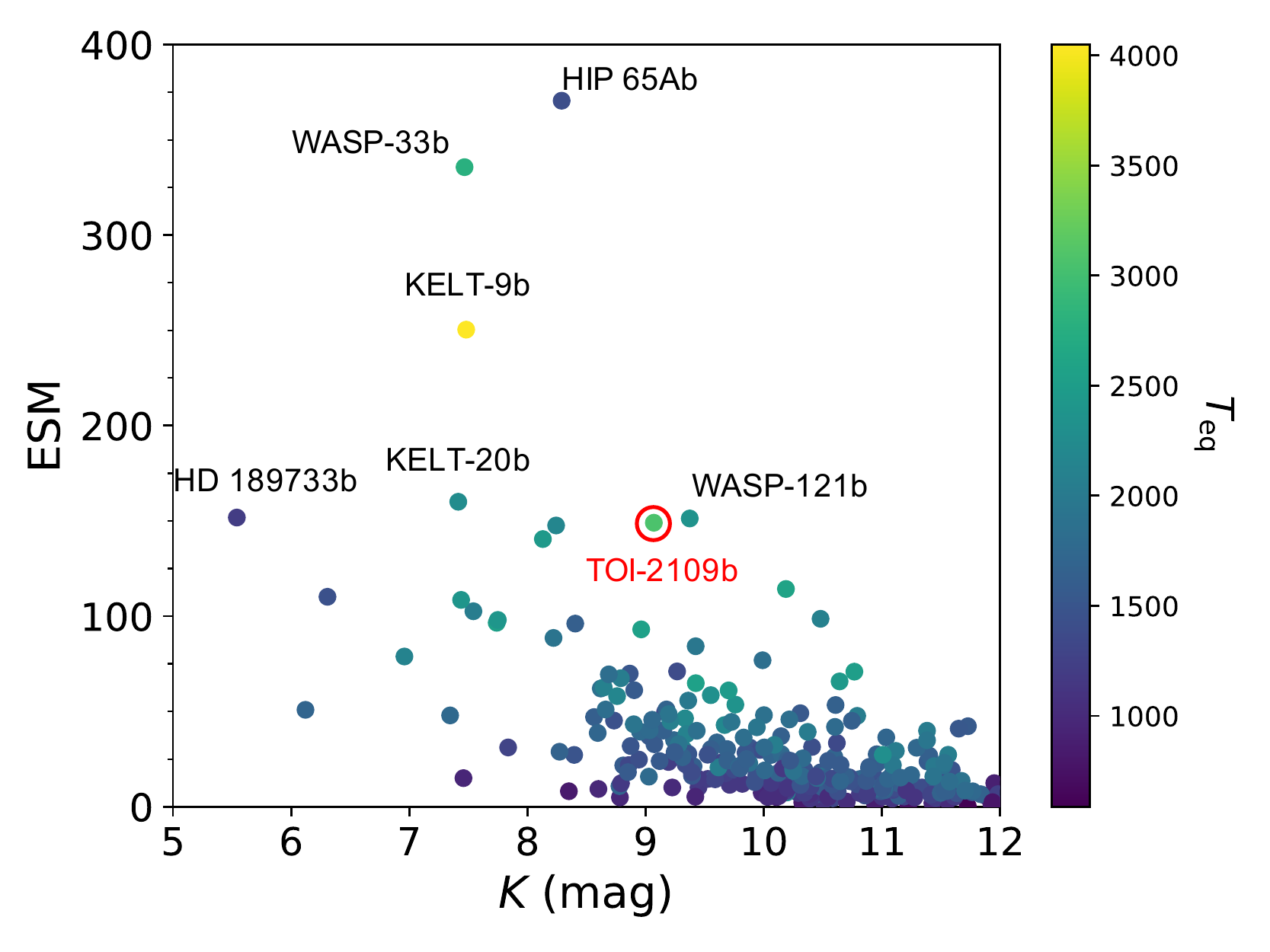}
\caption{Plot of the $K$-band ESM for all hot Jupiters (see selection criteria in Figure~\ref{fig:context} caption). The points are colored according to their $T_{\mathrm{eq}}$ values. The planets with the seven highest ESM values are labeled; TOI-2109b has the seventh highest ESM}.
\label{fig:esm}
\vspace{-0.5cm}
\end{figure}

We can quantify a planet's amenability to emission spectroscopy by calculating the Emission Spectroscopy Metric (ESM; \citealt{kempton2018}). We tailored the formulation of the ESM to be more suitable for hot gas giants by evaluating the planet--star contrast ratio in the $K$ band, instead of at 7.5 $\mu$m as originally defined. Figure~\ref{fig:esm} shows the ESM values for our sample of hot Jupiters. TOI-2109b has the seventh highest ESM, behind HIP 65Ab (an inflated hot Jupiter around a K-dwarf with $R_p/R_s \sim 0.25$; \citealt{nielsen2020}), three ultrahot Jupiters orbiting A-stars (WASP-33b, KELT-9b, and KELT-20b), HD 189733b, and WASP-121b.

\begin{figure}
\includegraphics[width=0.95\linewidth]{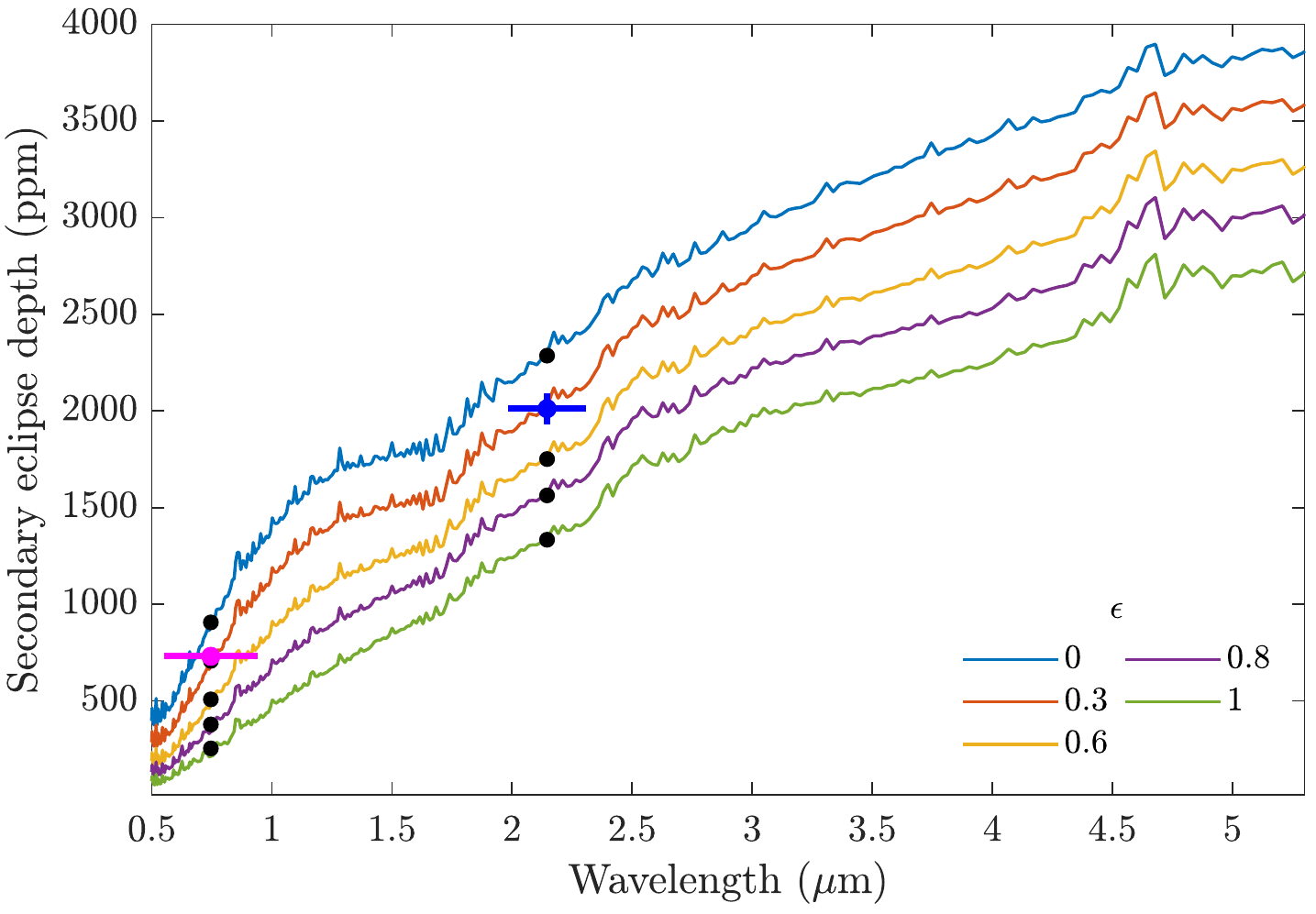}
\includegraphics[width=0.95\linewidth]{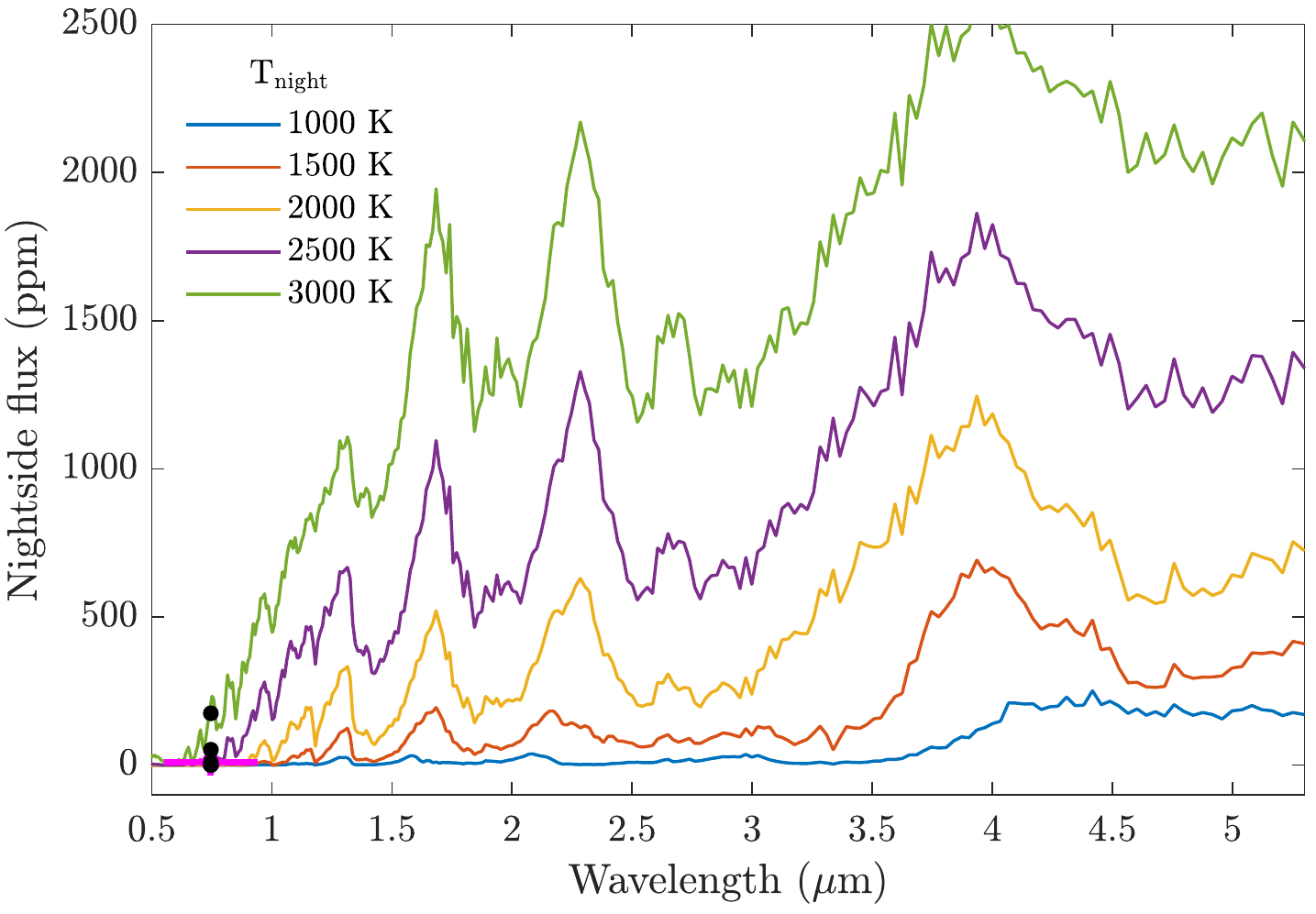}
\includegraphics[width=0.95\linewidth]{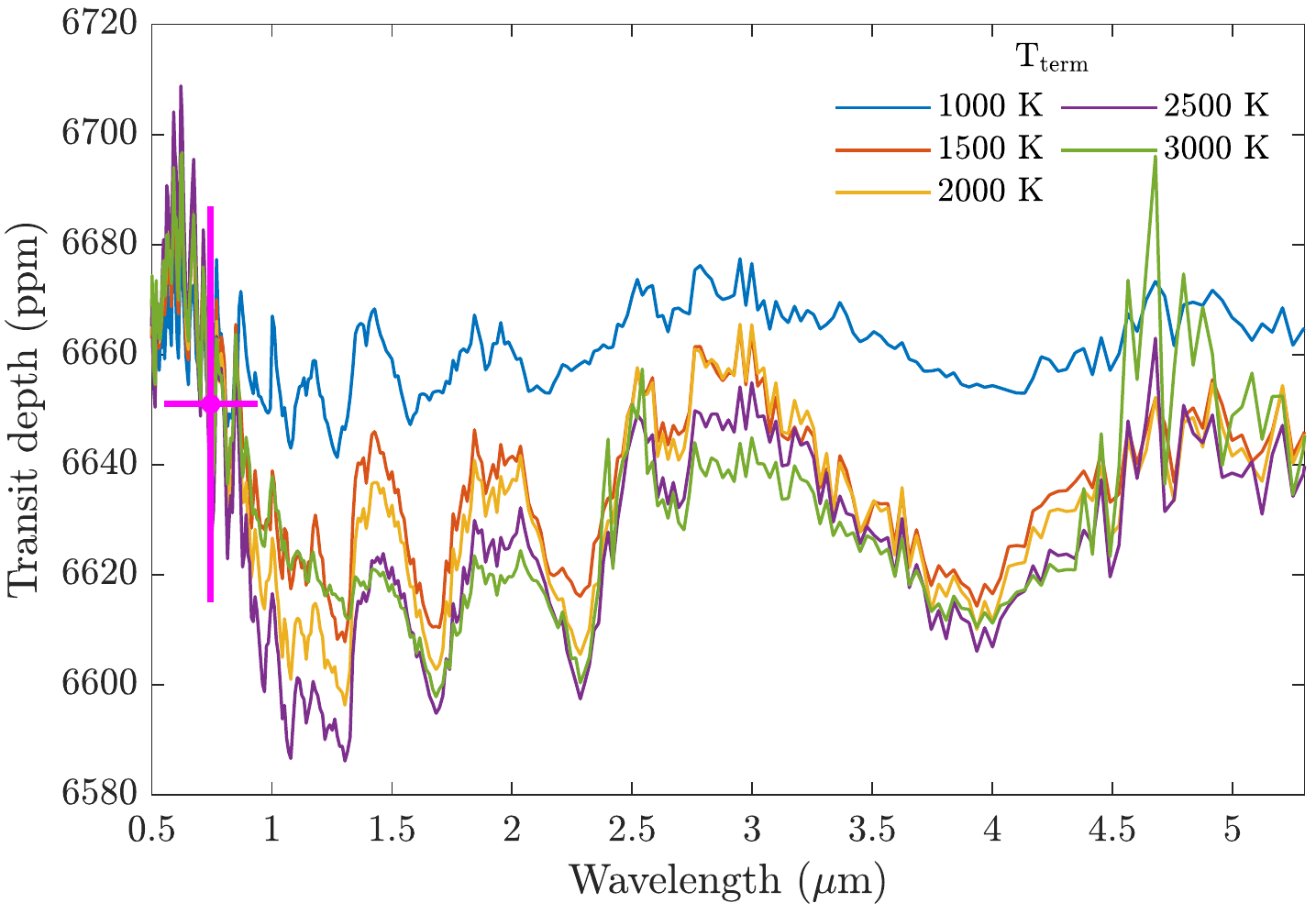}
\caption{A collection of 0.5--5.3 $\mu$m model spectra for TOI-2109b generated by the \texttt{HELIOS} code. Top: predicted dayside emission spectra for various assumed values of the day--night heat recirculation parameter $\epsilon$. The measured \tess- and \Ks-band secondary eclipse depths are shown in pink and blue, respectively, while the band-averaged model fluxes are given by the black points. Middle: nightside emission spectra for different nightside interior temperatures $T_{\mathrm{night}}$, showing large absorption features throughout the near-infrared. Bottom: model transmission spectra for a range of uniform terminator temperatures $T_{\mathrm{term}}$, normalized to the measured \tess-band transit depth.}
\label{fig:helios}
\end{figure}

To explore the prospects for atmospheric characterization in detail, we generated model dayside emission, nightside emission, and transmission spectra of TOI-2109b using the open-source radiative transfer code \texttt{HELIOS} \citep{malik17,malik19}. This model framework takes as input the stellar spectrum (interpolated from PHOENIX models) and the planet--star radius ratio and self-consistently computes the planet's temperature--pressure (TP) profile and emission/transmission spectra under radiative--convective equilibrium. Atmospheric opacities are calculated using the \texttt{HELIOS-K} module \citep{gh15,grimm2021}, while the equilibrium chemistry is computed using \texttt{FastChem} \citep{stock18}. Line lists for over 600 species are included,\footnote{ \url{http://www.opacity.world}} along with continuum opacities from H$^{-}$ \citep{john1988} and collision-induced absorption from H$_2$--H$_2$, H$_2$--He, and H--He \citep{abel11,karman19}. Following the analogous application of \texttt{HELIOS} to modeling the atmosphere of KELT-9b \citep{wong2020kelt9}, we fixed the atmospheric metallicity to solar, set the Bond albedo to zero, and assumed uniform dayside, nightside, and terminator TP profiles. To produce different versions of the spectra for comparison, we varied the day--night heat recirculation parameter $\epsilon$ (ranging from 0 to 1, as defined in \citealt{cowanagol}), nightside interior temperature $T_{\mathrm{night}}$, and terminator temperature $T_{\mathrm{term}}$.

Figure~\ref{fig:helios} shows a compilation of \texttt{HELIOS} model spectra for TOI-2109b assuming various values of $\epsilon$, $T_{\mathrm{night}}$, and $T_{\mathrm{term}}$. The dayside emission spectra are largely featureless throughout the wavelength range 0.5--5.3 $\mu$m. The notable trend is the increasing H$^{-}$ continuum opacity with decreasing $\epsilon$ (i.e., increasing dayside temperature), which is manifested by the emergence of the lobe-shaped excess emission feature shortward of $\sim$1.8 $\mu$m. This behavior has been identified and extensively characterized in previous theoretical work \citep[e.g.,][]{kitzmann2018,lothringer2018,parmentier2018}. The measured \tess-band and \Ks-band secondary eclipse depths are both consistent with the $\epsilon = 0.3$ model, indicating a moderate level of dayside--nightside heat recirculation. The H$^-$ emission feature helps explain the somewhat higher \tess-band brightness temperature when compared to the \Ks-band value (Section~\ref{sec:atm}). The $\epsilon = 0.3$ model has an effective temperature of $\sim$3700 K, in agreement with the measured dayside brightness temperature $T_{\mathrm{day}} = 3631 \pm 69$ K.

Given the poorly constrained nightside brightness temperature of TOI-2109b from our \tess-band nightside flux measurement ($T_{\mathrm{night}} < 2500$ K at $2\sigma$), we considered a wide range of plausible $T_{\mathrm{night}}$ values when generating the \texttt{HELIOS} emission spectra. As illustrated in the middle panel of Figure~\ref{fig:helios}, even moderate-precision spectroscopy in the near-infrared region ($\delta \sim 100$ ppm at $\sim$0.1 $\mu$m resolution) is expected to reveal prominent absorption features due to H$_2$O ($\sim$1.4--1.6, 1.7--2.2, 2.4--3.4 $\mu$m) and CO ($\sim$4--5 $\mu$m) for $T_{\mathrm{night}}$ values greater than $\sim$1500 K. Detecting and modeling these spectral features at high S/N and fine spectral resolution will provide robust constraints on the atmospheric metallicity, C/O, and O/H ratios, with broad implications for the formation and evolution of TOI-2109b (see, for example, the review by \citealt{madhu}).

\begin{figure*}[t]
\centering
\includegraphics[width=0.88\linewidth]{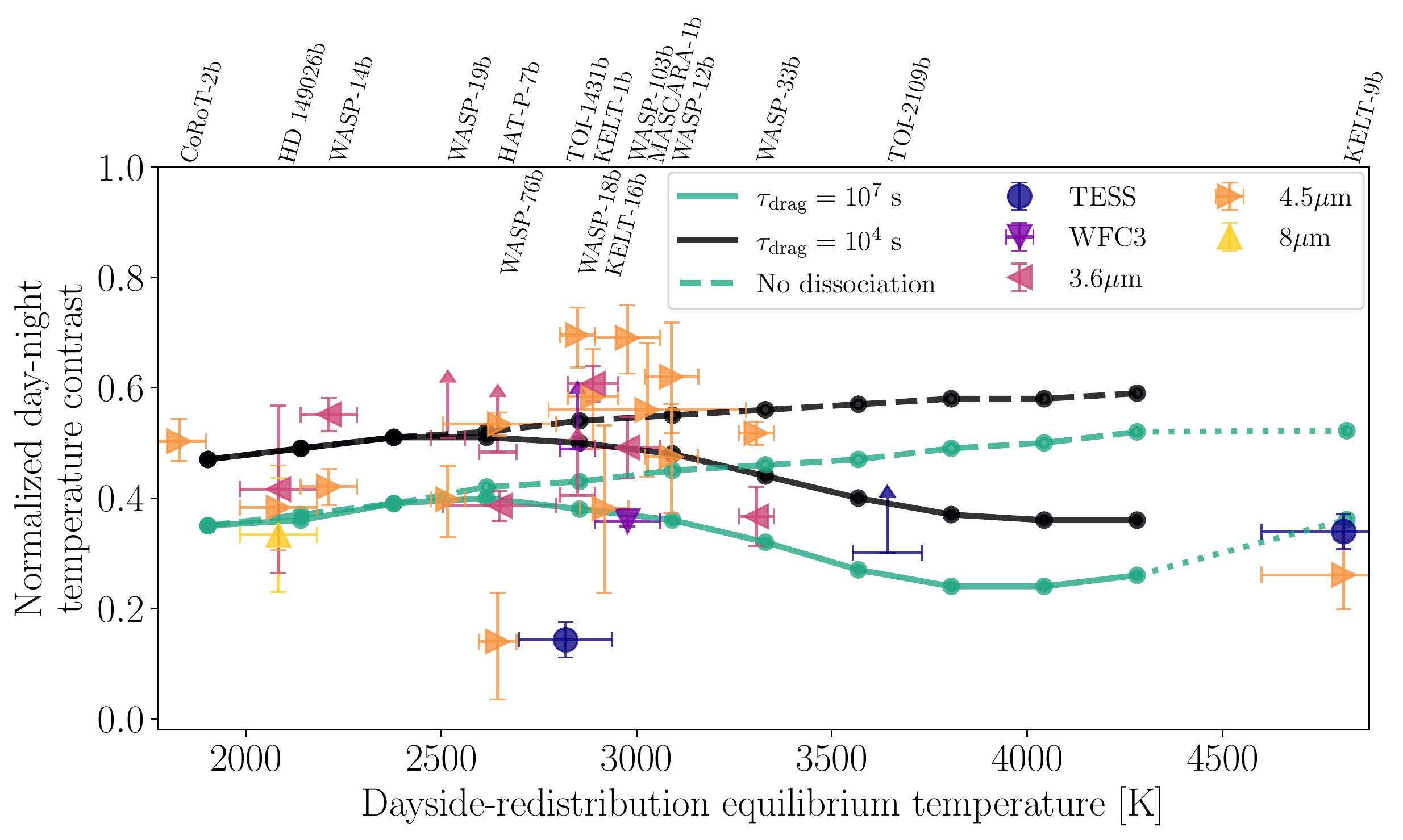}
\caption{Normalized day--night brightness temperature contrast for the sample of hot and ultrahot Jupiters with dayside-redistribution equilibrium temperatures $T_{\mathrm{eq}} > 1700$ K that have published full-orbit phase curves (data points with error bars). Also shown are 3D GCM predictions from the \texttt{MITgcm} (black and green points and curves) with varying drag timescales $\tau_\mathrm{drag}$. Models both including (solid lines) and not including (dashed lines) the effects of hydrogen dissociation/recombination are presented. The GCM simulations for dayside-redistribution equilibrium temperatures of $T_{\mathrm{eq}} < 4500 \mathrm{K}$ are from \cite{tan2019}, while the results for KELT-9b (connected to the suite of cooler GCM runs by dotted lines) were published in \citet{mansfield2020}. The observational data are updated from \citet{tan2019} to include the uniform analysis of Spitzer 4.5 $\mu$m phase curves by \citet{bell2021}, Spitzer 3.6 $\mu$m observations of KELT-1b \citep{beatty2019}, Spitzer 3.6 and 4.5 $\mu$m observations of WASP-76b \citep{may2021}, and the \tess\ phase-curve observations of KELT-9b \citep{wong2020kelt9}, TOI-1431b \citep{toi1431}, and TOI-2109b (this work).}
\label{fig:daynight_teq_mitgcm}
\end{figure*}

The bottom panel of Figure~\ref{fig:helios} shows a set of model transmission spectra for TOI-2109b ranging in $T_{\mathrm{term}}$ from 1000 to 3000 K; all of the curves are normalized to match the measured \tess-band transit depth. Generally, the molecular absorption features observed in the nightside emission spectra are also seen in transmission, though with significantly smaller amplitudes of $\sim$30 ppm or less. The 3000 K model has less prominent H$_2$O absorption features in the near-infrared, reflecting the marked decrease in water abundance due to dissociation at high temperatures. Meanwhile, the negative spectral slope throughout the red-optical and near-infrared is primarily determined by the wavelength-dependent H$^-$ opacity \citep{john1988}. Although the high surface gravity of TOI-2109b leads to relatively weak transmission features, the absorption features in the model spectra are nonetheless detectable by instruments such as NIRSpec on the James Webb Space Telescope (JWST). Moreover, for limb temperatures above 2000--2500 K, cloud formation is strongly disfavored \citep[e.g.,][]{wakeford2017,lothringer2018}, allowing for an unfettered view into the atmosphere. Obtaining a snapshot of the high-altitude chemistry and TP profile along the day--night terminator can provide additional insight into the overall atmospheric composition and dynamics.

In recent years, near-infrared emission spectroscopy across the entire orbital phase has been successfully carried out using the Hubble Space Telescope (HST) for a small handful of exoplanets, including WASP-18b \citep{arcangeli2019}, WASP-43b \citep{stevenson2014}, and WASP-103b \citep{kreidberg2018}. The ability to track the composition and TP profile of the atmosphere at all longitudes makes these studies immensely impactful for our understanding of global atmospheric properties. These data sets enable more sophisticated modeling to constrain the detailed characteristics of day--night atmospheric circulation and chemical gradients across the surface. Looking to the near future, the enhanced capabilities of JWST have placed spectroscopic phase curves at the forefront of efforts at atmospheric characterization for gas-giant exoplanets. TOI-2109b is an extremely promising candidate for phase-resolved emission spectroscopy. In particular, the planet's equilibrium temperature makes it an attractive target for probing fundamental trends in gas-giant atmospheric dynamics. 

Figure~\ref{fig:daynight_teq_mitgcm} shows the measured day--night brightness temperature contrast $1 - T_\mathrm{night}/T_\mathrm{day}$ for hot and ultrahot Jupiters, alongside theoretical predictions from the three-dimensional global circulation model (GCM) \texttt{MITgcm}, as computed in \citet{tan2019} and \citet{mansfield2020}. Two different levels of atmospheric drag were considered, parameterized by the Rayleigh drag timescale $\tau_{\mathrm{drag}}$: weak ($\tau_\mathrm{drag} = 10^7~\mathrm{s}$) and strong ($\tau_\mathrm{drag} = 10^4~\mathrm{s}$). Scenarios with and without hydrogen dissociation/recombination were modeled in order to assess its dynamical impact on the global heat transport. These effects were incorporated into the GCM by tracking the local atomic hydrogen mixing ratio and computing the recombination-driven heating/cooling that results when the atomic hydrogen mixing ratio decreases/increases, along with local changes in specific heat and mean molecular weight due to the spatially varying ratio of atomic to molecular hydrogen \citep{tan2019}. For the GCM simulations at dayside-redistribution equilibrium temperatures below 4500 K, which are taken directly from \citet{tan2019}, we uniformly assumed a planetary radius of 1.47 \rjup, a surface gravity of 11 \mss, and a rotation period of 2.43 days. Meanwhile, the GCM simulations of KELT-9b used a radius of 1.89 \rjup, a surface gravity of 19.95 \mss, and a rotation period of 1.48 days \citep{mansfield2020}.

Crucially, Figure~\ref{fig:daynight_teq_mitgcm} shows that the majority of extant phase-curve observations of hot and ultrahot Jupiters cannot distinguish between GCM predictions with and without hydrogen dissociation/recombination. This is because the impacts of hydrogen dissociation/recombination on day--night heat transport are largest for gas giants with dayside-redistribution equilibrium temperatures in the range $3500~\mathrm{K} \lesssim T_\mathrm{eq} \lesssim 4500~\mathrm{K}$, where the hydrogen fraction varies most sharply with temperature and the spatial variations in the atomic hydrogen mixing ratio are large \citep{bell2018,tan2019}. At the extreme irradiation level of KELT-9b, the effects of hydrogen dissociation/recombination are largely ``saturated,'' yielding a small day--night brightness temperature contrast that is consistent with observations \citep{mansfield2020,wong2020kelt9,bell2021}. As discussed in Section \ref{sec:atm}, TOI-2109b is the first ultrahot Jupiter discovered in the temperature range between KELT-9b and the cooler gas giants with $T_{\mathrm{eq}} < 3500$ K, i.e., in the transitional zone where the differential effects of hydrogen dissociation/recombination on global heat transport are largest. Future spectroscopic phase-curve observations of TOI-2109b will empirically test our current understanding of the dynamical impacts of hydrogen dissociation/recombination on the global heat transport of ultrahot Jupiters.

TOI-2109b also presents an enticing case for studying atmospheric variability. A handful of previous works have reported modulations in the phase-curve amplitude and phase offset on $\sim$10 day timescales for a few exoplanets (e.g., \citealt{armstrong2016}; \citealt{jackson2019}; but see also \citealt{lally2020}). The rapid rotation of TOI-2109b due to its ultrashort orbit can make its atmosphere particularly susceptible to hydrodynamic instabilities, which generate zonal propagation of waves at mid-to-high latitudes \citep{tan2019,tan2020}. Other mechanisms that may yield orbit-to-orbit evolution in the global atmospheric properties include transient waves near the equator \citep{komacek2020} and time-varying magnetohydrodynamic drag arising from the coupling of the planet's magnetic field with the partially ionized dayside atmosphere \citep{rogers2017}. The possibility of phase-curve variability can be explored when \tess\ returns to observe this system in the extended mission, and potentially in additional extended missions if approved by NASA.

\section{Summary}
\label{sec:sum}
In this paper, we have presented the discovery of the shortest-period gas-giant exoplanet yet known. TOI-2109b is a $5.02 \pm 0.75$ \mjup, $1.347 \pm 0.047$ \rjup\ ultrahot Jupiter on a $0.67247414 \pm 0.00000028$ day orbit around a rapidly rotating F-type star with $\teff = 6530^{+160}_{-150}$ K, $M_{*} = 1.447^{+0.075}_{-0.078}$ \msun, and $R_{*} = 1.698^{+0.062}_{-0.057}$ \rsun. By combining the \tess\ light curve with a large number of ground-based transit observations, we measured the transit geometry to high precision: $b = 0.7481 \pm 0.0073$, $a/R_* = 2.268 \pm 0.021$, and $i = 70\overset{\circ}{.}74 \pm 0\overset{\circ}{.}37$. Spectroscopic transit observations carried out using the TRES instrument recovered the Doppler shadow of the planet, definitively rejecting blended binary false positive scenarios and revealing a well-aligned orbit with a sky-projected obliquity of $\lambda = 1\overset{\circ}{.}7 \pm 1\overset{\circ}{.}7$.

Our analysis of the full-orbit \tess\ photometry produced a detailed characterization of the phase-curve modulations and a measured secondary eclipse depth of $731 \pm 46$ ppm. In addition, we detected two stellar variability signals with characteristic periods of $0.61395 \pm 0.00055$ and $0.9674 \pm 0.0013$ days. TOI-2109's atmospheric brightness modulation shows a peak-to-peak amplitude of $724 \pm 38$ ppm, a marginal $\sim$4$^{\circ}$ eastward shift in the location of the dayside hotspot, and a nightside flux consistent with zero. Meanwhile, the host star displays a strong ellipsoidal distortion signal with a semiamplitude of $245 \pm 19$ ppm, which is statistically identical to the theoretically predicted value based on the measured planet mass. Combining the \tess-band secondary eclipse depth with a \Ks-band measurement obtained with the Palomar/WIRC instrument, we measured a dayside brightness temperature of $3631 \pm 69$ K, making TOI-2109b the second-hottest known exoplanet behind KELT-9b.

The extremely short orbit of TOI-2109b and the intense planet--star gravitational interaction make the system an ideal target for searches for tidal orbital decay. For a given stellar tidal quality factor, TOI-2109b has by far the fastest predicted decay rate of any planetary mass companion, and future transit and secondary eclipse measurements may detect a decaying orbit within a few years. Likewise, the strong tidal forces exerted on the planet's atmosphere may lead to significant mass loss via Roche lobe overflow.

Given the high levels of stellar irradiation and the brightness of its host star, TOI-2109b is a prime candidate for intensive atmospheric characterization, particularly with secondary eclipse and phase-resolved emission spectroscopy. With an equilibrium temperature that falls within the wide gap between the hottest end-member KELT-9b and the cooler well-characterized ultrahot Jupiters (e.g., WASP-12b and WASP-33b), TOI-2109b is optimally situated to provide potentially transformative insights into the impacts of hydrogen dissociation/recombination on the global energy budget and atmospheric circulation of these extreme worlds. The legacy established by previous HST spectroscopic observations of gas giants in transmission and emission and the advent of the JWST era present a plethora of fruitful opportunities for follow-up studies to explore the detailed composition and dynamics of this unique ultrahot Jupiter.
\\

Funding for the \tess\ mission is provided by NASA's Science Mission directorate.
This paper includes data collected by the \tess\ mission, which are publicly available from the Mikulski Archive for Space Telescopes (MAST).
Resources supporting this work were provided by the NASA High-end Computing (HEC) Program through the NASA Advanced Supercomputing (NAS) Division at Ames Research Center for the production of the SPOC data products.

This work makes use of observations from the LCOGT network. Part of the LCOGT telescope time was granted by NOIRLab through the Mid-Scale Innovations Program (MSIP). MSIP is funded by the National Science Foundation (NSF).

This article is partly based on observations made with the MuSCAT2 instrument, developed by ABC, at Telescopio Carlos S\'anchez operated on the island of Tenerife by the Instituto de Astrof\'isica de Canarias in the Spanish Observatorio del Teide. This paper is also based on observations made with the MuSCAT3 instrument, developed by the Astrobiology Center and under financial support by JSPS KAKENHI (JP18H05439) and JST PRESTO (JPMJPR1775), at Faulkes Telescope North on Maui, HI, operated by the Las Cumbres Observatory. This work is partly supported by JSPS KAKENHI grant numbers JP17H04574, JP18H05439, JP18H05442, JP15H02063, JP21K13975, JP22000005, JST PRESTO grant number JPMJPR1775, and the Astrobiology Center of the National Institutes of Natural Sciences (grant numbers AB031010 and AB031014).

We thank the Palomar Observatory team, particularly Paul~Nied and Kevin~Rykoski, for enabling the \Ks-band secondary eclipse observations and facilitating remote operations on the Hale 200\arcsec\ Telescope.

Some of the observations in this paper made use of the High-Resolution Imaging instrument ‘Alopeke. ‘Alopeke was funded by the NASA Exoplanet Exploration Program and built at the NASA Ames Research Center by Steve~B.~Howell, Nic~Scott, Elliott~P.~Horch, and Emmett~Quigley. Data were reduced using a software pipeline originally written by Elliott~Horch and Mark~Everett. ‘Alopeke was mounted on the Gemini-North telescope of the international Gemini Observatory, a program of NSF’s OIR Lab, which is managed by the Association of Universities for Research in Astronomy (AURA) under a cooperative agreement with the National Science Foundation on behalf of the Gemini partnership: the National Science Foundation (United States), National Research Council (Canada), Agencia Nacional de Investigación y Desarrollo (Chile), Ministerio de Ciencia, Tecnología e Innovación (Argentina), Ministério da Ciência, Tecnologia, Inovações e Comunicações (Brazil), and Korea Astronomy and Space Science Institute (Republic of Korea).

This paper is partially based on observations made with the Nordic Optical Telescope, operated by the Nordic Optical Telescope Scientific Association at the Observatorio del Roque de los Muchachos, La Palma, Spain, of the Instituto de Astrof\'isica de Canarias.

This research has made use of the NASA Exoplanet Archive, which is operated by the California Institute of Technology, under contract with NASA under the Exoplanet Exploration Program.

This work has made use of data from the European Space Agency (ESA) mission Gaia (\url{https://www.cosmos.esa.int/gaia}), processed by the Gaia Data Processing and Analysis Consortium (DPAC, \url{https://www.cosmos.esa.int/web/gaia/dpac/consortium}). Funding for the DPAC has been provided by national institutions, in particular the institutions participating in the Gaia Multilateral Agreement.

I.W. and T.D.K. acknowledge funding from the 51 Pegasi b Fellowship in Planetary Astronomy sponsored by the Heising--Simons Foundation. S.V. is supported by an NSF Graduate Research Fellowship and the Paul \& Daisy Soros Fellowship for New Americans. H.A.K. acknowledges support from NSF CAREER grant 1555095. M.T. is supported by JSPS KAKENHI grant numbers JP18H05442, JP15H02063, and JP22000005.

\facilities{\tess, FLWO/KeplerCam, ULMT, LCOGT/McD, LCOGT/SSO, LCOGT/SAAO, LCOGT/CTIO, MLO, TCS/MuSCAT2, WBRO, Grand-Pra Observatory, FTN/MuSCAT3, Palomar/WIRC, Gemini-North/'Alopeke, FLWO/TRES, NOT/FIES}

\software{
\texttt{AstroImageJ} \citep{collins17},
\texttt{astropy} \citep{astropy2018},
\texttt{BANZAI} \citep{mccully2018},
\texttt{batman} \citep{batman},
\texttt{emcee} \citep{emcee},
\texttt{EXOFASTv2} \citep{eastman2013,eastman2019,eastman2017},
\texttt{ExoTEP} \citep{benneke2019,wong2020hatp12},
\texttt{FastChem} \citep{stock18},
\texttt{HELIOS} \citep{malik17,malik19},
\texttt{HELIOS-K} \citep{gh15},
\texttt{matplotlib} \citep{hunter2007}, 
\texttt{numpy} \citep{numpy}, 
\texttt{photutils} \citep{photutils},
\texttt{radvel} \citep{radvel},
\texttt{scipy} \citep{scipy},
\texttt{Tapir} \citep{jensen2013}.}



\end{document}